\documentclass[11pt]{article}
\usepackage{amsfonts,amsbsy}
\begin{document}

%\usepackage{amsfonts,amsbsy}
%\voffset-40mm
%\hoffset-20mm
%\topmargin30mm
%\headheight6mm
%\headsep9mm
%\textheight227mm
%\oddsidemargin27mm
%\evensidemargin27mm
%\textwidth160mm
%\overfullrule3pt
%\hbadness=100000

%%%%%%%%%%%%%%%%%%%%%%%%%
% Macros for harvmac -> LaTeX
%
%
\catcode`\@=11 % This allows us to modify PLAIN macros.

%% Use of \eqn and ignore \ref and \cite of latex
%% Use of Latex convention of sectioning.
%
%\global\newcount\chapno \global\chapno=0
\global\newcount\secno \global\secno=0
\global\newcount\meqno \global\meqno=1

\def\newsec#1{\global\advance\secno by1
\global\subsecno=0\eqnres@t
\section{#1}}
\def\eqnres@t{\xdef\secsym{\the\secno.}\global\meqno=1}
\def\sequentialequations{\def\eqnres@t{\bigbreak}}\xdef\secsym{}
\global\newcount\subsecno \global\subsecno=0
\def\subsec#1{\global\advance\subsecno by1
\subsection{#1}}
\def\subsubsec#1{\subsubsection{#1}}

\def\draftmode{\message{ DRAFTMODE }
%\def\draftdate{{\rm preliminary draft:
%\number\month/\number\day/\number\yearltd\ \ \hourmin}}%
%\headline={\hfil\draftdate}
\writelabels
{\count255=\time\divide\count255 by 60 \xdef\hourmin{\number\count255}
\multiply\count255 by-60\advance\count255 by\time
\xdef\hourmin{\hourmin:\ifnum\count255<10 0\fi\the\count255}}}
%   use \nolabels to get rid of eqn, ref, and fig labels in draft mode
\def\nolabels{\def\wrlabeL##1{}\def\eqlabeL##1{}\def\reflabeL##1{}}
\def\writelabels{\def\wrlabeL##1{\leavevmode\vadjust{\rlap{\smash%
{\line{{\escapechar=` \hfill\rlap{\tt\hskip.03in\string##1}}}}}}}%
\def\eqlabeL##1{{\escapechar-1\rlap{\tt\hskip.05in\string##1}}}%
\def\reflabeL##1{\noexpand\llap{\noexpand\sevenrm\string\string\string##1}}}

\nolabels

\def\eqn#1#2{
\xdef #1{(\secsym\the\meqno)}%\writedef{#1\leftbracket#1}%
\global\advance\meqno by1
$$#2\eqno#1\eqlabeL#1
$$}
% We want \eqalign back

\def\eqalign#1{\null\,\vcenter{\openup\jot\m@th
\ialign{\strut\hfil$\displaystyle{##}$&$\displaystyle{{}##}$\hfil
\crcr#1\crcr}}\,}

%-------------------
% footnote
\def\foot#1{\footnote{#1}}
%--------------------

%-----------------------------
% Generate refs.tmp and   \listrefs generate
%the standard latex referencing
%
\global\newcount\refno \global\refno=1
\newwrite\rfile
\def\ref{[\the\refno]\nref}
\def\nref#1{\xdef#1{[\the\refno]}%\writedef{#1\leftbracket#1}%
\ifnum\refno=1\immediate\openout\rfile=refs.tmp\fi
\global\advance\refno by1\chardef\wfile=\rfile\immediate
\write\rfile{\noexpand\bibitem{\string#1}}\findarg}
% horrible hack to sidestep tex \write limitation
\def\findarg#1#{\begingroup\obeylines\newlinechar=`\^^M\pass@rg}
{\obeylines\gdef\pass@rg#1{\writ@line\relax #1^^M\hbox{}^^M}%
\gdef\writ@line#1^^M{\expandafter\toks0\expandafter{\striprel@x #1}%
\edef\next{\the\toks0}\ifx\next\em@rk\let\next=\endgroup\else\ifx\next\empty

\else\immediate\write\wfile{\the\toks0}\fi\let\next=\writ@line\fi\next\relax

}}
\def\striprel@x#1{} \def\em@rk{\hbox{}}
\def\lref{\begingroup\obeylines\lr@f}
\def\lr@f#1#2{\gdef#1{\ref#1{#2}}\endgroup\unskip}
\def\semi{;\hfil\break}
\def\addref#1{\immediate\write\rfile{\noexpand\item{}#1}}
\def\listrefs{

}
\def\startrefs#1{\immediate\openout\rfile=refs.tmp\refno=#1}
\def\xref{\expandafter\xr@f}\def\xr@f[#1]{#1}
\def\refs#1{\count255=1[\r@fs #1{\hbox{}}]}
\def\r@fs#1{\ifx\und@fined#1\message{reflabel \string#1 is
undefined.}%
\nref#1{need to supply reference \string#1.}\fi%
\vphantom{\hphantom{#1}}\edef\next{#1}\ifx\next\em@rk\def\next{}%
\else\ifx\next#1\ifodd\count255\relax\xref#1\count255=0\fi%
\else#1\count255=1\fi\let\next=\r@fs\fi\next}
\newwrite\lfile
{\escapechar-1\xdef\pctsign{\string\%}\xdef\leftbracket{\string\{}
\xdef\rightbracket{\string\}}\xdef\numbersign{\string\#}}
\def\writedefs{\immediate\openout\lfile=labeldefs.tmp
\def\writedef##1{%
\immediate\write\lfile{\string\def\string##1\rightbracket}}}
\def\writestop{\def\writestoppt{\immediate\write\lfile{\string\pageno%
\the\pageno\string\startrefs\leftbracket\the\refno\rightbracket%
\string\def\string\secsym\leftbracket\secsym\rightbracket%
\string\secno\the\secno\string\meqno\the\meqno}\immediate\closeout\lfile}}

\def\writestoppt{}\def\writedef#1{}

\catcode`\@=12 % at signs are no longer letters
%

%Finally the standard harvmac macros
%
\def\noblackbox{\overfullrule=0pt}
\hyphenation{anom-aly anom-alies coun-ter-term coun-ter-terms
}
\def\inv{^{\raise.15ex\hbox{${\scriptscriptstyle -}$}\kern-.05em 1}}
\def\dup{^{\vphantom{1}}}
\def\Dsl{\,\raise.15ex\hbox{/}\mkern-13.5mu D}%this one can be subscripted
\def\dsl{\raise.15ex\hbox{/}\kern-.57em\partial}
\def\del{\partial}
\def\Psl{\dsl}
\def\tr{{\rm tr}} \def\Tr{{\rm Tr}}
\font\bigit=cmti10 scaled \magstep1
\def\biglie{\hbox{\bigit\$}} %pound sterling
\def\lspace{\ifx\answ\bigans{}\else\qquad\fi}
\def\lbspace{\ifx\answ\bigans{}\else\hskip-.2in\fi} % $$\lbspace...$$
\def\boxeqn#1{\vcenter{\vbox{\hrule\hbox{\vrule\kern3pt\vbox{\kern3pt
\hbox{${\displaystyle #1}$}\kern3pt}\kern3pt\vrule}\hrule}}}
% matters of taste
\def\tilde{\widetilde} \def\bar{\overline} \def\hat{\widehat}
%
% some sample definitions
\def\e#1{{\rm e}^{^{\textstyle#1}}}
\def\grad#1{\,\nabla\!_{{#1}}\,}
\def\gradgrad#1#2{\,\nabla\!_{{#1}}\nabla\!_{{#2}}\,}
\def\ph{\varphi}
\def\psibar{\overline\psi}
\def\om#1#2{\omega^{#1}{}_{#2}}
\def\vev#1{\langle #1 \rangle}
\def\lform{\hbox{$\sqcup$}\llap{\hbox{$\sqcap$}}}
\def\darr#1{\raise1.5ex\hbox{$\leftrightarrow$}\mkern-16.5mu #1}
\def\lie{\hbox{\it\$}} %pound sterling
\def\ha{{1\over2}}
\def\half{{\textstyle{1\over2}}} %puts a small half in a displayed eqn
\def\roughly#1{\raise.3ex\hbox{$#1$\kern-.75em\lower1ex\hbox{$\sim$}}}

%
% End of Macro
%%%%%%%%%%%%%%%%%%%%%%%%%%%%%%%%%%
%%%%%%%%%%%%%%%%%%%%%%%%%%%%%%%%%%%
%%%%%%%%%%%%%%%%%%%%%%%%%%%%%%%%%

% Bold Math Italic

\font\tencmmib=cmmib10  \skewchar\tencmmib='177
\font\sevencmmib=cmmib7 \skewchar\sevencmmib='177
\font\fivecmmib=cmmib5 \skewchar\fivecmmib='177
\newfam\cmmibfam
\textfont\cmmibfam=\tencmmib
\scriptfont\cmmibfam=\sevencmmib
\scriptscriptfont\cmmibfam=\fivecmmib
\def\cmmib#1{{\fam\cmmibfam\relax#1}}
%%%%%%%%%%%%%%

%---------------------------------------------------
\def\a{\alpha}
\def\b{\beta}
\def\c{\chi}
\def\d{\delta}  \def\D{\Delta}
\def\e{\varepsilon} \def\ep{\epsilon}
\def\f{\phi}  \def\F{\Phi}
\def\g{\gamma}  \def\G{\Gamma}
\def\k{\kappa}
\def\l{\lambda}  \def\L{\Lambda}
\def\m{\mu}
\def\n{\nu}
\def\r{\rho}
\def\vr{\varrho}
\def\o{\omega}  \def\O{\Omega}
\def\p{\psi}  \def\P{\Psi}
\def\s{\sigma}  \def\S{\Sigma}
\def\th{\theta}  \def\vt{\vartheta}
\def\t{\tau}
\def\w{\varphi}
\def\x{\xi}
\def\z{\zeta}
\def\CA{{\cal A}}
\def\CB{{\cal B}}
\def\CC{{\cal C}}
\def\CD{{\cal D}}
\def\CE{{\cal E}}
\def\CF{{\cal F}}
\def\CG{{\cal G}}
\def\CH{{\cal H}}
\def\CI{{\cal I}}
\def\CJ{{\cal J}}
\def\CK{{\cal K}}
\def\CL{{\cal L}}
\def\CM{{\cal M}}
\def\CN{{\cal N}}
\def\CO{{\cal O}}
\def\CP{{\cal P}}
\def\CQ{{\cal Q}}
\def\CR{{\cal R}}
\def\CS{{\cal S}}
\def\CT{{\cal T}}
\def\CU{{\cal U}}
\def\CV{{\cal V}}
\def\CW{{\cal W}}
\def\CX{{\cal X}}
\def\CY{{\cal Y}}
\def\CZ{{\cal Z}}
\def\EJ{\mathfrak{J}}
\def\EM{\mathfrak{M}}
\def\ES{\mathfrak{S}}
\def\BB{\mathib{B}}
\def\BC{\mathib{C}}
\def\BH{\mathib{H}}
%%%%%%%%%%%%%%%%
\def\V{\mathbb{V}}
\def\E{\mathbb{E}}
\def\R{\mathbb{R}}
\def\C{\mathbb{C}}
\def\Z{\mathbb{Z}}
%-----------------------------------------------------
%Math
\def\rd{\partial}
\def\grad#1{\,\nabla\!_{{#1}}\,}
\def\gradd#1#2{\,\nabla\!_{{#1}}\nabla\!_{{#2}}\,}
\def\om#1#2{\omega^{#1}{}_{#2}}
\def\vev#1{\langle #1 \rangle}
\def\darr#1{\raise1.5ex\hbox{$\leftrightarrow$}
\mkern-16.5mu #1}
\def\Ha{{1\over2}}
\def\ha{{\textstyle{1\over2}}}
\def\fr#1#2{{\textstyle{#1\over#2}}}
\def\Fr#1#2{{#1\over#2}}
\def\rf#1{\fr{\rd}{\rd #1}}
\def\rF#1{\Fr{\rd}{\rd #1}}
\def\df#1{\fr{\d}{\d #1}}
\def\dF#1{\Fr{\d}{\d #1}}
\def\DDF#1#2#3{\Fr{\d^2 #1}{\d #2\d #3}}
\def\DDDF#1#2#3#4{\Fr{\d^3 #1}{\d #2\d #3\d #4}}
\def\ddF#1#2#3{\Fr{\d^n#1}{\d#2\cdots\d#3}}
\def\fs#1{#1\!\!\!/\,}   %Feynman dagger
\def\Fs#1{#1\!\!\!\!/\,} %Feynman dagger
\def\roughly#1{\raise.3ex\hbox{$#1$\kern-.75em
\lower1ex\hbox{$\sim$}}}
\def\ato#1{{\buildrel #1\over\longrightarrow}}
\def\up#1#2{{\buildrel #1\over #2}}
\def\opname#1{\mathop{\kern0pt{\rm #1}}\nolimits}
\def\tr{\opname{Tr}}
\def\Re{\opname{Re}}
\def\Im{\opname{Im}}
\def\End{\opname{End}}
\def\dim{\opname{dim}}
\def\vol{\opname{vol}}
\def\group#1{\opname{#1}}
\def\SU{\group{SU}}
\def\U{\group{U}}
\def\SO{\group{SO}}
\def\pr{\prime}
\def\ppr{{\prime\prime}}
\def\bs{\mathib{s}}
\def\bbs{\bar\mathib{s}}
\def\Dp{\rd_{\!A}}
\def\Dpp{\bar\rd_{\!A}}
\def\Da{d_{\!A}}
\def\bari{\bar\imath}
\def\barj{\bar\jmath}
% arrows
\def\mapr{\!\smash{\mathop{\longrightarrow}\limits^{\bs_+}}\!}
\def\mapl{\!\smash{\mathop{\longleftarrow}\limits^{\bs_-}}\!}
\def\mapbr{\!\smash{\mathop{\longrightarrow}\limits^{\bbs_+}}\!}
\def\mapbl{\!\smash{\mathop{\longleftarrow}\limits^{\bbs_-}}\!}
\def\mapd{\Big\downarrow\rlap{$\vcenter{\hbox{$\scriptstyle\bbs_-$}}$}}

\def\mapu{\Big\uparrow\rlap{$\vcenter{\hbox{$\scriptstyle\bbs_+$}}$}}
\def\maprd{\rlap{\lower.3ex\hbox{$\scriptstyle\bs_+$}}\searrow}
\def\mapld{\swarrow\!\!\!\rlap{\lower.3ex\hbox{$\scriptstyle\bs_-$}}}
\def\ne{\nearrow}
\def\se{\searrow}
\def\nw{\nwarrow}
\def\sw{\swarrow}
\def\etal{et al.}
\def\git{/\kern-.25em/}
\def\Ker{\hbox{Ker}\;}
%%%%%%%%%%%%%%%%%%%%%%%%%%%%%%%%%%%%%%%
%Journals
\def\cmp#1#2#3{Comm.\ Math.\ Phys.\ {{\bf #1}} {(#2)} {#3}}
\def\pl#1#2#3{Phys.\ Lett.\ {{\bf #1}} {(#2)} {#3}}
\def\np#1#2#3{Nucl.\ Phys.\ {{\bf #1}} {(#2)} {#3}}
\def\prd#1#2#3{Phys.\ Rev.\ {{\bf #1}} {(#2)} {#3}}
\def\prl#1#2#3{Phys.\ Rev.\ Lett.\ {{\bf #1}} {(#2)} {#3}}
\def\ijmp#1#2#3{Int.\ J.\ Mod.\ Phys.\ {{\bf #1}} {(#2)} {#3}}
\def\jmp#1#2#3{J.\ Math.\ Phys.\ {{\bf #1}} {(#2)} {#3}}
\def\prp#1#2#3{Phys.\ Rep.\ {{\bf #1}} {(#2)} {#3}}
\def\ap#1#2#3{Ann.\ Phys.\ {{\bf #1}} {(#2)} {#3}}
\def\cqg#1#2#3{Class. Quantum Grav.\ {{\bf #1}} {(#2)} {#3}}
\def\ptrsls#1#2#3{Philos.\ Trans.\  Roy.\ Soc.\ London{{\bf #1}}
{(#2)}
{#3}}
\def\prsls#1#2#3{Proc.\ Roy.\ Soc.\ London Ser.\ {{\bf #1}} {(#2)}
{#3}}
\def\pnas#1#2#3{Proc.\ Nat.\ Acad.\ Sci.\ USA.\ {{\bf #1}} {(#2)}
{#3}}
\def\jdg#1#2#3{J.\ Differ.\ Geom.\ {{\bf #1}} {(#2)} {#3}}
\def\top#1#2#3{Topology {{\bf #1}} {(#2)} {#3}}
\def\zp#1#2#3{Z.\ Phys.\ {{\bf #1}} {(#2)} {#3}}
\def\am#1#2#3{Ann.\ Math.\ {{\bf #1}} {(#2)} {#3}}
\def\mm#1#2#3{Manuscripta \ Math.\ {{\bf #1}} {(#2)} {#3}}
\def\ma#1#2#3{Math.\ Ann.\ {{\bf #1}} {(#2)} {#3}}
\def\ivm#1#2#3{Invent.\ Math.\ {{\bf #1}} {(#2)} {#3}}
\def\plms#1#2#3{Proc.\ London Math.\ Soc.\ {{\bf #1}} {(#2)} {#3}}
\def\dmj#1#2#3{Duke Math.\  J.\ {{\bf #1}} {(#2)} {#3}}
\def\bams#1#2#3{Bull.\ Am.\ Math.\ Soc.\ {{\bf #1}} {(#2)} {#3}}
\def\jams#1#2#3{J.\ Am.\ Math.\ Soc.\ {{\bf #1}} {(#2)} {#3}}
\def\jgp#1#2#3{J.\ Geom.\ Phys.\ {{\bf #1}} {(#2)} {#3}}
\def\ihes#1#2#3{Publ.\ Math.\ I.H.E.S. \ {{\bf #1}} {(#2)} {#3}}

%%%%%%%%%%%%%%%%%%%%

\def\submit{\baselineskip=20pt plus 2pt minus 2pt}
\def\lin#1{\medskip\noindent {${\hbox{\it #1}}$}\medskip }
\def\linn#1{\noindent $\bullet$ {\it #1} \par}
%%%%%%%%%%%%%%%%%%%%%%%%%%%%%%%%%%%%%%%%%%%%%%%%%%%%%%%%%%%%%%%%%%%%%%%%%%

\makeatother

%%%%%%%%%%%%%%%% References %%%%%%%%%%%%%%%%%%%%%%%%%%%%%%%%%%%%%%%

%%%%%%%%%%%%%%%% End References %%%%%%%%%%%%%%%%%%%%%%%%%%%%%%%%%%%%

%%%%%%%%%%%%%%%% Titlepage %%%%%%%%%%%%%%%%%%%%%%%%%%%%%%%%%%%%%%%%%

\def\etag#1{\eqnn#1\eqno#1}
\def\subsubsec#1{{\medbreak\smallskip\noindent{\it #1 } }}

\def\inbar{\,\vrule height1.5ex width.4pt depth0pt}
\def\IC{\relax{\hbox{$\inbar\kern-.3em{\rm C}$}}}
\def\IR{\relax{\rm I\kern-.18em R}}
\font\cmss=cmss10 \font\cmsss=cmss10 at 7pt
\def\IZ{\relax\ifmmode\mathchoice
{\hbox{\cmss Z\kern-.4em Z}}{\hbox{\cmss Z\kern-.4em Z}}
{\lower.9pt\hbox{\cmsss Z\kern-.4em Z}}
{\lower1.2pt\hbox{\cmsss Z\kern-.4em Z}}\else{\cmss Z\kern-.4em Z}\fi}

%%%%%%%%%%%%%%%%%%%%%%%%%%%%%%%%%%%%%%%%%%%%%%%%%%

\thispagestyle{empty}
\begin{flushright}
\textsc{CU-TP-990}\\
%\textsc{The Final Draft}
\end{flushright}
\vskip5mm

\begin{center}
{\LARGE \sc Topological Open $p$-Branes}\\[2mm]

\vskip 1cm
{Jae-Suk Park}
%\foot{Supported by DOE Grant \# DE-FG02-92ER40699}
\\[8mm]
{\itshape
Department of Physics, Columbia University\\
New York, N.Y.~10027, U.S.A.}\\
{\tt jspark@phys.columbia.edu}\\
[20mm]

\end{center}       

By exploiting the BV quantization of topological bosonic open membrane,
we argue that flat $3$-form $C$-field leads to deformations
of the algebras of multi-vectors on the Dirichlet-brane world-volume
as $2$-algebras.
This would  shed some new light on geometry of
M-theory  $5$-brane and associated decoupled theories.
We show that, in general,  
topological  open $p$-brane has a structure of $(p+1)$-algebra 
in the bulk, while a structure of $p$-algebra in the boundary.
The bulk/boundary correspondences are exactly as of the generalized
Deligne conjecture (a theorem of Kontsevich) in the algebraic world
of $p$-algebras. It also imply that the algebras of quantum observables 
of $(p-1)$-brane are ``close to'' the algebras of  its classical observables
as $p$-algebras.  We interpret above as deformation
quantization of  $(p-1)$-brane, generalizing the $p=1$ case. 
We  argue that there is such quantization based on  the direct relation
between BV master equation and Ward identity of the bulk topological
theory. The path integral of the theory will lead to the explicit formula.
We also discuss some applications to topological strings  and conjecture 
that the homological mirror symmetry has further generalizations 
to the categories of $p$-algebras.

\vskip 2cm

\leftline{Dec. 15, 2000}

\setcounter{page}{0}

%\draftmode
\vfill
\eject

\def\bos#1{\boldsymbol{#1}}
\def\tQ{\tilde{\bos{Q}}}

\renewcommand{\baselinestretch}{1.45}

\newsec{Introduction}

The discovery of D-branes -- extended objects carrying RR charge,
has greatly enhanced our understanding
of string theory \cite{P}. D-branes can be realized as certain
Dirichlet boundary condition of the fundamental open string.
One can also say that the fundamental open strings describe excitations
of the D-brane. The low energy dynamics of a D-brane is described by
the maximally supersymmetric Yang-Mills (SYM)
theory on the D-brane worldvolume.
The open string can naturally be coupled to flat NS 2-form B-fields.
It is by now well-known that there are suitable decoupling limits
of the bulk degrees of freedom, and the dynamics of the D-brane
worldvolume is described eitherby  non-commutative SYM or
by the non-commutative open string \cite{SW,SST,GMMS}.
The basic picture is that the B-field induces a non-commutative
deformation of the algebra $\CO(X)$ of functions on the
D-brane world-volume $X$; it makes
the world-volume non-commutative \cite{CDS,Sm,SW}.

For the non-commutativity the supersymmetry and the metric are
secondary. We can consider open bosonic strings in an arbitrary number
of Euclidean dimensions coupled only with the B-field
(we may allow a Poisson bi-vector in general).
Then the problem becomes equivalent to the quantization
of the boundary particle theory, which has
the ``D-brane'' world-volume as its classical phase space \cite{CF}.
This consideration leads to the path integral derivation of
the celebrated solution of deformation quantization by Kontsevich
\cite{K2}. A crucial physical insight in the above approach
is that the physical consistency of the bulk open string theory
implies associativity of the non-commutative deformation
of the algebra of functions
on the D-brane world-volume \cite{CF}.  A simple generalization
of the  above leads to a physical proof of the formality theorem of
Kontsevich.
This implies a deep connection between open strings and the world
of associative algebras.

Strominger \cite{Sr} and Townsend \cite{To}
showed that the M theory $5$-brane can be interpreted as
a D-brane of the open super-membrane and all D-branes of Type IIA
string can be obtained by $S^1$-compactification.
They also argued that the boundary dynamics of the above
system is controlled by a six-dimensional self-dual string \cite{W7}.
The open membrane can naturally be coupled with a
flat $3$-form $C$-field. The presence of the M$5$-brane requires
self-duality for the parallel $C$-field.
Recently the authors of \cite{AOS,SW,BBSS,KaS,GMSS} showed that
there is suitable decoupling limit such that the
bulk theory becomes topological and only the modes
in the brane are left. The resulting theory is now
called OM theory, which is related to other decoupled theories by
a web of dualities \cite{GMSS}.
An open question is the algebraic or geometrical
meaning of turning on such a background,
the resulting boundary dynamics, etc.

The main purpose of this paper is to uncover
the basic picture on the role of $C$-field
in more mundane situations.
For this  it is suffices to study the bosonic open membrane coupled
with the $C$-field only. We call the resulting theory
the topological open membrane theory.
The topological open membrane makes sense in arbitrary dimensions,
as does the topological open string.
Actually we will start from one further step
back by considering the open membrane without background.
Then we interpret the topological open membrane
theory as a certain deformation of the theory without background.
The theory without background will  tell us that the underlying
algebraic structure of  the boundary string theory
is the  Gerstenhaber algebra (G-algebra
in  short \cite{G})  of polyvectors on $X$.
More appropriately it is the algebra $\CO(\Pi T^{*}X)$
of functions on the superspace $\Pi T^{*}X$,
which is the total space of the cotangent bundle over $X$
after a parity change of the fiber.
Then quantum consistency of the theory requires that
the $C$-field must be flat, which
corresponds to the infinitesimal deformation of the
above G-algebra as a strongly homotopy G-algebra (a
$G_{\infty}$-algebra in short \cite{Ta1, TT} or $2$-algebra
\cite{K2}). We call the resulting algebra the $2$-algebra of $X$.
Then one may specialize to the $6$-dimensional case and
consider the deformation by a self-dual $C$-field only.
We may call this the self-dual $2$-algebra of
six dimensions $X$.
An interesting point of OM theory is that it requires
a non-vanishing constant self-dual $C$-field \cite{GMSS}.
Thus the theory from the beginning should
involve the deformed $2$-algebra or self-dual
$2$-algebra of $X$.
We will leave the detailed study of path integrals (deformed
algebra) and applications to physics for a future publication \cite{HLP}.

Our approach also has a natural generalization to higher
dimensional topological open $p$-branes. We shall see
that open $p$-branes have a deep connection with the world of
$p$-algebras. The bulk and the boundary correspondence
of open $p$-brane theory follows exactly the generalized Deligne
conjecture involving $(p+1)$ and $p$-algebra \cite{K3}.
The crucial tools for our approach are the Feynman path integrals
\'a la BV quantization \cite{BV, W1, S1, S2, W2, AKSZ,S3}.
We will also discuss some applications to the homological mirror
conjecture \cite{K1}.
We should mention that the $2$-algebra
is not an alien to string theory. It already appeared
in the closed and open-closed string field theories of Zwiebach \cite{Z1,
Z2},
VOA, TCFT and  $D=2$  string theory \cite{WZ, V, WuZ, LZ, BCP, PS, Gt}
(see also \cite{St3} for a nice review).

Now we begin a rather detailed  introduction or sketch of our program,
treating all topological open  $p$-branes uniformly and
emphasizing the more mathematical side of our story.

\medskip
\noindent
{\it A sketch of our program}
\medskip

The basic idea of deformation quantization
is that the algebra of observables in quantum mechanics
is  close, as an associative algebra,  to the  commutative algebra
of functions on the classical phase space \cite{BFFLS,K2}.
Thus the program reduces to finding formal deformations of the
commutative algebra along non-commutative directions as an associative
algebra. This is realized as deformations of the usual products
of functions to star products, whose
associativity automatically implies
that an infinitesimal should be  a Poisson
bi-vector.  A  surprise
of Kontsevich's result is that the quantization of the particle somehow
requires open string theory \cite{K2}.
Catteneo and Felder showed
that Kontsevich's formula is the perturbative expansion of the path
integral of bosonic topological open string theory \cite{CF}.

A novelty of this approach is that the problem of deformation
quantization of particles (thus the deformation of an associative
algebra) is  equated to quantum consistency of
the bosonic open string theory.
This  maps the set of equivalence classes
of Poisson structures on the target space $X$ to the set
of equivalence classes of  deformations of the open string theory
satisfying the BV master equation.
Then the BV master equation automatically implies, via the Ward identity,
that a suitable path integral of the theory on the disk defines a
bijection from the above
equivalence class to the set of isomorphism classes of associative
star products.  
In general this approach leads to a string
theoretic derivation of the formality theorem of Kontsevich.
The unifying mathematical notion behind
the above correspondences is the operads of little intervals and
associated Swiss-Cheese operads, which relate associative
algebras with 2-algebras (Deligne conjecture) \cite{Ta1,Ta2,K3}.
On the other hand
the Swiss-Cheese operad \cite{Vr1} is closely related to the tree
level open-closed string field theory of Zwiebach \cite{Z2}.
Actually the path integral approach shows that the Deligne conjecture
is just the bulk/boundary correspondence.
Recently Hofman and Ma discussed those interrelations
in a more general class of topological open-closed string theory \cite{HM}
(see also \cite{La,Moore} on some recent development on
topological open-closed string theory).

We shall see that the topological open $p$-brane for any $p>0$
is closely related 
with the world of $(p+1)$ and $p$-algebras.
This generalizes the relation in the $p=1$ case,
where $1$-algebra is another name for associative algebra.
We may use topological open $p$-brane theory
to define deformation quantization of the boundary closed bosonic
$(p-1)$-brane.
We shall see that the problem is equivalent to the problem of deformation
of $p$-algebra as a $p$-algebra.
Our basic tool is  the method of  BV quantization for
Feynman path integrals.
Here we sketch the general principle of our program.

Bosonic open $p$-brane theory is a theory of maps
$$\phi : N_{p} \rightarrow X,$$ where
$N_{p}$  is a $(p+1)$-dimensional  manifold with boundary,
regarded as the open $p$-brane world volume,
and $X$ is the target space.
The bosonic topological open $p$-brane coupled
with a $(p+1)$-form $c$ in $X$ (open $p$-brane in a
closed NS $p$-brane background)
is described by the
action functional
\eqn\aaa{
I^{\pr} = \int_{N_{p}}\phi^{*}(c) + \int_{\rd N_{p}} \CV
}
where $\CV$ denotes a possible boundary interaction.
We will regard the above theory as describing deformations
of bosonic open $p$-brane theory {\it without background}
defined by certain action functional $I_{o}$ being first
order in derivatives and invariant under
affine transformations of $X$.
After BV quantization
one may obtain the BV master action functional $\bos{S}_{o}$
of $I_o$.
Then we examine consistent  deformations, modulo
equivalence, of $\bos{S}_{o}$ using the BV master equation.
The resulting theory $\bos{S}^{\pr}$
will be identified with the BV quantization of the theory $I^{\pr}$
if the deformation preserves the ghost number symmetry.

In general the undeformed theory $\bos{S}_{o}$
tells us which mathematical structure (associated with the boundary
degrees) we want to deform.  It can be determined by
correlation functions of observables
inserted on the boundary $\rd N_{p}$.
Now the bulk deformation term in the action functional
tells us how to deform the mathematical structure by
specifying only the infinitesimal deformation.
And the perturbative expansion with the bulk term determines
the deformation of the mathematical structure to all orders.
Note that the consistent bulk deformation
term is determined by the BV master
equation, while the correlation functions should satisfy the BV
Ward identity of the theory, which is
a direct consequence of the BV master equation.

It turns out that the BV quantized topological open $p$-brane theory
is a theory of  maps
\eqn\aac{
\bos{\phi} :\Pi TN_p \rightarrow M_p(X)
}
between two superspaces $\Pi TN_{p}$ and
$M_{p}(X)$ associated with $N_{p}$ and $X$, respectively.

The superspace $\Pi TN_p$ is the total space of the tangent bundle
of $N_{p}$ after parity change of the fiber.
We denote a set of local coordinates on $\Pi TN_p$
by $(\{x^\m\}| \{\th^\m\})$, $\m =1,\ldots,p+1$, where
$\th^\m$ are odd constants. We introduce
the ghost number or degree $U\in \Z$. We assign
$U =1$ to $\th^{\m}$.

The target superspace $M_{p}(X)$ of the $p$-brane for
$p\geq 1$ can most easily be described recursively.
$M_p$ for $p\geq 1$ is the total space of the twisted by
$[p]$ cotangent bundle $T^{*}[p]M_{p-1}$ over $M_{p-1}$
and $M_0=X$ is the target space $X$.
For example
\eqn\aade{\eqalign{
M_{1} &=T^*[1]X \equiv \Pi T^{*}X, \cr
M_{2} &= T^{*}[2]M_{1}=T^{*}[2]\left( \Pi T^{*}X\right),\cr
M_{3} &=T^{*}[3] M_{2}= T^{*}[3]\left(T^{*}[2]\left( \Pi
T^{*}X\right)\right),
}
}
etc.
Note that the base space
$M_{p-1}$ of the target superspace $M_{p}$ of the $p$-brane
is the target superspace of the $(p-1)$-brane. Physically
for open the $p$-brane, $M_{p-1}$ corresponds to the target superspace
of boundary the $(p-1)$-brane. The iterative nature
of target superspace is due to the degeneracy of the first
order formalism, which requires to introduce ghosts
for ghosts etc.

Now we explain the notation $T^{*}[p]M_{p-1}$.
We introduce the following set of local coordinates
$(base|fiber)$
on $M_{p}\rightarrow  M_{p-1}$;
\eqn\aafa{
(\{q^{\a}\}|\{ p_{\a}\}),
}
where $\a=1,\ldots, 2^{p-1}\times \hbox{dim}(X)$.
We assign ghost number $U$ or degree of such coordinates
by the formula
\eqn\aaf{
U(q^{\a}) +  U(p_{\a}) = p, \qquad U(q^{\a})\geq 0,\quad U(p_{\a})\geq 1.
}
Thus the twisted by $[p]$ cotangent bundle  $T^{*}[p]M_{p-1}$
over $M_{p-1}$ is the cotangent bundle over $M_{p-1}$
with the above assignment of ghost number.
A coordinate is commuting or even if the ghost number is even.
A coordinate is anti-commuting or odd if the ghost number is odd.
An index $\a$ can be
either a tangent or a cotangent index of $X$ and indices for
$q^{\a}$ and $p_{\b}$ should be (up, down) or (down,up)
for $\a=\b$.
The ghost number of various coordinates can be determined
recursively by assigning $U=0$ to local coordinates on $M_{0}=X$.
For instance the ghost numbers of, say, base coordinates
$q^{\a}$ of $M_{p}$ can be different in general for each $\a$.
We also note that $M_p(X)$ for $p\geq 2$ can be identified
with the total space of $(p-1)$th iterated supertangent
bundle over $\Pi T^*X$, i.e.,
$M_p(X) \simeq \Pi T(\Pi T(\ldots (\Pi T(\Pi T^* X))\ldots ))$.

We describe
a map $\bos{\phi} :\Pi TN_p \rightarrow M_p$
locally by local coordinates on $M_{p}$
\eqn\aai{
(\bos{q}^{\a}, \bos{p}_{\a}) :=(q^{\a}(x^{\m},\th^{\m}) ,
p_{a}(x^{\m},\th^{\m}))
}
which are functions on $\Pi TN_{p}$.
The superfields  $(\bos{q}^{\a}, \bos{p}_{\a})$ combine
all the ``fields'' and ``anti-fields'' of the theory.
The assignment of ghost
numbers \aaf\ are a consequence of BV quantization.

We note that the target superspace $M_{p}(X)$, for $p \geq 1$,
always has the following non-degenerate canonical symplectic
form $\o_{p}$ for any manifold $X$
as the total space of the (twisted by $[p]$) cotangent bundle over
$M_{p-1}$;  
\eqn\aah{
\o_{p} = dp_{\a}\wedge d q^{\a}.
}
The symplectic form $\o_p$ carries degree $U=p$.
The parity of $\o_{p}$ is the same as the parity of $p$.
Now the degree $U=p$ symplectic structure $\o_{p}$ on $M_{p}$
defines a degree $U=-p$ (odd or even) graded Poisson bracket
$[.,.]_{p+1}$ on functions on $M_{p}$.
The BV bracket $(.,.)_{BV}$ of the $p$-brane theory is an
odd Poisson bracket with degree $U=1$ among local functions
on the space $\CA$ of all maps $\bos{\phi} :\Pi T N_{p} \rightarrow M_{p}$.
The corresponding odd symplectic form $\bos{\o}$ with degree $U=-1$
on $\CA$ originates from a degree $U=p$ (odd or even as the same  parity
of $p$) symplectic form $\o_{p}$ on the (super)-target space $M_{p}$
by the formula
\eqn\aaj{
\bos{\o} := \int_{N_{p}} d^{p+1}\th \;\bos{\phi}^{*}(\o_{p}).
}
The super-integral shifts the degree $U$ by $(-p-1)$.
The above considerations
lead us to the following crucial relation
\eqn\aak{
\left( \int_{N_{p}} d^{p+1}\th \;\bos{\phi}^{*}(\g),
\int_{N_{p}} d^{p+1}\th \;\bos{\phi}^{*}(\g)\right)_{BV}
= \int_{N_{p}} d^{p+1}\th \;\bos{\phi}^{*}([\g,\g]_{p+1}),
}
where $\g$ is a local function on $M_{p}$.

The BV action functional $\bos{S}_{o}$ of the topological open
$p$-brane without background is defined as follows
\eqn\aal{
\bos{S}_{o}
=\int_{N_{p}} d^{p+1}\!\th\;\biggl( \bos{p}_{\a} D \bos{q}^{\a}
+\bos{\phi}^{*}\left( h({q}^{\a},{p}_{a})\right)\biggr),
}
where $D =\th^{\m}\rd_{\m}$ and $h({q}^{\a},{p}_{a})$
is a degree $U=p+1$ function on $M_{p}$.
The function $h({q}^{\a},{p}_{a})$ is invariant under
affine transformations on $X$ and constant on $X$.
It satisfies 
\eqn\aala{
[h,h]_{p+1}=0, \qquad  h(q^{\a},  p_{\a})|_{M_{p-1}}=0.
}
Thus $h$ generates
a differential $Q_{o}$ with $U=1$ and $Q^{2}_{o}=0$ via
the bracket,
\eqn\aan{
[h,\ldots]_{p+1} =Q_{o}.
}
Futhermore $Q_o |_{M_{p-1}}=0$.
It turns out that
$h=0$ if $p=1$ and $h$ for general $p$ can be determined
recursively.
Note also that
\eqn\aam{
\left(\int_{N_{d}} d^{d+1}\!\th\;\left( \bos{p}_{\a} D
\bos{q}^{\a}\right), \ldots\right)_{BV} = D.
}
The BV BRST charge $\bos{Q}_{o}$ carrying $U=1$
corresponds to an odd Hamiltonian
vector of $\bos{S}_{o}$ on the space $\CA$ of all fields, i.e.,
$\bos{Q}_{o}=\left(\bos{S}_{o},\ldots\right)_{BV}$.
Then we obtain another crucial relation;
\eqn\aao{
\bos{Q}_{o}= D + \bos{\phi}^{*}(Q_{o}).
} 
Combining \aak, \aak, and \aao,
we see that the action functional satisfies the quantum master equation
if the boundary conditions are such that $\bos{p}_{\a}(x) =0$
``in directions tangent'' to $\rd N_{p}$
for $x\in\rd N_{p}$.\foot{
It means that the worldvolume ($N_p$)
scalar components $p_\a(x^\m)$ of $\bos{p}_\a$
vanish at the boundary, the vector components
in directions tangent  to $\rd N_{p}$ vanish at
the boundary, etc.
}

Now we describe the possible bulk deformations.
We consider any function (or sum of functions in general)
$\g(q^{\a},p_{\a})$ of  $M_{p}$, whose degree $U=|\g|$
has the same parity as $p+1$.
The action functional $\bos{S}_{\g}$ deformed by $\g$
is given by
\eqn\aar{
\bos{S}_{\g} = \bos{S}_{o}
+ \int_{N_{p}}d^{p+1}\th\;  \bos{\phi}^{*}\left(\g\left(p^{\a},
q_{\a}\right)\right).
}
The above corresponds to an even function on $\CA_{p}$.
Combining \aak\ and \aam, we see that
the deformed action functional satisfies the master equation
$(\bos{S}_\g,\bos{S}_\g)_{BV}=0$\foot{In the present model
the classical master equation also implies the quantum
master equation.}
if and only if
\eqn\aat{
\eqalign{
\int_{\rd N_{p}}d^{p}\th\; \bos{\phi}^{*}(\g(q^{\a},p_{\a})  = 0,\cr
\left[h+\g,h+\g\right]_{p+1}=0.
}
}
Note that the first condition
%is from the requirement
%$D\int_{N_{p}}d^{p+1}\th\; \bos{\phi}^{*}\left(\g\right) =0$.
%It also 
gives a reason to call
$\int_{N_{p}}d^{p+1}\th\; \bos{\phi}^{*}\left(\g\right)$ the bulk term.
Now the boundary conditions and the definition \aan\
imply that the above conditions are equivalent to
\eqn\aas{
\eqalign{
\g|_{M_{p-1}}  = 0,\cr
%Q_{o}\g  +\Fr{1}{2}[\g,\g]_{p+1}=0.
}
}
Thus
the set of equivalence classes of
bulk deformations of the topological $p$-brane theory,
satisfying the BV master equation,
is isomorphic to the set of equivalence classes of
solutions of the Maurer-Cartan (MC) equation
for functions $\g$ on $M_{p+1}$ satisfying the condition $\g|_{M_{p-1}}=0$.
We denote the Hamiltonian vector of $h+\g$  by $Q_{\g}$;
\eqn\aast{
Q_{\g}:= [h +\g,\ldots]_{p+1},
}
which satisfies $Q_{\g}^{2}=0$ due to the second equation in \aat.
We note that the restriction $Q_{\g}|_{M_{p-1}}$ of $Q_\g$
to the base space $M_{p-1}$ corresponds to a first order
differential acting on functions on the base space
$M_{p-1}$.
%We may replace the condition in \aas\ by more stronger
%version
%\eqn\aasu{
%Q_{\g}|_{M_{p-1}}=0.
%}
%With the above condition any function $f(q^{\a})$ on the base space
%$M_{p-1}$
%of $M_{p}$ satisfy $Q_{\g}f(q^{\a})=0$.

The BRST charge $\bos{Q}_{\g} \equiv (\bos{S}_\g,...)$
or the odd Hamiltonian vector of $\bos{S}_{\g}$
is given by
\eqn\aasr{
\bos{Q}_{\g} = D +\bos{\phi}^{*}(Q_{\g}),
}
which satisfies $\bos{Q}^2_{\g}=0$ due to the master equation.
This implies that, for any function $\k$
on $M_{p}$ satisfying $Q_{\g}\k=0$, we have
\eqn\aaox{
\bos{Q}_{\g}\bos{\phi}^{*}(\k) = D\bos{\phi}^{*}(\k).
}
This gives us so called the descent equations.
Related to the above we note that
the action functional $\bos{S}_{\g}$
has another fermionic symmetry generated by an odd vector
$\CK_{\m}= -\Fr{\rd}{\rd \th^{\m}}$ with $U=-1$ since it
is written as a superspace integral. Then \aao\ implies
that
\eqn\aaoa{
\{\bos{Q}_{\g}, \bos{Q}_{\g}\} = 0,\qquad
\{\bos{Q}_{\g},\CK_{\m}\} = -\rd_{\m},
\qquad
\{\CK_{\m},\CK_{\n}\} = 0.
}

Now we turn to the boundary interactions and boundary observables.
The boundary interaction due to the boundary conditions
is given by a certain local functional
$\CV(\bos{q}^{\a}, D\bos{q}^{\a})$ of
$\bos{q}^{\a}$ and $D\bos{q}^{\a}$;
\eqn\aap{
\bos{S}^{\pr}_{o} = \bos{S}_{o} + \int_{\rd N_{p}}d^{d}\th\;
\CV\left(\bos{q}^{\a}, D\bos{q}^{\a}\right).
}
The above action functional satisfies the quantum master equation
for any such $\CV$ satisfying $(\bos{\phi}^{*}(Q_{\g}))\CV=0$.
%The above is always true if we impose stronger condition \aasu\ for
%bulk deformation.
Clearly $\CV$ should have $U=p$ to preserve
the ghost number symmetry.
We consider $N_{p}$ as a $(p+1)$-dimensional disk with boundary
$\rd N_{p} = S^{p}$.  On the boundary we have $n+1$ punctures
$x_{0},\ldots x_{n}$.
We also consider $S^{p-1}_{i}$ surrounding
a puncture. We consider a local function
$f(q^{\a})$ of the base of $M_{p}\rightarrow M_{p-1}$
satisfying $Q_{\g}|_{M_{p-1}}f(q_{\a})=0$.
Now we let $\bos{f} := f(\bos{q}_\a) := f(q_\a(x^\m,\th^\m))$
be the corresponding function of superfields $\bos{q}^\a$.
The descent equation \aaox\ implies that
$\bos{Q}_\g \bos{f} = D\bos{f}$ and
we obtain the following
non-trivial BV observables
\eqn\aaq{
\eqalign{
\CO^{0}_{f}(x_{i})&= f(x_{i}),\cr
\CO^{(p-1)}_{f}(x_{i})
&=\oint_{S^{p-1}_{i}} d^{p-1}\th\; \bos{f},\cr
\CO^{(p)}_f
&=\int_{\rd N_{p}} d^{p}\th \bos{f}.
}
}
The last one above
may be regarded as part of the boundary  interaction.

%This is the end of a description of BV quantized topological
%open $p$-brane theory.

Now we turn to the role of the path integral.
For simplicity we ignore boundary interactions  and consider
the action functional $\bos{S_{\g}}$
\eqn\aba{
\bos{S}_{\g}
=\int_{N_{d}} d^{d+1}\!\th\left( \bos{p}_{\a} D \bos{q}^{\a} \right)
+ \int_{N_{d}} d^{d+1}\!\th\;\biggl(
h(\bos{q}^{\a},\bos{p}_{a})
+\g(\bos{q}^{\a},\bos{p}_{a})
\biggr) .
}
The first term is the kinetic term and the remaining  terms  may be
regarded
as ``interaction'' terms. Note that the interaction terms are coming
from the function $(h +\g)$ on $M_{p}$ with $(h+\g)|_{M_{p-1}}=0$,
which we called bulk terms.
After a suitable gauge fixing one may
evaluate correlation functions
of observables supported on the boundary $\rd N_{p}$
using perturbation theory.
Note that such observables
originate from functions on the base space
$M_{p-1}(X)$ of $M_{p}\rightarrow M_{p-1}(X)$.

Now we introduce the following definition
\begin{quote}
 
We call the algebra
$\CO(M_{p}(X))$
of functions on $M_{p}(X)$ with the bracket $[.,.]_{p+1}$,
and ordinary (super-commutative and associative) product
the classical $(p+1)$-algebra $Cl_{p+1}(X)$ of $X$.
Thus, by definition, the  classical $p$ algebra
$Cl_{p}(X)$ of $X$ is the algebra
$\CO(M_{p-1}(X))$ of functions on the base space of $M_{p}\rightarrow
M_{p-1}$.

\end{quote}

Note that the classical $1$-algebra $Cl_{1}(X)$ $(p=0)$  is the algebra
$\CO(X)$ 
of functions on $X$ without a bracket (since we do not have $\o_{0}$).
The classical $2$-algebra $Cl_{1}(X)$ $(p=1)$  is the algebra of
polyvectors on $X$ with the wedge product and Schouten-Nijenhuis
bracket.

Combining altogether, we showed that

\begin{quote}

The BV structure of  the topological bosonic open $p$-brane theory
originates from the classical $(p+1)$-algebra
$Cl_{p+1}(X)=\CO(M_{p}(X)$.
A BV master action functional of the theory is determined
by a bulk term associated with $(h +\g)$,
which is a function on $M_{p}$ with $(h+\g)|_{M_{p-1}}=0$
solving the MC equation of $Cl_{p+1}(X)$.
The classical algebra of observables  in the boundary
is the classical $p$-algebra $Cl_{p}(X)$,
which is the algebra  $\CO(M_{p-1}(X))$ of functions
on the base space  $M_{p-1}$ of $M_{p}\rightarrow M_{p-1}$.

\end{quote}

The perturbative expansion of the theory
can be viewed as a certain morphism
$Cl_{p+1}(X) \rightarrow Qh_{p+1}(X)$
satisfying suitable Ward identities controlled by
the bulk theory. We call $Qh_{p+1}(X)$ the quantum
$(p+1)$ algebra.  Note that the Ward identity is
a direct consequence of the BV master equation.
Thus the set of equivalence classes of dolutions to the BV master equation
is isomorphic to the set of equivalence classes of of solutions to the
Ward identity. This implies that the morphism is
a quasi-isomorphism.
Now the quantum algebra of observables (defined by the correlation
functions) should be a deformation of the
classical $p$-algebra $Cl_{p}(X)$. We call the resulting
algebra of quantum observables the quantum $p$-algebra $Qh_{p}(X)$.
This is again controlled by the bulk Ward identity.

It turns out that the classical $p$-algebra $Cl_{p}(X)$
is an example of the so called
cohomological $p$-algebra $H^{*}(A_{p})$  \cite{GJ, K2}.
Kontsevich defined a  $p$-algebra as an algebra over the operad
${\tt Chains}({\tt C}_{p})$, where ${\tt C}_{p}$ is the $p$-dimensional
little disk operad.
According to Kontsevich an algebra over the cohomology
$H_{*}({\tt C}_{p})$
(cohomological $p$-algebra in short)
is a twisted Gerstenhaber algebra with Lie bracket with degree
$1-p$ for $p=2k$, where $k$ is a positive integer,
and  a twisted Poisson algebra with Lie bracket with
degree $1-p$ for $p=2k+1$, both with the commutative associative
product of degree $0$ and zero differential \cite{K3}.
This agrees with our definition of a classical $p$-algebra
$Cl_{p}(X)$.\foot{Note, however, that
we generally have a differential.}

The generalized Deligne conjecture says that for
every $p$-algebra there exists an universal $p+1$ algebra acting on it.
For example there is a structure of $(p+1)$ algebra on the
generalized Hochschild complex ${\tt Hoch}(A_{p})$ of
a $p$-algebra $A_{p}$ \cite{K3,Ta2}. The picture is that there is a
$(p+1)$-algebra
controlling deformations of a $p$-algebra as a $p$-algebra.
Kontsevich, generalizing Tarmarkin \cite{Ta1},
proved the above
conjecture as well as the formality of $p$-algebra, i.e.,
${\tt Chains}({\tt C}_{p})\otimes \R$ is quasi-isomorphic
to its cohomology  $H_{*}({\tt C}_{p})\otimes \R$ endowed with zero
differential. 
In particular the two sets of equivalence classes of
solutions to the Maurer-Cartan (MC) equations on
${\tt Chains}({\tt C}_{p})\otimes \R$ and $H_{*}({\tt C}_{p})\otimes\R$
are isomorphic.
Kontsevich suggested that   the generalized Deligne conjecture
seems to be  related to quantum field theories with
boundaries. 
He also conjectured the existence of a structure
of $p$-algebra on $p$-dimensional conformal field theories \cite{K3}.

It is amusing to see that all the above is beautifully realized
in the topological open $p$-brane theory.\foot{We
note that such a possibility was considered before \cite{HM,D},
though without actual realizations.}
Recall that the bulk theory is determined by a
cohomological $(p+1)$-algebra as the algebra of functions on $M_{p}(X)$,
while the boundary observables  are associated
with a cohomological $p$-algebra as the algebra of functions
on the base space $M_{p-1}$ of $M_{p}\rightarrow M_{p-1}$.
Note that the algebra of polynomials $\R[\{q^{\a}\}]$ -- the polynomials
in the coordinates of the base space $M_{p-1}(X)$ -- can be
viewed as a cohomological $p$-algebra without differential.
It is easy to see (following  sect.3.4. of \cite{K2}) that the
Hochschild
cohomology of the above cohomological $p$-algebra
is the algebra  $\R[\{q^{\a}\},\{p^{\a}\}]$ -- the polynomials
in the coordinates of the total space $M_{p}$, without bracket.
Note also that our assignment of ghost numbers is consistent with
\cite{K2}.  One may endow  $\R[\{q^{\a}\}]$ with a bracket.
Then the the Hochschild cohomology should have
a differential.  It turns out that the differential is exactly the
differential $Q_{o}$ associated with the term $h$ that appeared in
the definition of the open p-brane without background.
In fact the term  $\int_{N_{p}}d^{p+1}\th\; \phi^{*}(h)$ in the action
functional 
is responsible for the appearance of the bracket of the $p$-algebra
by correlation functions.

It is also not difficult to see the appearance of ${\tt Hoch}(A_{p})$
and the structure of $(p+1)$-algebra.
Applying a theorem in \cite{K3} it is easy to see
that the Hochschild cohomology $H^{*}({\tt Hoch}(Cl_{p}(X)))$
of the classical algebra $Cl_{p}(X)$ is the classical
$(p+1)$-algebra $Cl_{p+1}(X)$.
Thus a bulk ``interaction'' term of the
$p$-brane
theory is an element of  $H^{*}({\tt Hoch}(Cl_{p}(X)))$.\foot{It seems 
to be more natural
to modify Kontsevich's definition (Sect. 3.4 in \cite{K2}).
In our case $H^{*}({\tt Hoch}(Cl_{p}(X)))$
is an algebra of functions on $M_{p+1}(X)$
but with an additional condition that
such a function should vanish
after restriction to $M_{p-1}(X)$ .
}
The bulk ``interaction'' term (in the path integral) generates
elements of the Hochschild complex
${\tt Hoch}(Cl_{p}(X))$ of the classical $p$-algebra
$Cl_{p}(X)$ by perturbative expansions.

Thus all together we may conclude that the path integral
of the topological open $p$-brane serves
as a morphism between the two $(p+1)$-algebras associated
with the target space $X$;
$H^{*}({\tt Hoch}(Cl_{p}(X))$ and ${\tt Hoch}(Cl_{p}(X))$.
As we argued before  the morphism must be a quasi-isomorphism
due to the BV master equation of the topological open $p$-brane theory.
One should check this by working out the BV Ward identity of the
theory by carefully working out the compactification of the moduli space,
related with the higher dimensional Swiss-Cheese operads.
Furthermore the Ward identity will tell us that the quantum algebra
$Qh_{p}(X)$ of observables has the structure of a $p$-algebra.
This motivates us to define the deformation quantization of the
$(p-1)$-brane
as the deformation of the algebra
$\CO(M_{p-1}(X))\equiv Cl_{p}(X)$ of functions
on $M_{p-1}(X)$ as a $p$-algebra.

We note the path integral derivation of the formality theorem
for the $p=1$ case goes along the same lines as the WDDV equation \cite{W6,
DVV},
and as that of 2D strings \cite{WZ, V} involving compactification of
the moduli space.  Similarly the topological open $p$-brane
is determined essentially by the bulk BV master action functional
involving $Cl_{p+1}(X)$, boundary observables involving $Cl_{p}(X)$
and the moduli space of disks with boundary punctures.
The rest is determined by the Feynman path integral. It is highly unlikely
that the quantum algebra $Cl_{p}(X)$ is not a $p$-algebra defined
by Kontsevich  based on the same data \cite{K3}.
Now there are several mathematical
proofs of the generalized Deligne conjecture of formality of $p+1$ algebra
\cite{Ta1, K2, TT, KS}. However none of those proofs
seem to give an explicit quasi-isomorphism
except for the $p=1$ case \cite{K1}.
The topological open $p$-brane theory
will lead to such an explicit formula.
The details will appear elsewhere \cite{HLP}.
It will be also interesting to see if the path
integrals of topological open $p$-brane ``confirm''
another conjecture of Kontsevich on the action of
motivic Galois group \cite{K3}.

We like to mention that our approach is quite similar to
that of Witten in his attempt to formulate background independent
open string field theory, suggesting a generalization for the
closed string case \cite{W5}.  It seems to be also closely
related to the use of Chern-Simons theory in $3$ dimensions
on rational conformal field theory and vise versa \cite{W9}.
The model for $p=2$ is related with
the so called extended or BV Chern-Simons theory \cite{K0, AS, AKSZ, S3}
and the BV quantized higher dimensional BF theory \cite{IKb,CR}.
The interplay between bulk and boundary obviously reminds
of the Maldacena conjecture \cite{M, GKP, W10}, which is perhaps
a purely algebraic (or geometrical) remnant.
There is long history on deep relations between
open string theory and $1$-algebras ($A_{\infty}$-algebras \cite{St1}).
A central example is  the open string field theory
based on a non-commutative and strictly associative algebra \cite{W}
and its generalization up to homotopy  \cite{GZ}.
The associativity up to homotopy allows one to construct an action
functional of open string field theory satisfying the Batalin-Vilkovisky (BV
in short)  master  equation,  therefore admitting a consistent quantization.
Similar structures also appear as
Fukaya's $A_{\infty}$-category \cite{F1} in the open string version of
the so called A model of the topological sigma model \cite{W4}.
The open-closed string field theory \cite{Z2}
is related to the algebra over Swiss-Cheese operads,
while the closed string field theory is based on the $L^{\infty}$ part of
the homotopy $2$-algebra \cite{Z1, KSV1, KSV2}.
It is also natural to expect that the homological mirror conjecture
\cite{K1}
can eventually  be the physical equivalence of open-closed string field
theory.
It seems to be also reasonable
to suspect that one may define open-closed $p$-brane
field theory using higher dimensional Swiss-Cheese operads with
additional decoration as in the $p=1$ case.
Actually the above is true only for the  string tree
level. The string field theory is based on a certain loop generalization
of the above. For higher dimensional branes it is practically impossible
to consider any ``loops'' as summing over all topologies
and geometries. We may content with the theory
defined on tree level.

\lin{The structure of this paper}

This paper is organized as follows.
In Sect.~$2$ we review the BV quantization method,
emphasizing the underlying super-geometrical structure
and relations with deformation  theory.
Then we reformulate the Catteno-Felder model
and the A model
in a language suitable for our purpose.
This section will set up notations and our general strategy.
In Sect.~$3$ we study
BV quantization of the topological open membrane.
This is a detailed example of the general structure
discussed in the introduction for $p=2$.
We also present the leading deformations of
$Cl_{2}(X)$.  We also discuss different choices
of boundary conditions and their implications
to possible generalizations of homological
mirror symmetry to the category of homotopy $2$-algebras
(open membrane).
In  Sect.~$4$ we return to the $p=1$ case (string).
We introduce an unified topological sigma model,
which has the A, B and Catteneo-Felder models as special limits.
We construct an extended B model parametrized by the extended
moduli space of complex structures and show how the
non-commutativity appears for the open string case.
We briefly discuss applications of the extended B model to
the homological mirror conjecture.
We also conjecture that the homological mirror conjecture
can be generalized to the category of any $p$-algebra
of $X$ or the physical equivalence of any topological open
$p$-brane theory.

\newsec{Preliminary}

In this section we begin by reviewing the method of
BV quantization. For details we refer to \cite{BV,W2, S1,S2,W5,AKSZ}.
We compare it with the modern deformation theory and argue,
as a general statement,
that deformation problems of certain mathematical structures may be
represented as a BV quantization problem of a suitable quantum
field theory and vice versa. Then we reconsider the formulation of
Kontsevich-Catteneo-Felder as a deformation problem of bosonic
string theory. We also discuss BV quantization of the A model,
generalizing \cite{AKSZ}, for a later purpose.
We do not assume any originality
in this section, but perhaps some new interpretations.

\subsec{BV quantization and deformation theory}

A path integral is a formal integral of certain observables
over the space of all fields of a classical theory weighted
by the exponential of the classical action functional $I$. An observable of
the theory is a function on the space of all fields invariant
under the symmetries of the classical action functional. Consequently
one needs to mod out volume of the orbit of symmetry group. Thus we need
to construct a ``well-defined'' quotient measure for the path integral.
The BRST-BV quantization is a systematic and versatile way
for doing this.
The BV-BRST quantization can be done by the following steps.

The first step is to fermionize the symmetry by introducing
anticommuting ghost fields for the infinitesimal parameters of the
symmetry. We call the corresponding charge of the fermionic
symmetry the BRST charge $\bos{Q}$.
One introduces an additive quantum
number $U$ called ghost number or degree and assign
$U=1$ to $\bos{Q}$.
We call
the set of original fields and ghosts the set of ``fields''.
One  introduces a
set of ``anti-fields''  for the set
of ``fields''
such that in the space $\CA$ of
all fields (thus ``fields'' and ``anti-fields'') one  has
a natural odd symplectic structure $\bos{\o}$.
Then we have a corresponding odd Poisson bracket $(.,.)_{BV}$
called the BV bracket among the functions on $\CA$.
The idea is
that one regards  ``fields'' as coordinates, while
``anti-fields'' are regarded as corresponding conjugate  momenta but
with the opposite parities (commuting$\equiv$even and
anti-commuting$\equiv$odd) in a
certain infinite dimensional phase space. More precisely,
one assigns integral ghost number $U$, or degree, to each
field such that $U(\bar\phi) = -1 - U(\phi)$, where $\bar\phi$ is the
``anti-field'' of a ``field'' $\phi$. The parity of a field is
the same as the parity of its ghost number $U$.
It follows that the odd symplectic form $\bos{\o}$ carries $U=-1$,
while the BV bracket carries $U=1$.
A BV bracket has the following properties
\eqn\zaa{
\eqalign{
(A,B)_{BV} &=
-(-1)^{(|A|+1)(|B|+1)}(B,A)_{BV},\cr
(A,(B,C)_{BV})_{BV}
&= ((A,B)_{BV},C)_{BV}
+(-1)^{(|A|+1)(|B|+1)}(B,(A,C)_{BV})_{BV},
}
}
where $A,B,C$ are (even or odd) local functions on
$\CA$ and $|A|=U(A)$, etc.
We also have the
Leibniz law, stating that
the BV bracket behaves as a derivation on the ordinary product of
functions on $\CA$;
\eqn\zzaa{
(A, BC)_{BV} = (A,B)_{BV}C +
(-1)^{(|A|+1)|B|}B(A,C)_{BV}.
}
Such a product is (super)-commutative and associative and has degree $0$.

One requires, in addition to the odd symplectic structure  $\bos{\o}$,
that $\CA$ has a volume element specified by a density $\r$ compatible with
$\bos{\o}$. Then one has a BV Laplacian $\bos{\triangle}_{\r}$
defined by $\bos{\triangle}_{\r} A = \Fr{1}{2}\hbox{div}_{\r} V_{A}$, where
$A$ is a function on $\CA$, $V_{A}$ is the Hamiltonian vector and
$\hbox{div}_{\r}$ is the divergence calculated with respect to $\r$.
The compatibility is the condition  $\bos{\triangle}_{\r}^2=0$.
Note that $\bos{\triangle}_{\r}$ carries ghost number $U=1$.
{}From now one we shall not specify the density $\r$ explicitly. In local
coordinates
$(\{\phi^{\a}\}, \{\bar\phi_{\a}\})$ on $\CA$ we have
$\bos{\triangle} = \rd^{2}/\rd \phi^{\a}\rd\bar\phi_{\a}$.
Now
the BV bracket can be defined by the failure of $\bos{\triangle}$
being a derivation of the product of functions on $\CA$ by
the formula
\eqn\zac{
(A,B)_{BV}= 
(-1)^{|A|}\bos{\triangle}(AB)  -(-1)^{|A|}\bos{\triangle}(A)B
+A\bos{\triangle}(B).
}
The algebra of functions on $\CA$ endowed, by the relations
\zaa, \zzaa\ and \zac, with
the bracket $(.,.)_{BV}$ with $U=1$  generated by $\bos{\triangle}$ as
well as with the (super)-commutative and associative product with $U=0$,
is called a BV algebra.

The BV action functional $\bos{S}$ is an even function on $\CA$
with vanishing ghost number\foot{One may allow the
action functional to be any even function. Our action functional
will always have vanishing ghost number unless specified
otherwise.},
such that (i) its restriction to the subspace of
``fields''  is the original classical action functional $I$,
(ii) it generates the BRST symmetry via the BV bracket
i.e.,
\eqn\zabb{
(\bos{S},\ldots)_{BV} = \bos{Q}.
}
Equivalently $\bos{Q}$ is the odd Hamiltonian vector of $\bos{S}$.
(iii) it satisfies
the quantum
master equation
\eqn\zab{
(\bos{S},\bos{S})_{BV} - 2\hbar
\bos{\triangle}\bos{S}=0 \equiv \bos{\triangle} e^{-
\Fr{\bos{S}}{\hbar}}=0,
}
where $\hbar$ denotes Planck constant.
%Equivalently $(\bos{Q} - \hbar \bos{\triangle})^{2} =0$.
A BRST-BV observable $\CO$ is a function on $\CA$ annihilated by
$\bos{Q} - \hbar\bos{\triangle}$.
The BV master equation is the condition that the expectation
value 
\eqn\zae{
\left<\CO\right> =\int_{\CL}d\m\; \CO e^{-\bos{S}/\hbar}
}
of a BV observable $\CO$ is invariant under continuous deformations
of the Lagrangian subspace $\CL$ with respect to $\bos{\o}$ in $\CA$.
It also implies that
the following path integral identically vanishes
\eqn\zad{
\left< (-\hbar \bos{\triangle} + \bos{Q})A\right>
:=\int_{\CL}d\m\; (-\hbar \bos{\triangle}A + \bos{Q}A)
e^{-\Fr{1}{\hbar}\bos{S}}=
0,
}
for any product of functions $A$ on $\CA$.
Picking a homology class of a Lagrangian subspace $\CL$ of $\CA$ is
called a gauge fixing.

There are cases in which the classical
master equation $(\bos{S},\bos{S})=0$ implies the quantum master
equation \zab, i.e., $\bos{\triangle}\bos{S}=0$.
Then the master equation implies that $\bos{S}$ has a fermionic symmetry
generated by a nilpotent, $\bos{Q}^2=0$,  BRST charge $\bos{Q}$,
which acts on functions on $\CA$ as an odd derivation.
That is, from \zaa, \zabb, and the classical master equation,
\eqn\zag{
\eqalign{
\bos{Q}^{2}&=0,\cr
\bos{Q}(A,B)_{BV} &= (\bos{Q}A, B)_{BV}
+(-1)^{|A|+1}(A, \bos{Q}B)_{BV}.\cr
}
}
The above structure induces the on the BV algebra a structure of
differential BV (dBV) algebra.
Remark that $\bos{Q}$ can be identified with an odd nilpotent vector on
$\CA$.\foot{We will always denote, as is the convention, a
derivation with the
corresponding vector field by the same symbol.}
% A supermanifold
% with such an odd vector is called $Q$-manifold. A supermanifold
%with odd symplectic structure is called $P$-manifold. A $P$-manifold
%with compatible volume element is called $SP$-manifold
%Thus in the present case
%the space $\CA$ of all fields
%is an infinite-dimensional $QSP$-manifold.
Now we may further specialize to the
case that there are classes of BV observables $A_{i}$
satisfying $\bos{\triangle}A_{i}=0$.
Such observables satisfy $\bos{Q}A_{i}=0$.
Note  that the BRST charge $\bos{Q}$ transforms as a scalar
under the rotation group of the manifold on which the field theory is
defined.
In other words the fermionic symmetry is global. Then the BV quantized
theory, under the above assumption, is a cohomological field theory,
first introduced by Witten.
The fixed point theorem
of Witten \cite{W3}
implies that the path integral of $\bos{Q}$-invariant
observables
is further localized to an integral over the $\bos{Q}$-fixed point
locus $\CM$ in $\CL$.

Now we turn to the BV Ward identity.
We consider expectation values
$\left<A_1\ldots A_n\right>$  of products of  functions $A_i$ on
$\CA$.
Note that the space $\CA$ of all fields
is a graded (by the ghost number) superspace.
Thus observables also carry ghost numbers.
In general any correlation function has vanishing ghost number.
Usually the path integral measure carries a ghost number anomaly
although the BV action functional has vanishing ghost number.
Consequently the net ghost number of $A_{1}\ldots A_{n}$ should
cancel the ghost number anomaly in order to have non-vanishing correlation
function.
Now the identity \zad\ implies that
\eqn\zah{
\hbar\left<\bos{\triangle}(A_1\ldots A_n)\right>
= \left<\bos{Q}(A_1\ldots
A_n)\right> .
%= \int_{\overline\CM} d(A^\pr_1\ldots A^\pr_n)
}
The above identity is called the BV Ward identity,
which is non-empty if the net ghost number of $A_{1}\ldots A_{n}$
plus $1$ is the total ghost number anomaly.
Now consider the case that $\bos{\triangle}A_i=0$.
{}From \zaa\ and \zac,
we have
\eqn\zai{
\hbar\sum_{1\leq j < k\leq n}\s_{jk}\left<
\left(A_j, A_k\right)_{BV}\prod_{i\neq j,k}A_i\right> =
\left<\bos{Q}(A_1\ldots
A_n)\right>, 
}
where $\s_{jk}$ is a sign factor.
Now we assume that the classical master equation implies
the quantum master equation.
Then $\bos{Q}$ behaves like the exterior derivative
on field space. Thus the right hand side of above receives
contributions only from the boundary in field space.
For this one should introduce an appropriate compactification.\foot{
For example, all the WDDV type equations (the
associativity
of quantum cohomology rings, the path integral derivation of
Kontsevich's $L^{\infty}$-quasi-isomorphism, and the
$A^{\infty}$-structure of Fukaya category) are based
on elaborations of the above idea.}

\subsubsection{Deformations of BV quantized theory}

Now we consider deformations of the BV quantized field theory, defined
by the triple $(\CA, \bos{\o},\bos{S})$.
Note that such a field theory
is typically defined on a manifold $N$ and  $\CA$ is the space of all fields
under consideration  (the space of all sections of a certain bundle over
$N$),
and $\bos{\o}$ is an odd symplectic structure on $\CA$ with $U=-1$.
The action functional $\bos{S}$ is a function with $U=0$ defined
as an integral of a top-form on $N$ over $N$, satisfying the BV master
equation. For simplicity we assume
that $(\bos{S},\bos{S})=\bos{\triangle}\bos{S}=0$.

Two BV quantized theories
$(\CA_{1},\bos{\o}_{1},\bos{S}_{1})$ and
$(\CA_{2},\bos{\o}_{2},\bos{S}_{2})$ are physically equivalent
if there is a diffeomorphism $F:\CA_{1}\rightarrow \CA_{2}$
such that $F^{*}\bos{\o}_{2} =\bos{\o}_{1}$ and $F^{*}\bos{S}_{2}
=\bos{S}_{1}$. It also follows that the odd Hamiltonian vectors
$\bos{Q}_{1}$ and $\bos{Q}_{2}$ of $\bos{S}_{1}$ and $\bos{S}_{2}$,
respectively, are related as $\bos{Q}_{2} = F_{*}\bos{Q}_{1}$ \cite{GZ}.
Now we consider physically inequivalent deformations of a given theory
$(\CA, \bos{\o},\bos{S})$.

We consider a basis
$\{\G_{a}\}$ of $\bos{Q}$-cohomology among functions on $\CA$,
which are integrals of top-forms on $N$ over $N$. Then we can
consider a family of action functionals given by
$\bos{S} + t^{a}\G_{a}$, where $\{t^{\a}\}$ are formal parameters
on a dual basis of $\bos{Q}$-cohomology. Since $\bos{Q}\G_{a}\equiv
(\bos{S}, \G_{a}) =0$ the family of action functionals again satisfies
the BV master equation up to first order in $t^{a}$. To go beyond
the first order we should imagine a certain function $\G(t)$, satisfying
$\bos{\triangle}\G(t)=0$, which has
the following formal expansions
\eqn\zba{
\G(t) = t^{a}\G_{a} + \sum_{n > 1}t^{a_{1}}\ldots
t^{a_{n}}\G_{a_{1}\ldots a_{n}}.
}
Then we consider the family of action functionals $\bos{S}(t)$ defined by
\eqn\zbb{
\bos{S}(t) = \bos{S} + \G(t).
}
The above deformation is said to be well-defined if $\bos{S}(t)$
satisfies the BV master equation
\eqn\zbc{
(\bos{S}(t),\bos{S}(t)) =0 \;\equiv\; \bos{Q}\G(t) +
\Fr{1}{2}(\G(t),\G(t))=0.
}
Obviously $\G_{a_{1}\ldots a_{n}}$ for $n>1$ can not be an element
of $\bos{Q}$-cohomology.
We call the deformation \zbb\ unobstructed if the solution
of $\bos{Q}\G + \Fr{1}{2}(\G,\G)=0$ has the expansion \zba.
We call the deformation \zbb\ classical
(or unextended) if $t^{a}\neq 0$ only for $\G_{a}$ with
ghost number $U=0$, while otherwise $t^{a}=0$.
We define the extended moduli space $\mathfrak{M}$ of
(the well-defined
deformations of) the theory  by the solution space of \zbc\ modulo
equivalences. We note that the deformed theory \zbb\ has a new
BRST charge $\bos{Q}(t)$ defined by $(\bos{S}(t),\ldots)\equiv\bos{Q}(t)$.
Then the equation \zbc\ is equivalent to $\bos{Q}(t)^{2}=0$ and
\eqn\zbd{
\bos{Q}(t)\G(t) =0.
}
We denote call thesubspace $\mathfrak{M}_{cl}\subset
\mathfrak{M}$ consisting of solutions with $U=0$ the classical moduli space.
Obviously the tangent space of $\mathfrak{M}$ at a classical point
is isomorphic to the $\bos{Q}$ cohomology group, provided that
the deformations are unobstructed.

The above discussion on
BV quantization is closely related to
modern deformation theory. Details and precise definitions of
deformation theory can be found in \cite{K2,BK}. Relations
with BV quantization are also discussed in \cite{St2}.
Modern deformation theory associates a certain differential
graded Lie algebra (dgLa in short)
$\mathfrak{g}=\oplus \mathfrak{g}^{k}$,
modulo quasi-isomorphism,
with a mathematical structure being deformed.
This gives rise to a formal supermoduli space
$\mathfrak{M}_{\mathfrak{g}}$
defined by the solution space of the Maurer-Cartan (MC in short) equation
modulo equivalences,
\eqn\zbe{
d\g +\Fr{1}{2}[\g,\g]=0/\sim
}
where $\g \in \mathfrak{g}$ and $(d, [.,.])$ denote the differential
and the bracket of $\mathfrak{g}$ with degree $1$ and $0$, respectively;
\eqn\zbf{
d:\mathfrak{g}^{k}\rightarrow \mathfrak{g}^{k+1},
\qquad 
[.,.]: \mathfrak{g}^{k}\otimes \mathfrak{g}^{\ell}\rightarrow
\mathfrak{g}^{k+\ell}.
}
The differential and the bracket have the following properties
\eqn\zbg{
\eqalign{
d^{2} &=0,\cr
d[\g_{1},\g_{2}] &= [d\g_{1},\g_{2}] +(-1)^{|\g_{1}|}[\g_{1},
d\g_{2}],\cr
[\g_{1},\g_{2}] &=-(-1)^{|\g_{1}|\g_{2}|}[\g_{2},\g_{1}],\cr
[\g_{1},[\g_{2},\g_{3}]] &= [[\g_{1},\g_{2}],\g_{3}]
+(-1)^{|\g_{1}||\g_{2}|}[\g_{2},[\g_{1},\g_{3}]].\cr
}
}

A crucial idea is that one imagines an
underlying formal supermanifold $M_{\mathfrak{g}}$ with
a nilpotent odd vector $Q_{\mathfrak{g}}$ with degree $1$
such that $d$ and $[.,.]$ are the first and second coefficient in
the Taylor expansion of $Q_{\mathfrak{g}}$.
Then the condition that $Q^{2}_{\mathfrak{g}}=0$
is equivalent to those in \zbg\ \cite{AKSZ}.
Thus
\eqn\zbh{
Q_{\mathfrak{g}}\g = d\g +\Fr{1}{2}[\g,\g].
}
Then the moduli space $\mathfrak{M}_{\mathfrak{g}}$ can be identified
with the space of $Q_{\mathfrak{g}}$-cohomology
$\CM_{\mathfrak{g}}= \hbox{Ker }
Q_{\mathfrak{g}}/\hbox{Im } Q_{\mathfrak{g}}$, where the quotient
by $\hbox{Im } Q_{\mathfrak{g}}$ plays the role of dividing out the
equivalence in \zbe.
In general, $Q_{\mathfrak{g}}$ may have Taylor components
beyond the quadratic terms and those define an $L_{\infty}$-structure.
We may have
\eqn\zbi{
Q_{\mathfrak{g}}\g = d\g +\Fr{1}{2}[\g,\g] + \Fr{1}{3!}[\g,\g,\g] +\ldots
}
A quasi-isomorphism between two dgLas $\mathfrak{g}_{1}$ and
$\mathfrak{g}_{2}$
is a $Q$-equivariant map between the two associated
formal supermanifolds
with base points such that the first Taylor coefficients of the map
induces an isomorphism between the cohomologies of $(
\mathfrak{g}_{1},d_{1})$
and $(\mathfrak{g}_{2},d_{2})$. It follows that the $Q_{\mathfrak{g}_{1}}$
cohomology is isomorphic to that of $Q_{\mathfrak{g}_{1}}$. Thus the
two formal supermoduli spaces $\mathfrak{M}_{\mathfrak{g}_{1}}$ and
$\mathfrak{M}_{\mathfrak{g}_{2}}$ are isomorphic.

We note that 
after shifting the degree by $-1$, the
relations in \zbg\ are identical to those in \zaa\ and \zag.
This implies that one may ``represent'' a deformation theory
by a suitable BV quantized field theory. That is, for a given
mathematical structure one wants to deform and the associated
dgLa $\mathfrak{g}$, one may consider a BV quantized field theory
whose moduli space is isomorphic to the formal super-moduli space
$\mathfrak{M}_{\mathfrak{g}}$.\foot{
We also note that the BV quantization can be generalized
to deal with deformation problems involving the full
$L_{\infty}$-algebra \cite{BDA}.}
Then the quantum field theory
gives {\it additional} information about the deformation theory
via correlation functions. What would be the additional
information?  We do not know the general answer.
It may, for example,
be possible that
deformation problems of different mathematical structures
lead to quantum field theories which are physically equivalent,
etc. etc. Such phenomena is usually called ``dualities''
or ``mirror symmetry''.

\subsec{Bosonic string theory as a deformation theory}

String theory in the sigma model approach is
a theory of maps ${u}:\S\rightarrow X$ from a Riemann
surface $\S$ to a target space $X$. Picking local
coordinates $\{u^I\}$ on $X$ such a map is described
by a set of  functions $\{{u}^{I}(x^{\m})\}$ on $\S$,
where $x^{\m}$, $\m=1,2$, denotes a set of local coordinates on $\S$.
The bosonic string action in a NS $B$-field background
is given by
\eqn\iaa{
I^\pr = \Fr{1}{2}\int_{\S}\left({b}_{IJ}d{u}^{I}\wedge d{u}^{J}\right)
}
where ${b}_{IJ}$ denote the components  of the NS $2$-form B-field,
i.e. $B = \Fr{1}{2}{b}_{IJ}d x^{I}\wedge dx^{J}$.
Classically  a $B$-field gives a certain %geometrical
structure to the target space $X$. For instance a non-degenerate
$B$-field corresponds to an almost symplectic structure on $X$,
while a non-degenerate closed (flat $dB=0$) $B$-field corresponds to
a symplectic structure on $X$.
For our purpose it is convenient to introduce a first order formalism
starting from 
a bivector field $\pi=\Fr{1}{2}b^{IJ}\rd_{I}\wedge\rd_{J}\in
\G(\wedge^{2}TX)$, which is not necessarily non-degenerate. We have
\eqn\iab{
I= \int_{\S} \left( H_{I}\wedge d{u}^{I}
+\Fr{1}{2} b^{IJ}H_{I}\wedge H_{J}
%+\pi\a^{\pr}g^{IJ}H_{I}\wedge * H_{J}
\right),
}
where $H_{I}$ is an ``auxiliary'' $1$-form field in $\S$ taking values
in ${u}^{*}(T^{*}X)$. The boundary condition for $\rd\S \neq 0$
is that $H_{I}$ vanishes along the tangential direction in $\rd\S$.
For a non-degenerate bi-vector, we can integrate out
the auxiliary field $H_{I}$ and we obtain the original action functional
\iaa.
We note that the action functional $I$ was originally
studied for closed Rieman surface in \cite{IKa,SS}. 
The authors of \cite{ASS} also found relation wiht
geometric quantization of the symplectic leaves in the target.
The path integral approach of Catteo-Felder is
based on BV quantization of the above action
functional \cite{CF}.
Here we find  it more convenient to make a detour.

The first order formalism above allows us to consider a notion
of string theory without an NS background,
defined by the following action functional
\eqn\iad{
I_o = \int_{\S}  H_{I}\wedge d{u}^{I}.
}
The resulting theory is obviously topological in two dimensions.
It is also invariant under affine
transformations of $X$ and does not depends on any other structure
in the target space.
We may view the  string theory  in the NS $B$-field background
as a certain deformation of the theory $ I_o$ along the ``direction''
of the bi-vector.
We have a criterium for well-defined deformations
by requiring that a deformed theory should be a consistent
quantum theory.
For this we first quantize the theory with action functional
$I_{0}$ using the BRST-BV formalism. Then we deform the theory
along a certain direction and examine if the deformed theory
satisfies the quantum master equation.

\subsubsection{BV quantization}

The quantization of $I_{0}$
is rather simple. We note that the action functional
$I_{0}$ has the following symmetry
\eqn\cia{
\d u^{I} =0,\qquad
\d H_{I} = - d\c_{I},
}
where $\c_{I}$ is an infinitesimal gauge parameter taking values
in $u^{*}(T^{*}X)$, which vanishes on $\rd\S$ if the boundary
of $\S$ is non-empty. In the BRST quantization one promotes the
symmetry to a fermionic one by taking $\c_{I}$ anticommuting with
ghost number $U=1$. This
is equivalent to regarding $\c_{I}$ as taking value in $u^{*}(\Pi
T^{*}X)$, where $\Pi T^{*}X$ denotes the parity change of the
fiber of $T^{*}X$. Now we have the following fermionic symmetry
with charge $\bos{Q}_{o}$ carrying $U=1$
\eqn\cib{
\bos{Q}_{o} u^{I} =0,\qquad
\bos{Q}_{o} H_{I} = - d\c_{I},\qquad
\bos{Q}_{o} \c_{I} =0,
}
and satisfying $\bos{Q}_{o}^{2}=0$.
In the present case we introduce a
set of ``anti-fields'' $(\eta_{I}, \r^{I}, v^{I})$
with the ghost numbers $U=(-1, -1, -2)$
for the set
of ``fields'' $(u^{I}, H_{I}, \c_{I})$.
Let $\CA$ denote the space of all fields.
The ``fields'' and the ``anti-fields'' should have a pairing
defining
a two-form in $\S$ such that one can integrate over $\S$ to get
a two-form in $\CA$. Thus the  ``anti-fields'' $(\eta_{I}, \r^{I}, v^{I})$
are $(2, 1, 2)$-forms on $\S$.
We can easily find 
the BV action functional $\bos{S_{o}}$ satisfying all the requirements
stated in Sect.~$2.1$ as follows
\eqn\cic{
\bos{S}_{o} = \int_{\S} \left(H_{I}\wedge du^{I} -\r^{I}\wedge
d\c_{I}\right).
}
Thus the BRST transformation laws are
\eqn\cid{
\eqalign{
\bos{Q}_{o} u^{I} &=0,\cr
\bos{Q}_{o} \c_{I} &=0,\cr
}\qquad
\eqalign{
\bos{Q}_{o} \r^{I} &= -d u^{I},\cr
\bos{Q}_{o} H_{I} &= - d\c_{I},\cr
}
\qquad
\eqalign{
\bos{Q}_{o} v^{I} &= -d \r^{I},\cr
\bos{Q}_{o} \eta_{I} &= - d H_{I},\cr
}
}
satisfying $\bos{Q}_{o}^{2}=0$. It is trivial to check
the BV master equation
\eqn\cie{
(\bos{S}_{o},\bos{S}_{o})_{BV}=0,\qquad
{\bos{\triangle}}\bos{S}_{o}=0,
}
provided that we have the right boundary conditions
mentioned earlier.

A more conceptual and compact formulation can be obtained
by combing
``fields'' and ``anti-fields'' into two superfields
$(\bos{u}^{I},\bos{\c}_{I})$ carrying $U=(0,1)$;
\eqn\cif{
\eqalign{
\boldsymbol{u}^{I}:= u^I(x^\m,\th^\m) &= u^{I}(x^\m)  +
\r^{I}_{\n}(x^\m)\th^{\n}
-\Fr{1}{2} v^{I}_{\n_1\n_2}(x^\m)\th^{\n_1}\th^{\n_2},\cr
\boldsymbol{\chi}_{I}:=\c_I(x^\m,\th^\m) &= \c_{I}(x^\m)  +
H_{I\n}(x^\m)\th^{\n}
+\Fr{1}{2} \eta_{I\n_1\n_2}(x^\m)\th^{\n_1}\th^{\n_2},\cr
}
}
where $\th^{\m}$ is an anti-commuting vector with $U=1$.
In terms of the superfields we have
\eqn\cig{
\boldsymbol{S}_{o} = \int d^{2}\th\;
\boldsymbol{\c}_{I} D\boldsymbol{u}^{I},
}
and
\eqn\cih{
\bos{Q}_o\bos{u}^{I}=
D \bos{u}^{I},\qquad
\bos{Q}_o \bos{\c}_{I}=
D \boldsymbol{\c}_{I},
}
where $D = \th^{\m}\rd_{\m}$.
The natural odd symplectic form
${\bos{\o}}$
in $\CA$ is defined by
\eqn\cii{
{\boldsymbol{\o}} = \int_{\S}d^{2}\th \left(
\bos{\d} \bos{u}^{I}
\bos{\d}\bos{\c}_{I}\right),
}
where $\bos{\d}$ denotes the exterior differential on $\CA$.
The odd symplectic form ${\bos{\o}}$ has degree $U=-1$
since $d^2\th$ shift the degree by $U=-2$.
The BV bracket is the odd Poisson bracket among
functions in $\CA$ with respect to ${\bos{\o}}$.
The BRST charge $\bos{Q}_o$ can be identified
with the odd Hamiltonian vector of $\boldsymbol{S}_{o}$;
\eqn\cik{
\bos{i}_{\bos{Q}_{o}}{\bos{\o}} =
\bos{\d}\bos{S}_{o},\hbox{ {\it equivalently} }
\left(\bos{S}_o,\ldots\right)_{BV} = {\bos{Q}}_o.
}

Now we observe that the superfields $(\bos{u}^{I},\bos{\c}_{I})$
parametrize maps 
\eqn\dnvb{
\bos{\phi}:\Pi T\S \rightarrow \Pi T^{*}X
} 
between
the two superspaces. Here $\Pi T\S$ denotes the total
space of the tangent bundle of $\S$ after parity change of
the fiber. We regard   $(\{x^{\m}\},\{\th^{\m}\})$
as a set of local coordinates on $\Pi T\S$.
With the target superspace $\Pi T^{*}X$ we mean the total
space of the cotangent bundle of $X$ after parity
change of the fiber. We denote a set of local coordinates
on $\Pi T^*X$ by $(\{u^I\},\{\c_I\})$ carrying
the ghost number $U=(\{0\},\{1\})$.\foot{By abuse
of notation, $u^I$ may denote both a coordinate on $X$
or a function $u^I(x^\m)$ on $\S$ etc. This should
not cause any confusion in the present context.}
Thus the space $\CA$ of all fields is the
space of all maps above.
The odd symplectic form ${\boldsymbol{\o}}$ \cii\ on $\CA$
is the unique extension of the odd symplectic form $\o=du^I d\c_I$
with degree $U=1$ on $\Pi T^{*}X$ to $\CA$.
More precisely 
\eqn\cna{
\bos{\o} \equiv \int_{\S}d^{2}\th\; \bos{\phi}^{*}(\o),
}
where $\bos{\phi}^{*}$ is the pull-back of a map
$\bos{\phi}:\Pi T\S
\rightarrow \Pi T^{*}X$.
The odd Poisson bracket $[.,.]_S$ with degree $U=-1$
among functions $\g_i$
on
$\Pi T^{*}X$ is called the Schouten-Nijenhuis bracket;
\eqn\cnb{
[\g_1,\g_2]_S := 
\Fr{\rd \g_1}{\rd \c_I}\Fr{\rd \g_2}{\rd u^I}
-(-1)^{(|\g_1|-1)(|\g_2|-1)}\Fr{\rd \g_2}{\rd \c_I}\Fr{\rd \g_1}{\rd u^I}
}
where $|\g|$ denotes the degree of $\g$.
We remark that a function on $\Pi T^{*}X$ with $U=p$
is a $p$-vector (an element of $\G(X, \wedge^{p}TX)$)
after parity change. We also remark that
functions on $\Pi T^{*}X$ after shifting the degree by $1$
together with the Schouten-Nijenhuis bracket form a dgLa with
zero differential. The product of functions on $\Pi T^{*}X$
corresponds to wedge products of multivectors on $X$.
We also note that a function $\g$ on $\Pi T^{*}X$ induces
a function $\int_{\S}d^{2}\th\; \bos{\phi}^{*}(\g)$ on $\CA$
with $U = |\g| -2$. It follows that
\eqn\impo{
\left(\int_{\S}d^{2}\th\; \bos{\phi}^{*}(\g),
\int_{\S}d^{2}\th\; \bos{\phi}^{*}(\g)\right)_{BV}
= \int_{\S}d^{2}\th\; \bos{\phi}^{*}([\g,\g]_{S}).
}

\subsubsection{Deformations}

\def\tT{\tilde{\bos{\triangle}}}

Now we examine deformations of the theory.
A $0$-dimensional observable with ghost number $U=p$ may be any
function on $\CA$ of the form
\eqn\cpa{
\g^{(0)} =\Fr{1}{p!}\g^{I_{1}\ldots I_{p}}(u^{I}(x^{\m}))
\c_{I_{1}}(x^{\m})\ldots\c_{I_{p}}(x^{\m}).
}
Any such $\g^{(0)}$ satisfies
$\bos{Q}_{o}\g^{(0)} =0$ since $
\bos{Q}_{o}u^{I}=\bos{Q}_{o}\c_{I}=0$.
It is obvious that no such $\g^{(0)}$ can be $\bos{Q}_{o}$-exact.
Thus an arbitrary functional $\g(u^{I}(x^{\m}),\c_{J}(x^{\m}))$ belongs to
the $\bos{Q}_{o}$ cohomology group. It is also obvious that
$\bos{\triangle} \g^{(0)} =0$.
The coefficients $\g^{I_{1}\ldots I_{p}}$ can be identified
with coefficients of a multivector $\G(X,\wedge^{p}TX)$.
Thus the space of $0$-dimensional observables is isomorphic to the
space
of all multivectors on $X$.
Now we denote by
$\bos{\g}:= \g^{(0)}(\bos{u}^{I},\bos{\c}_{I})$
the corresponding function of
superfields $(\bos{u}^{I},\bos{\c}_{I})$.
Equivalently $\bos{\g}:= \bos{\phi}^{*}(\g)$.
Then
we have the following expansions
\eqn\cjd{
\bos{\g} = \g^{(0)} + \left(\CK_{\m}\g^{(0)}\right) \th^{\m}
+\Fr{1}{2}\left(\CK_{\m}\CK_{\n}\g^{(0)}\right)\th^{\m}\th^{\n},
}
where $\CK_{\m}:= -\Fr{\rd}{\rd\th^{\m}}$.
We note that $\CK_{\m}$ is anticommuting and carries $U=-1$.
Together with $\bos{Q}_{o}$ we have the following anticommutation
relations
\eqn\cja{
\{{\boldsymbol{Q}}_{o} ,{\boldsymbol{Q}}_{o}\}=0,\qquad
\{{\boldsymbol{Q}}_{o} ,\CK_{\m}\}= - \rd_{\m},\qquad
\{\CK_{\m},\CK_{\n}\}=0.
}
The above relations and \cjd\ imply that
$\g^{(n)} := \Fr{1}{n!}
\left(\CK_{\m_{1}}\ldots \CK_{\m_{n}}\g\right)
dx^{\m_{1}}\wedge\ldots\wedge dx^{\m_{n}}$, where
$n=0,1,2$, satisfy the descent equations
\eqn\cjb{\eqalign{
\bos{Q}_{o}\g^{(0)} &=0,\cr
\bos{Q}_{o}\g^{(1)} &= - d \g^{(0)},\cr
\bos{Q}_{o}\g^{(2)} &= - d \g^{(1)}.\cr
}
}
Consequently $\int_\S\g^{(2)}\equiv \int_{\S}d^{2}\th\; \bos{\g}$
is invariant under $\bos{Q}_{o}$,
if $\g^{(1)}=0$ at the boundary $\rd\S$ of
$\S$, i.e., $\bos{Q}_{o}\int_\S\g^{(2)} =-\int_{\rd\S}
\g^{(1)}$.  
It is not difficult to check
$\bos{\triangle}\int_{\S}d^{2}\th\;\bos{\g}=0$,
with suitable regularization \cite{CF}.

Provided that $\g^{(1)}=0$ at the boundary $\rd\S$,
$\int_{\S}d^{2}\th\; \bos{\g}$
is a $2$-dimensional observable, which can be used to deform
the theory;
\eqn\cjc{
\bos{S}_{\g} = \bos{S}_{o} + \int_\S d^{2}\th\; \bos{\g}.
}
The above action functional satisfies
$\bos{Q}_{o}\bos{S}_{\g}
=\bos{\triangle}\bos{S}_{\g}=\CK_{\m}\bos{S}_{\g}=0$.
It follows, from the relation \impo, that
the deformed action functional satisfies the BV master equation if
$[\g,\g]_{S}=0$.
Note that the ghost number of
$\int_{\S}d^{2}\th\;\bos{\g}$
is shifted by $2$ from the ghost number of $\bos{\g}$.
We can view $\bos{\g}$ as the pull-back of a function on
$\Pi T^{*}X$ by a map $\Pi T\S \rightarrow \Pi T^{*}X$.
Thus an even 
(odd) function on $\Pi T^{*}X$ leads to an even (odd) function on $\CA$.

Now the boundary condition introduced earlier implies
that $\g^{(1)}=0$ at the boundary iff $\g(u^I,\c_I)|_{\c_I=0} =0$
for all $I$. Equivalently $\g|_{X}=0$. Then we may use
$\int_{\S}d^{2}\th\; \bos{\g}$ to deform the theory.
On the other hand such a function $\g$ on $\Pi T^*X$
does not give non-trivial observables supported
on the boundary. Non-trivial boundary observables must
come from functions on the base space $X$ of  $\Pi T^*X$,
which can not be used to deform the bulk theory.

The above considerations lead us to
consider the following action functional with $U=0$
\eqn\cjj{
\bos{S} = \bos{S}_{o} +\int_\S d^{2}\th\; \bos{\pi},
}
where $\bos{\pi}= \Fr{1}{2}b^{IJ}(\bos{u}^K)\bos{\c}_I\bos{\c}_J$
satisfying
\eqn\cpb{
\left(\int_{\S}d^{2}\th\;\bos{\pi},
\int_{\S}d^{2}\th\;\bos{\pi}\right)_{BV}=0.
}
The above condition implies
that the action functional $\bos{S}$ satisfies the
master equation $\left(\bos{S},\bos{S}\right)_{BV}=0$.
A parity changed bivector $\pi = \Fr{1}{2}b^{IJ}(u^{K})\c_{I}\c_{J}$
on $X$ satisfying $[\pi,\pi]_{S}=0$ is called a Poisson bi-vector.
Since $\bos{S}$ is an even function on the space of all fields
(the space of all maps $\Pi T\S\rightarrow \Pi T^{*}X$) $\CA$
we have an odd Hamiltonian vector $\bos{Q}$
defined by
$\bos{i}_{\bos{Q}}\bos{\o} =\bos{\d}\bos{S}$ or, equivalently
$\left(\bos{S},\ldots\right)_{{BV}}=\bos{Q}$,
regarding the Hamiltonian vector $\bos{Q}$ as an odd derivation.
Thus
$\bos{Q}$ is the BV-BRST charge of the action functional $\bos{S}$.
Explicitly
\eqn\snc{
\bos{Q}
= \left(\Fr{\rd {\bos{\pi}}}{\rd\bos{\c}_{I}}
+ D\bos{u}^{I}\right)\Fr{\rd}{\rd \bos{u}^{I}}
+\left(\Fr{\rd {\bos{\pi}}}{\rd\bos{u}^{I}}
+ D\bos{\c}_{I}\right)\Fr{\rd}{\rd \bos{\c}_{I}}.
}
%or
%\eqn\sncc{
%\eqalign{
%\tQ \bos{u}^{I}&=\Fr{\rd {\bos{\hat f}}}{\rd\bos{\c}_{I}}
%+ D\bos{u}^{I},\cr
%\tQ \bos{\c}_{I}&=
%\Fr{\rd {\bos{\hat f}}}{\rd\bos{u}^{I}}
%+ D\bos{\c}_{I}.
%}
%}
The BV master equation is equivalent to
the condition that $\bos{Q}$ is nilpotent.
Finally we note that
\eqn\cjakj{
\{{\boldsymbol{Q}} ,{\boldsymbol{Q}}\}=0,\qquad
\{{\boldsymbol{Q}} ,\CK_{\m}\}= - \rd_{\m},\qquad
\{\CK_{\m},\CK_{\n}\}=0.
}

The bosonic part of the action functional $\bos{S}$ is given by
the classical action functional $I$ in \iab. By construction,
compare with \cite{CF}, $\bos{S}$ is obtained by BV quantization of $I$.
We also mention that the anti-commutation relation \cjakj\
should be compared with the relations $L_{-1}=\{Q, b_{-1}\}$ and
$\bar L_{-1}=\{Q, \bar b_{-1}\}$ in $2$-dimensional conformal
field theory (see in particular \cite{WZ} on closely related issues
with this paper) or with a twisted $N_{ws} =(2,2)$ world-sheet
supersymmetry \cite{W1}.

\subsec{Open string and formality}

Now we recall the solution of deformation
quantization based on the path integral approach.

We first recall some basic properties of associative algebras (see
\cite{K2,Ts}).
On any associative algebra $A$ we have the Hochschild
co-chain complex $(\d, \oplus_{n} C^{n}(A,A))$
where $C^{n}(A,A)$ is the space of linear maps
$A^{\otimes n}\rightarrow A$ and $\d:C^{\bullet}(A,A)\rightarrow
C^{\bullet+1}(A,A)$ with $\d^{2}=0$ is the Hochschild differential.
For $A=C^{\infty}(X)$ the Hochschild co-chains are identified with the space
of multidifferential operators.
It is known that the Hochschild cohomology
$HH^{\bullet}(A,A)$ is isomorphic to the space of multivectors
$\G(X,\wedge^{\bullet}TX)$.\foot{Thus the
$\bos{Q}_{o}$-cohomology is isomorphic
to the total Hochschild cohomology $\oplus_{n} HH^{n}(A,A)$.}
We recall that the space $\G(X,\wedge^{\bullet+1}TX)\equiv
HH^{\bullet +1}(A,A)$ together
with the Schouten-Nijenhuis bracket $[.,]_{S}$  and zero-differential form
forms the dgLa called $T^{\bullet}_{poly}(X)$. This dgLa originates
from another dgLa called $D_{poly}(X)$
on the space $C^{\bullet +1}(A,A)$ with the Gerstenhaber (G in short)
bracket
$[.,.]_{G}$ and differential $\d$ and
associative cup product. The G bracket
\eqn\zva{
[.,.]_{G} : C^{n}(A,A)\otimes C^{m}(A,A) \rightarrow C^{m+n-1}(A,A)
}
is defined as follows. For $\Phi_{1}\in C^{n}(A,A)$
and $\Phi_{2}\in C^{m}(A,A)$
\eqn\zvb{
[\Phi_{1},\Phi_{2}]_{G} = \Phi_{1}\circ \Phi_{2}
-(-1)^{(n-1)(m-1)}\Phi_{2}\circ\Phi_{1},
}
where
\eqn\zvc{
\eqalign{
&\Phi_{1}\circ \Phi_{2}(a_{1},\ldots,a_{n+m-1})\cr
&\qquad
= \sum^{n-1}_{j=0}(-1)^{(m-1)j}\Phi_{1}(a_{1},\ldots,a_{j},
\Phi_{2}(a_{j+1},\ldots,a_{j+m}),\ldots).
}
}
The differential $\d:C^{\bullet}(A,A)\rightarrow C^{\bullet +1}(A,A)$
is defined by $\d =[m,\ldots]_{G}$ where $m(a,b)=ab$ for $a,b\in A$.
Then $[m,m]_{G}=0$ is equivalent to $m$ being associative,
which implies that $\d^{2}=0$.

Now for a given manifold $X$ a star product $f\star g$ among two
functions $f$ and $g$ is defined by
\eqn\zvf{
f\star g =f g + \Pi(f,g),
}
where $\Pi \in D^{1}_{poly}(X)$ has a formal expansion
\eqn\rrrdsas{
\Pi(f,g) = \Pi_{1}(f,g)\hbar + \Pi_{2}(f,g)\hbar^{2} +\ldots.
}
The associativity of the star product can be written
in terms of the MC equation for $D^{1}_{poly}(X)$ as
\eqn\zve{
\d \Pi +\Fr{1}{2}[\Pi,\Pi]_{G} = 0 \equiv [m +\Pi,m+\Pi]_{G}=0.
}
To first approximation the associativity implies
that $\Pi(f,g)_{1} = \left< \pi, df\otimes dg\right>\equiv
\Fr{1}{2}\{ f, g\}$ and the bracket $\{.,.\}$ satisfies
the Jacobi identity. Namely the bivector $\pi$ is Poisson and
$\{.,.\}$ is the Poisson bracket.
Equivalently
\eqn\zvd{
[\pi,\pi]_{S}=0,
}
which is 
the MC equation for $T^{1}_{poly}(X)$.

Consequently the problem of deformation quantization
for an arbitrary Poisson manifold $X$ is equivalent to
proving isomorphism between two moduli spaces
defined by the two MC equations \zve\ and \zvd.
Kontsevich proved the more general theorem ($L_{\infty}$ formality) that
the dgLa of  multi-vectors on $X$ with vanishing differential and
Schouten-Nijenhuis bracket
is  quasi-isomorphic to the dgla of multi-differential
operators on $X$. He also gave an explicit expression
for his quasi-isomorphism in the case of $X =\R^{n}$,
essentially summing over contributions from certain
graphs resembling Feynman path integral.

Catteneo-Felder studied the path integral
of  the theory  defined by the action
functional $\bos{S}$ on the disk $\S =D$ with boundary
punctures, where observables constructed from functions on $X$
are inserted. They obtained
Kontsevich's explicit formula  by perturbative
expansions around
constant maps \cite{CF}.
This formulation essentially maps the space of polyvectors on
the target space $X$ with vanishing Schouten-Nijenhuis bracket modulo
equivalence to the space  of solution of the BV master equation of the
theory deformed along the direction of polyvectors modulo equivalences.
The BV master equation then  implies according to \zai\
that the perturbative expansion of the theory defines
a quasi-isomorphism.
That is, the MC equation \zvd\ for $T^{1}_{poly}(X)$ is the BV master
equation, while the MC equation \zve\ is the Ward identity.

We remark that an associative algebra $A$ is
a $1$-algebra and its Hochschild cochain complex
$(C^{\bullet}(A,A),\d)$  has the structure of a $2$-algebra
by the cup product and the G-bracket $[.,.]_{G}$.
We note that the cup product is (super)-commutative
only up to homotopy.
The cohomology of the Hochschild complex is isomorphic to the space
of polyvectors on $X$ and also has a structure of a $2$-algebra
with the Schouten-Nijenhuis bracket (induced from $G$-bracket)
and the wedge product (induced from the cup product),
which is (super)-commutative.
We call the latter the cohomological $2$-algebra. The formality theorem
of Kontsevich means that the $2$-algebra and its cohomology are
equivalent up to homotopy as a $L_{\infty}$ algebra, thus forgetting
the product structure, for $A = C(X)$.
Tarmakin generalized the formality to the category of $G_{\infty}$
algebras for $X=\R^{n}$ \cite{Ta1}. Note that the Schouten-Nijenhuis
bracket can be generated by $\triangle = \Fr{\rd^{2}}{\rd
u^{i}\rd\c_{i}}$ for $X=\R^{n}$. Thus the cohomological $2$-algebra
is a BV algebra. The $BV_{\infty}$-formality is conjectured in
\cite{Kr}.
We do not want to review here all those purely algebraic approaches
(see however \cite{K3}).
We should also mention that those are closely related with
the open-closed string field theory of Zwiebach.
For us it should suffice to meditate OS.

\subsec{A model}

In this subsection we reexamine the
A model of the topological sigma model.
The A model was originally defined for an arbitrary almost complex
manifold \cite{W1}.
Our presentation for the A model will be similar to that of the
Catteneo-Felder
model in the previous section. Here, however, we will take the opposite
direction by starting from a dGBV algebra associated with a symplectic
manifold $X$. Then we will construct the corresponding two-dimensional
sigma model, which leads to the A model after gauge fixing.
A similar approach to the A model in the K\"{a}hler case is
discussed in \cite{AKSZ}.
We recall that the A model for the K\"{a}hler case is a twisted
version of a $N_{ws}=(2,2)$ world-sheet supersymmetric
sigma model.

\subsubsection{Covariant Schouten-Nijenhuis bracket}

Consider a manifold $X$ with
a Poisson bi-vector $\pi$, which
corresponds to even function $\pi = \Fr{1}{2}b^{IJ}\c_{I}\c_{J}$
on $\Pi T^{*}X$ with $[\pi,\pi]_{S}=0$.
The associated odd nilpotent Hamiltonian vector $Q_{\pi}$
is given by
\eqn\tba{
Q_{\pi}=
b^{IJ}\chi_{J}\Fr{\rd}{\rd {u}^{I}}
+\Fr{1}{2}\rd_{K}b^{IJ}\c_{I}½\c_{J}\Fr{\rd}{\rd \c_{K}}.
}
Since $Q_{\pi}^{2}=0$, $Q_{\pi}$ defines a cohomology on
$\Pi \G(\wedge^{\bullet}TX)$. The resulting cohomology
is called the Poisson cohomology  $H^{\bullet}_{\pi}(X)$ of
$X$ \cite{Lich}. For the symplectic case we are considering
the Poisson cohomology is isomorphic to de Rham cohomology.

Now we consider the dual picture in $\Pi TX$.
We introduce natural local coordinates
$(u^{I},\p^{I})$ on $\Pi TX$.
For any differentiable manifold
$X$ we have a distinguished odd vector $Q$ in $\Pi TX$
\eqn\tbc{
Q = \p^{I}\Fr{\rd}{\rd u^{I}}
}
of degree $U=1$ with $Q^{2}=0$.
It is
obtained by the parity change of the exterior derivative $d$ on $X$.
Any differential form on $X$ corresponds to a
function on $\Pi TX$. The $Q$-cohomology is isomorphic to the
de Rham cohomology. We consider a Poisson bi-vector $\pi$ on $X$
and define the associated contraction operator $i_{\pi}$;
\eqn\tbd{
\eqalign{
i_{\pi}&= \Fr{1}{2} b^{IJ}
\Fr{\rd^{2}}{\rd\p^{I}\rd \p^{J}}.\cr
}
}
Then  we define an odd second order differential operator
$\triangle$ with degree $U=-1$ by
\eqn\tbe{
\triangle_{\pi}:=[i_{\pi},Q] = b^{IJ}\Fr{\rd^{2}}{\rd u^{I}\rd\p^{J}}
+\Fr{1}{2}\rd_{I}b^{JK}\p^{I}\Fr{\rd^{2}}{\rd\p^{I}\rd\p^{J}}.
}
It is not difficult to show that the condition $[\pi,\pi]_{S}=0$
implies
\eqn\tbf{
\triangle^{2}_{\pi}= [Q, \triangle_{\pi}]=0.
}
The operator $\triangle_{\pi}$ acting on polynomial functions in $\Pi TX$
is  the parity changed version of the
Koszul-Brylinski boundary operator, which defines the Poisson homology
$H^{\pi}_{\bullet}(X)$ \cite{Kos, Br}. If $X$ is an unimodular Poisson
manifold 
(including the symplectic case)
with dimensions $n$ we have the duality $H^{\pi}_{p}(X) = H^{n-p}_{\pi}(X)$.
It follows that an element $\a \in H^{\bullet}(X)$ satisfies
$Q \a =\triangle_{\pi} \a=0$ for a symplectic manifold $X$.
For a K\"{a}hler manifold we have $\triangle_{\pi} = i\rd^{*} -i\bar\rd^{*}$
by the K\"{a}hler identity. From now on we omit the subscript $\pi$
from $\triangle_{\pi}$.

The operator $\triangle$ allows us to define an odd Poisson
structure on $\Pi TX$, which is the covariant version of the
Schouten-Nijenhuis bracket \cite{Kos}.
Let $a, b$ denote  functions on $\Pi TX$ with degree $U=|a|,|b|$.
The ordinary product $a \cdot b$ originates
from the wedge product on $\G(\wedge^{\bullet} T^{*}X)$.
It follows that $Q$ is a derivation while $\triangle$ fails to be a
derivation.
One defines the covariant Schouten-Nijenhuis bracket by the formula
\eqn\tbg{
[a\bullet  b] = (-1)^{|a|}
\triangle(aü\cdot  bü) -(-1)^{|a|}\triangle
a\cdot bü -
a\cdot \triangle b.
}
It is not difficult to check that $(Q,\triangle, \cdot, \CO(\Pi
TX)$ form a dGBV algebra \cite{Me}.

{}From now on we assume that $X$ is a symplectic manifold.
A symplectic
manifold is a Poisson manifold with a non-degenerated Poisson
bi-vector $\pi$
Let $b_{IJ}$ be the inverse of $b^{IJ}$. Then the condition
$[\pi,\pi]_{S}=0$
is equivalent to $d \o=0$, where $\o = \Fr{1}{2}b_{IJ}du^{I}\wedge
du^{J}$. Let $B = \Fr{1}{2}b_{IJ}\p^{I}\p^{J}$ be the parity
changed symplectic form. It is not difficult
to show that
\eqn\tbj{
\triangle B=0,\qquad
[B\bullet B] =0,\qquad [B\bullet a] = Q a.
}
Thus $Q$ is the Hamiltonian vector of $B$.

\subsubsection{A BV sigma model}

Now we consider a two dimensional sigma model
which is a theory of maps $\Pi T\S\rightarrow \Pi TX$.
We denote
local coordinates in $\Pi T\S$ by $(x^{\m},\th^{\m})$, $\m=1,2$.
We denote $(\bos{u}^{I}:=u^{I}(x^{\m},\th^{\m}),
\bos{\p}^{I}:=\p^{I}(x^{\m},\th^{\m})$ as the extension of
$(u^{I},\c^{I})$ to functions on $\Pi T\S$.
Let $\CA$ denote the space of all maps
$\bos{\phi}:\Pi T\S\rightarrow \Pi T X$.
The odd vector $Q$ on $\Pi TX$
can naturally be extended to an odd vector $\bos{Q}$
on $\CA$;
\eqn\tca{
\eqalign{
\bos{Q} =\bos{\p}^{I}\Fr{\rd}{\rd \bos{u}^{I}}.
}
}
It is obvious that $\bos{Q}^{2}=0$. We define the
action functional $\bos{S}$ of the theory by
$\int_{\S}d^{2}\th\;\bos{\phi}^{*}(B)$;
\eqn\tcb{
\boldsymbol{S} = \Fr{1}{2}\int d^{2}\th\;\left(
\bos{b}_{IJ}\bos{\p}^{I}\bos{\p}^{J}\right).
}
The odd symplectic structure on $\Pi TX$
induces an odd Poisson structure on $\CA$.
The BV operator $\bos{\triangle}$ is defined by
the formula $\bos{\triangle} : =[i_{\bos{\pi}},\bos{Q}]$.
The associated BV bracket, denoted by $(.,.)_{BV}$, is
the BV bracket of the theory. It follows from \tbj\ that
\eqn\tcc{
\bos{\triangle}\bos{S}=0,\qquad
\left(\bos{S},\bos{S}\right)_{BV} =0,\qquad
\left(\bos{S},\ldots\right)_{BV} =\bos{Q}.
}
Thus the action functional satisfies the quantum master equation.

We may expand the two superfields
$(\bos{u}^{I},\bos{\p}^{I})$ carrying $U=(0,1)$;
as follows
\eqn\tcd{
\eqalign{
\boldsymbol{u}^{I} &= u^{I}  +
\r^{I}_{\m}\th^{\m} +\Fr{1}{2} v^{I}_{\m\n}\th^{\m}\th^{\n},\cr
\boldsymbol{\p}^{I} &= \p^{I}  +
H^{I}_{\m}\th^{\m} +\Fr{1}{2} \eta^{I}ü_{\m\n}\th^{\m}\th^{\n}.\cr
}
}
As before the ghost number of $\th^{\m}$ is assigned the ghost number $U=1$.
The $\bos{Q}$ transformation law is
\eqn\tce{
\bos{Q}\bos{u}^{I} =\bos{\p}^{I},\qquad \bos{Q}\bos{\p}^{I}=0.
}
The explicit form of the action functional $\bos{S}$ is
\eqn\tcf{
\eqalign{
\bos{S} = \int_{\S}d \th^{2}&\biggl(
\Fr{1}{2}b_{IJ}H^{I}\wedge H^{J} + 2 b_{IJ}\eta^{I}\p^{J}
+\rd_{K}b_{IJ}\r^{K}\wedge H^{I}\p^{J}\cr
&
+\Fr{1}{2}\rd_{K}\rd_{L}b_{IJ}\r^{K}\wedge\r^{L}\p^{I}\p^{J}
+\rd_{K}b_{IJ}v^{K}\p^{I}\p^{J}\biggr).
}
}

\subsubsection{Gauge fixing}

Recall that the quantum master equation
is the condition that the path integral of the theory
$
\left<\CO\right>=\int_{\CL} d\m\; \CO e^{-\Fr{1}{\hbar}\bos{S}}
$
restricted to a Lagrangian submanifold $\CL \subset \CA$
is invariant under smooth deformations of $\CL$,
provided that $\CO$ is a BV observable,
$(\bos{Q}-\hbar\bos{\triangle})\CO=0$.
Picking a homology class of $\CL$ is called BV gauge fixing.
In the present case all the observables $\CO_{\a}$ will satisfy
$\bos{\triangle} \CO_{\a}=0$.
Thus $\CO_{\a}$ should be invariant under $\bos{Q}$.
Note also that the action functional $\bos{S}$ is
also invariant under $\bos{Q}$. Then we can apply
the fixed point theorem of Witten \cite{W3}, since $\bos{Q}$ generates
a global fermionic symmetry on $\S$. According to this
theorem the path integral is localized to an integral over
the fixed point locus of $\bos{Q}$.
Consequently the path integral is localized to an integral
over the fixed point locus $\CL_{0}$ in the Lagrangian
subspace $\CL$. Note that the space of all fields $\CA$
and its Lagrangian subspace $\CL$ are both infinite
dimensional superspaces.
Thus the path integral is difficult to make sense of. However we can
reduce the integral to an finite dimensional subspace $\CL_{0}$
due to the fixed point theorem.
Now we look for an appropriate gauge fixing.

On any symplectic manifold $(X, \o)$ there is an almost
complex structure $J \in \G(End(TX))$ compatible with $B$.
The almost complex structure $J^{I}_{J}$ obeying
\eqn\waa{
J^{I}{}_{K}J^{K}{}_{J}= -\d^{I}{}_{K}
}
is compatible with $b_{IJ}$ in the sense that
\eqn\wab{
b_{IJ} = J^{K}{}_{I}J^{L}{}_{J}b_{KL}.
}
Thus $b^{IJ}$ is of type $(1,1)$.
The above relation is equivalent to the following condition
\eqn\wac{
g_{IJ} = g_{JI},
}
where $g_{IJ} = b_{IK}J^{K}{}_{J}$ is a Riemannian metric
with torsion free connection.
We want to gauge fix the theory such that
the path integral is localized to
the moduli space of pseudo-holomorphic
maps
\eqn\wad{
d {u}^{I}+ J^{I}{}_{K}*d{u}^{K}=0,
}
where $d$ and $*$, $*^{2}=-1$ denote the exterior derivative
and the Hodge star in $\S$.
Note that the transformation laws of the anti-ghost multiplet
$(\r^{I}, H^{I})$ are
\eqn\wae{
\bos{Q}\r^{I} = -H^{I}, \qquad \bos{Q}H^{I}=0.
}
Thus $H^{I}=0$ at the fixed point locus.
The appropriate gauge fixing should be
$ H^{I} = d {u}^{I}+ J^{I}{}_{K}*d{u}^{K}=0$.
For this purpose we should impose the ``self-duality'' condition
of Witten \cite{W1}
\eqn\wself{
\eqalign{
\r^{I}&=*J^{I}{}_{K} \r^{K},\cr
H^{I}&=*J^{I}{}_{K} H^{K}.\cr
}
}
Then, as explained by Witten, the $\bos{Q}$ transformation law
on the ``self-dual'' part $\r^{+I}$ of $\r^{I}$ should be
\eqn\waf{
\eqalign{
\bos{Q} \r^{+I} =&- H^{+I}
-\Fr{1}{2}\left(\CD_{K}J^{I}{}_{J}\right)\p^{K}*\r^{+J}
+\G^{I}{}_{JK}\p^{J}\r^{+K},
}
}
where $\CD_{K}$ denotes the covariant derivative.
In the above the term proportional to
$\CD^{K}J^{I}{}_{L}$ is needed for the compatibility of ${\bos{Q}}$
with the ``self-duality'' \wself.
\eqn\wag{
{\bos{Q}}\r^{+I} = *\left({\bos{Q}}
J^{I}{}_{K}\right)\r^{+K} +
J^{I}{}_{K}{\bos{Q}} \r^{+K}.
}
The last term is needed for covariance with respect to
reparameterizations of $u^{I}$.
For this we regard $(H^{-}_{I},\r^{+I})$ as fields, while
$(\r^{-I}, H^{+}_{I})$ are the corresponding anti-fields.
Now we pick the following
gauge fermion $\Psi$ with $U=-1$
\eqn\wah{
\Psi =\int_\S b_{IJ}\r^{+I}\wedge\left(d u^{J}
+\Fr{1}{2}\left(\CD_{K}J^{J}{}_{L}\right)\p^{K}*\r^{+L}
-\G^{J}{}_{KL}\p^{K}\r^{+L}
\right).
}
Now the Lagrangian submanifold $\CL$ is
determined 
by the
following equations;
\eqn\wai{
b_{IJ}v^J=\Fr{\d\Psi}{\d \p^{I}},
\quad
b_{IJ}\r^{-I}:=\Fr{\d\Psi}{\d H^{-I}},\quad
b_{IJ}H^{+I}:=\Fr{\d\Psi}{\d \r^{+I}},\quad
b_{IJ}\eta^{J}:=\Fr{\d\Psi}{\d u^I}.
}
Thus we have, for instance,
\eqn\waj{
\eqalign{
\r^{-I} &=0,\cr
H^{I} &= \left(d u^{I} + *J^{I}{}_{K}du^{K}\right)üüü
+\Fr{1}{2}\left(\CD_{K}J^{I}{}_{L}\right)\p^{K}*\r^{+L}
-\G^{I}{}_{KL}\p^{K}\r^{+L},
}
}
as desired. Then the gauge fixed action functional
(or the action restricted
as a function on $\CL$) is exactly the action functional of
Witten's topological sigma model \cite{W1}. The transformation laws
of $\r^{I}$ after the gauge fixing (restricted to $\CL$) is then
\eqn\wak{
\eqalign{
\bos{Q} \r^{I} =&- \left(du^{I} + * J^{I}{}_{K}du^{K}\right)üüü
-\Fr{1}{2}\left(\CD_{K}J^{I}{}_{J}\right)\p^{K}*\r^{J}
+\G^{I}{}_{JK}\p^{J}\r^{K}.
}
}
Since $\p^{I}=0$ at the fixed point the fixed point locus
$\CL_{0}$ in $\CL$ is the space of pseudo holomorphic maps.

\newsec{Topological Open Membrane}

We want to study the theory of membranes in the background of a
$C$-field but no metric. The theory is defined by maps  from a
$3$-dimensional
world-volume $N$ with boundary to the target space $X$.
We assume that the canonical line bundle $\det (T^{*}N)$ is
trivial so that we have a well-defined volume element.
We describe a map $u: N\rightarrow X$ locally
by functions $u^{I}(x^{\m})$ where $\{x^{\m}\}$, $\m=1,2,3$
are local coordinates on $N$.
We consider a  $3$-form $C =\Fr{1}{3!}c_{IJK}(u^L)du^{I}\wedge
du^{J}\wedge du^{K}$ in $X$, where $\{u^{I}\}$ denote local
coordinates on $X$.  Now the action functional
of the membrane coupled to a $C$-field is, for $\rd N=0$,
\eqn\paa{
\eqalign{
{I}^{\pr}_{1} =\int_{N} u^{*}(C)
= \Fr{1}{3!}\int_{N} c_{IJK}(u^L)du^{I}\wedge du^{J}\wedge du^{K}.
}
}
To adopt the same strategy as for the string theory, we rewrite
the action in the following first order formalism:
\eqn\mbm{
\eqalign{
{I}_{1} = \int_{N}&\biggl(
F^{(2)}_{I}\wedge \left(A^{I} - d u^{I}\right)
-A^{I}\wedge d H_{I}\
+\Fr{1}{3!}c_{IJK}(u^L)A^{I}\wedge A^{J}\wedge A^{K}
\biggr),
}
}
where $A^{I}$ are $1$-forms in $N$ taking values in $u^{*}(TX)$,
while $F^{(2)}_{I}$ and $H_{I}$ are $2$ and $1$-forms, respectively,
taking values in $u^{*}(T^{*}X)$. The algebraic equation of motion
of $F^{(2)}_{I}$ imposes the constraint $A^{I} -du^{I}=0$ and
the $H_{I}$ equation of motion $dA^{I}=0$ gives the integrability
condition. Thus we have a well-defined first order formalism.
On shell we recover the action functional $I^\pr_1$ in \paa.
We might try to do BV quantization of the first order action
functional
$I_{1}$. It is however more appropriate to consider the membrane theory
without background and study consistent deformations.

We may define the bosonic membrane theory
without background by the following action functional
\eqn\mcd{
I_{o} =\int_{N}\biggl(
F^{(2)}_{I}\wedge \left(A^{I} - d u^{I}\right)
-A^{I}\wedge d H_{I}
\biggr),
}
where all fields above carry ghost number $U=0$.
{}From now on we remove the restriction $\rd N=0$.
The boundary conditions are such that
$A^{I}(x)$ and 
$* F^{(2)}_{I}$ vanish along the directions tangent
to $\rd N$ while $H_{I}(x)$ vanishes along the direction normal to
$\rd N$ for $x\in \rd N$.
On shell we have the boundary string without background,
\eqn\mbn{
{I}_o|_{\hbox{on shell}} = \int_{\rd N} H_{I}\wedge du^{I},
}
as a bonus of the integrability of the first order formalism.
We note that $I_{o}$ is invariant, up to a total derivative,
under the following
BRST symmetry (the bosonic symmetry after fermionization)
\eqn\mcg{
\eqalign{
\bos{Q}_{o} u^{I} &=\p^{I},\cr
\bos{Q}_{o}  A^{I} &= -d\p^{I},\cr
}\qquad
\eqalign{
\bos{Q}_{o} F^{(2)}_{I}&= -d\eta_{I},\cr
\bos{Q}_{o} H_{I} &= -d\c_{I} +\eta_{I},\cr
}
}
where $\p^{I}$ and $\c_{I}$ are $0$-forms taking values
in $u^{*}(\Pi TX)$ and $u^{*}(\Pi T^{*}X)$, respectively, with $U=1$
while $\eta_{I}$ are $1$-forms taking value in  $u^{*}(\Pi T^{*}X)$
with $U=1$.
We have $\bos{Q}_{o}  I_{o} = \int_{\rd N } (\eta_{I}\wedge du^{I} +
\p^{I}dH_{I})$. The boundary conditions are such that $\p^{I}(x)=0$
and $\eta_{I}(x)$ vanish along the directions tangent to $\rd N$,
while $H_{I}(x)=0$ along the direction normal to $\rd N$ for
$x\in \rd N$.
The above BRST transformation laws should be completed by
demanding $\bos{Q}_{o}^{2}=0$;
\eqn\pab{
\bos{Q}_{o} \p^{I} =0,\qquad \bos{Q}_{o}\eta_{I} =  dF_{I},
\qquad \bos{Q}_{o} \c_{I} = F_{I},
\qquad
\bos{Q}_{o}  F_{I}=0.
}
Note that we introduce a new scalar bosonic field $F_{I}$ taking values
in $u^{*}(T^{*}X)$ with $U=2$.\foot{From $\bos{Q}_{o}^{2}F^{(2)}_{I} =
- d\left(\bos{Q}_{o}\eta_{I}\right)=0$, we see that the general solution
for $\bos{Q}_{o}\eta_{I}$ is an exact $1$-form. The moral is that
wherever there is an ambiguity, which is actually
a gauge symmetry, there should be a new field (or ghost for ghost).}
The boundary condition is that $F_{I}(x)=0$ for $x\in \rd \S$.

Now all those fields appearing above
are regarded as ``fields''. One then introduces ``anti-fields''
as follows,
\eqn\pad{
\matrix{
\hbox{Fields}
& u^{I} & F_{I}& \eta_{I}& F^{(2)}_{I}& \c_{I} & H_{I}  & \p^{I} &
A^{I}\cr
\hbox{Anti-Fields} &\eta^{(3)}_{I} & \r^{I}_{(3)} & u^{I}_{(2)} &
\r^{I}
&A^{I}_{(3)}&\p^{I}_{(2)} &H^{(3)}_{I} & \c^{(2)}_{I}.
}
}
Here we used the convention that Latin letters denote bosonic (or even)
fields while Greek letters denote fermionic (or odd) fields.
If a ``field'' is a $n$-form on $N$ its ``anti-field'' is a $(3-n)$-form.
The ghost numbers $U$ of a ``field'' $\phi$ and its ``anti-field''
$\bar\phi$ are relates as $U(\bar\phi) = -1 - U(\phi)$.
Now we follows the usual steps to find a BV master action functional.
The resulting theory is described in the following subsection.

\subsec{BV quantized membrane without background}

We start from the total space of  $\Pi T^{*}X$,
with local coordinates $\left(\{u^{I}\}, \{\c_{I}\}\right)$.
Next, we consider the total space of $\Pi T(\Pi T^{*}X)$,
with local coordinates $\left(\{\p^{I}\},\{F_{I}\}\right)$ on the
the fiber. We assign
degrees or ghost numbers $U=(0,1,1,2)$ to $(u^{I},\c_{I},
\p^{I},F_{I})$.
Now we define the following
{\it even} symplectic structure on the space $\Pi T(\Pi T^{*}X)$
with degree $U=2$:
\eqn\mab{
\o = dF_{I}du^{I} + d\p^{I}d\c_{I}.
}
We note that the total space of $\Pi T(\Pi T^{*}X)$
can be identified with the total space of $T^*[2](\Pi T^{*}X)$.
Then the degree $U=2$ symplectic structure $\o$ can be
identified with the canonical symplectic form on
$T^*[2](\Pi T^{*}X)$ as a ``cotangent bundle''.

Based on the even symplectic structure $\o$ we define an even Poisson
bracket among functions in
$\left(\{\phi^{I}\}, \{\c_{I}\}|\{\p^{I}\},\{F_{I}\}\right)$,
i.e. functions on $\Pi T(\Pi T^{*}X)$, by the formula
\eqn\pxa{
\{\g_{1},\g_{2}\}_{P} = \Fr{\rd \g_{1}}{\rd F_{I}}\Fr{\rd \g_{2}}{\rd
u^{I}} 
- (-1)^{|\g_{1}||\g_{2}|}\Fr{\rd \g_{2}}{\rd F_{I}}\Fr{\rd \g_{1}}{\rd
u^{I}} 
+ \Fr{\rd \g_{1}}{\rd \p^{I}}\Fr{\rd \g_{2}}{\rd \c_{I}}
-(-1)^{|\g_{1}||\g_{2}|}\Fr{\rd \g_{2}}{\rd \p^{I}}
\Fr{\rd \g_{1}}{\rd\c_{I}},
}
where $|\g|$ denotse the degree $U$ of $\g$.
We note that the graded Poisson bracket has degree $U=-2$, i.e.,
$|\{\g_{1},\g_{2}\}_{P}| = |\g_{1}| +|\g_{2}| -2$.
It is not difficult to check the following properties:
\eqn\pxx{
\eqalign{
\{\g_{1},\g_{2}\}_{P} &=
-(-1)^{|\g_{1}||\g_{2}|}\{\g_{2},\g_{1}\}_{P},\cr
\{\g_{1}, \g_{2}\g_{3}\}_{P} &= \{\g_{1},\g_{2}\}_{P}\g_{3} +
(-1)^{|\g_{1}||\g_{2}+1|}\g_{2}\{\g_{1},\g_{3}\}_{P},\cr
\{\g_{1},\{\g_{2},\g_{3}\}_{P}\}_{P}
&= \{\{\g_{1},\g_{2}\}_{P},\g_{3}\}_{P}
+(-1)^{|\g_{1}||\g_{2}|}\{\g_{2},\{\g_{1},\g_{3}\}_{P}\}_{P}.
}
}
%{\it
The second relation above (the Leibniz law) implies that
the bracket behaves as a derivation of the ordinary product of
functions. Such a product is (super)-commutative and associative
and carries degree $0$.
Thus functions on $\Pi T(\Pi T^{*}X)$ form an algebra with a
degree $-2$ Poisson bracket, a (super)-commutative associative
product with degree $0$ and a vanishing differential (no differential
operator). We call such an algebra a cohomological $3$-algebra.
%}

On the space $\Pi T(\Pi T^{*}X)$ we have a canonical
nilpotent odd vector $Q_{o}$ with $U=1$
\eqn\pxb{
Q_{o} = \p^{I}\Fr{\rd}{\rd u^{I}} + F_{I}\Fr{\rd}{\rd\c_{I}},
}
owhich riginates from the exterior derivative on $\Pi T^{*}X$.
We find that $Q_{o}$ is the Hamiltonian vector of the following
odd function
\eqn\pxc{
h = F_{I}\p^{I}
}
carrying degree $U=3$. Note that an odd (even) function on
$\Pi T(\Pi T^{*}X)$ has odd (even) Hamiltonian vector since
our symplectic form is even.
We  have
\eqn\pxd{
\{h,h\}_{P}=0,\qquad
\{h, \ldots\}_{P} = Q_{o},\qquad Q_{o}^{2}=0.
}

The BV quantized topological open membrane is
a theory of maps
\eqn\sws{
\bos{\phi}: \Pi TN\rightarrow
\Pi T(\Pi T^{*}X),
}
where $\Pi TN$ is the parity change of the
total space of tangent bundle $TN$ of $N$.
We denote local coordinates on $\Pi TN$
by $(\{x^{\m}\},\{\th^{\m}\})$, $\m=1,2,3$. We parameterize a map
by functions 
\eqn\pxe{
(\bos{u}^{I},\bos{\c}_{I},\bos{\p}^{I},\bos{F}_{I})
:= (u^{I},\c_{I},\p^{I}, F_{I})(x^{\m},\th^{\m}).
}
Now we consider the space $\CA$
of all maps \sws. 
For any function $\g$ on $\Pi T(\Pi T^{*}X)$ we denote
the corresponding function of $(\{x^{\m}\},\{\th^{\m}\})$
by $\bos{\g}$, i.e., $\bos{\g} =\bos{\phi}^{*}(\g)
=\g(x^{\m},\th^{\m})$. We
have an expansion
\eqn\pxf{
\bos{\g} =\g (x^{\s}) + \g_{\m}(x^{\s})\th^{\m}
+\Fr{1}{2}\g_{\m\n}(x^{\s})\th^{\m}\th^{\n}
+\Fr{1}{3!}\g_{\m\n\r}(x^{\s})\th^{\m}\th^{\n}\th^{\r}.
}
We denote $\g^{(n)}
= \Fr{1}{n!}\g_{m_{1}\ldots\m_{n}}dx^{\m_{1}}\wedge\ldots\wedge
dx^{\m_{n}}$, where $n=0,\ldots,3$. We obtain functions
on $\Pi T\CA$ by 
$\int_{c_{n}}\g^{(n)}\equiv \int_{c_{n}}d^{n}\th\; \bos{\g}$,
where $c_{n}$ is a $n$-dimensional cycle in $N$.
We see that $\int_{M}d^{3}\th\; \bos{\g}$ is an even (odd) function
on $\CA$ if $\g$ is an odd (even) function on $\Pi T(\Pi T^{*}X)$,
with the degree shifted by $-3$.

On the space $\CA$ of all fields
we have the following odd symplectic structure
$\bos{\o}$ 
\eqn\mbb{
\bos{\o} =\int_{N}d^{3}\th\left( \d\bos{F}_{I}\d\bos{u}^{I}
+ \d\bos{\p}^{I}\d\bos{\c}_{I}\right).
}
We note that a $n$-form field is paired with a $(3-n)$-form field.
For instance the $0$-form part of $\bos{u}^{I}$, which is even,
is paired with the $3$-form part of $\bos{F}_{I}$, which is odd.
We also remark that $\th^{\m}$ carry degree $U=-1$ and
the degree of $\bos{\o}$ is $U=-1$.
Then we can define the BV bracket $(.,.)_{BV}$
as the odd Poisson bracket with respect to $\bos{\o}$ among
functions on $\CA$. The degree of the BV bracket is $U=1$.
We observe that the {\it odd} symplectic form $\bos{\o}$
(of degree $U=-1$) on $\CA$
is induced from the {\it even} symplectic structure $\o$
(of degree $U=2$) on  $\Pi T(\Pi T^{*}X)$.
Similarly the BV bracket (of degree $U=1$) among functions
on $\CA$ is induced from the even
Poisson bracket $\{.,.\}_{PB}$ (of degree $U=-2$)
among functions on $\Pi T(\Pi T^{*}X)$.
We also note that the relations in \pxx\ after shifting the degree
by $-3$ become the usual relations for the BV bracket.
It follows that 
\begin{quote}

$\{\g,\g\}_{P}=0$ if and only if
$\left(\int_{N}d^{3}\th\;\bos{\g},
\int_{N}d^{3}\th\;\bos{\g}\right)_{BV}=0$.

\end{quote}

We note that the superfields in \pxe\ contain all the ``fields'' and
``antifields''  \pad\ in the BV  quantization of membrane.
For the explicit identifications
we expand the superfields as follows:
\eqn\pya{
\eqalign{
\bos{u}^{I} &= u^{I} + \r^{I}_{\m}\th^{\m}
+ \Fr{1}{2}u^{I}_{\m\n}\th^{\m}\th^{\n}
+\Fr{1}{3!}\r^{I}_{\m\n\r}\th^{\m}\th^{\n}\th^{\s},\cr
\bos{F}_{I} &= F_{I} + \eta_{I\m}\th^{\m}
+ \Fr{1}{2}F_{I\m\n}\th^{\m}\th^{\n}
-\Fr{1}{3!}\eta_{I\m\n\s}\th^{\m}\th^{\n}\th^{\s},\cr
\bos{\p}^{I} &= \p^{I} + A^{I}_{\m}\th^{\m}
- \Fr{1}{2}\p^{I}_{\m\n}\th^{\m}\th^{\n}
-\Fr{1}{3!} A^{I}_{\m\n\r}\th^{\m}\th^{\n}\th^{\s} ,\cr
\bos{\c}_{I} &= \c_{I} + H_{I\m}\th^{\m}
+\Fr{1}{2!} \c_{I\m\n}\th^{\m}\th^{\n}
+ \Fr{1}{3!}H_{I\m\n\s}\th^{\m}\th^{\n}\th^{\s}.\cr
}
}
The ghost number of the superfields
$(\bos{u}^{I},\bos{\c}_{I},\bos{\p}^{I},\bos{F}_{I})$
are $U=(0,1,1,2)$, the same as the ghost numbers of $(u^{I},\c_{I},\p^{I},
F_{I})$. Note that the ghost number of $\th^{\m}$ is $U=1$.
As a differential form we write, for example,
$\r^{I}=\r^{I}_{\m}dx^{\m}$, $u^{I}_{(2)}=
\Fr{1}{2}u^{I}_{\m\n}dx^{\m}\wedge dx^{\n}$ and
$\r^{I}_{(3)} =\Fr{1}{3!}\r^{I}_{\m\n\s}dx^{\m}dx^{\n}dx^{\s}$,
etc.
Thus the  assignments of ``fields'' and ``anti-fields'' in \pad\
and the ghost numbers
are consistent with our definition of the odd symplectic structure
$\bos{\o}$ \mbb\ and the decompositions of the superfields.
One can also check that for any function $\g$ on $\Pi T(\Pi T^{*}X)$,
we have
\eqn\pyb{
\eqalign{
\bos{\triangle}\int_{N}d^{3}\th\; \bos{\phi}^{*}(\g)
&=(-1 + 3 -3 +1)\CC \int_{N} d v
\left(\Fr{\d^{2}}{\d F_{I} \d u^{I}} +\Fr{\d^{2}}{
\d \p^{I} \d \c_{I}}\right)\g^{(0)} \cr
&=0,
}
}
where $\CC$ is an infinite constant and $dv$ is the volume form on $N$.

The BV quantized version $\bos{S}_{o}$ of the action functional
$I_{o}$ in \mcd\ is given by
\eqn\mba{
\bos{S}_{o} = 
\int_{M}d^{3}\th\left(
\bos{\p}^{I}D \bos{\c}_{I}
+\bos{F}_{I}D\bos{u}^{I}
+\bos{F}_{I}\bos{\p}^{I}\right).
}
The BV BRST charge $\bos{Q}_{o}$ corresponds to
the odd Hamiltonian vector $\bos{Q}_{o} =
\left(\bos{S}_{o},\ldots\right)_{BV}$
of $\bos{S}_{o}$. We have
\eqn\mbaa{
\bos{Q}_{o} = D + \bos{\phi}^{*}(Q_{o}),
}
where $D =\th^{\m}\rd_{\m}$.
Explicitly,
\eqn\mbc{
\bos{Q}_{0} =\left(D\bos{u}^{I}+ \bos{\p}^{I}\right)\Fr{\rd}{\rd
\bos{u}^{I}} 
+ D\bos{\p^{I}}\Fr{\rd}{\rd\bos{\p}^{I}}
+\left(D\bos{\c}_{I} + \bos{F}_{I}\right)\Fr{\rd}{\rd\bos{\c}_{I}}
+D\bos{F}_{I}\Fr{\rd}{\rd\bos{F}_{I}}.
}
In components we see that the above BRST charge leads
to the transformation laws \mcg\ and \pab\
for the ``fields' as well as for the ``anti-fields''.

It is trivial to check that $\bos{S}_{o}$ satisfies
the quantum master equation,
\eqn\mbd{
\left(\bos{S}_{0},\bos{S}_{0}\right)_{BV}=0,\qquad
\bos{\triangle}\bos{S}_{0}=0.
}
Now we consider the case that the boundary of $N$ is non-empty.
Then we should impose suitable boundary conditions
such that $D\bos{S}_{o}=0$.
\eqn\mca{
\eqalign{
%\int_{\rd N}d\th\; \bos{F}_{I}\bos{\p}^{I} &=0,\cr
\int_{\rd N}d^{2}\th\; \bos{F}_{I}\bos{\p}^{I} &=0,\cr
\int_{\rd N}  d^{2}\th\; \left( \bos{\p}^{I} D\bos{\c}_{I}
+ \bos{F}_{I} D\bos{u}^{I}\right)&=0.
}
}
The above equations are satisfied by the boundary conditions
we introduced earlier.
We note that the $F_{I}$ enter into  the action functional
linearly.  Thus the integration over $F_{I}$ leads to a delta
function like constraint
\eqn\mcb{
D\bos{u}^{I} +\bos{\p}^{I}=0.
}
Then the on-shell action functional reduces to
the BV quantized boundary closed string theory without background,
\eqn\mcc{
\eqalign{
\bos{S}_{o}|_{\hbox{on shell}} &=
\int_{\rd N}d^{2}\th\left( \bos{\c}_{I}D\bos{u}^{I}
\right).
}
}

\subsection{Bulk deformations and boundary observables}

Now we consider a deformation the BV action functional $\bos{S}_{o}$
preserving the ghost number symmetry.
A deformed action functional is of the following form
\eqn\moa{
\bos{S}_{\g} = \bos{S}_{o} + \int_{N}\!\!d^{3}\th\; \bos{\phi}^{*}(\g),
}
where $\g$ is a function on $\Pi T(\Pi T^*X)$.
The idea and procedure behind the above
deformation
are exactly the same as those of the  string case discussed
extensively in Sect.~$2.2.2$.
The Brezin integral
$\int d^{3}\th$ will decrease the ghost number by $U=3$.
Thus ${\g}$ should have degree $U=3$ to preserve the
ghost number symmetry. Here we will impose such a condition.

The above deformation is
well-defined or admissible if $\bos{S}_{\g}$ satisfies
the quantum master equation,
\eqn\mob{
-\hbar\bos{\triangle}\bos{S}_{\g}
+\Fr{1}{2}\left(\bos{S}_{\g},\bos{S}_{\g}\right)_{BV} =0.
}
Since
$
\bos{\triangle}\bos{S}_{\g} =
\bos{\triangle}\int_{N}\!\! d^{3}\th\;\bos{\g}=0
$, 
the quantum master equation \mob\ reduces to
\eqn\mobdwssr{
\left(\bos{S}_{\g},
\bos{S}_{\g}\right)_{BV}=0.
}
The above is equivalent to the following
two conditions:
\eqn\moc{
\eqalign{
\int_{N}\!\!d^{3}\th\;D\bos{\phi}^{*}(\g) =0,\cr
\int_{N}\!\!d^{3}\th\;\bos{\phi}^{*}(Q_{o}\g +\Fr{1}{2}\{\g,\g\}_{P})
=0.
}
}
Note that the boundary conditions are such that
$\bos{\p}^{I}(x), \bos{F}_{I}(x)
=0$
in directions tangent to $\rd N_{p}$ for $x\in\rd N_{p}$.
Thus a consistent
deformation of the theory is determined by a degree
$U=3$ function $\g$ on $\Pi T(\Pi T^*X)$ satisfying
\eqn\nba{
\eqalign{
\g|_{\Pi T^{*}X}& =0,\cr
Q_{o}\g +\Fr{1}{2}\{\g,\g\}_{P} &=0.
}
}
The first condition means that the restriction of $\g(u^{I},\c_{I},
F^{I},\p_{I})$ to the base space $\Pi T^*X$ of $\Pi T(\Pi T^*X)$ vanishes.
We denote the set of equivalence classes of all solution
of \nba\ by $\mathfrak{M}^{cl}$.
Thus the space $\mathfrak{M}^{cl}$ is isomorphic to the set of
equivalence classes of all consistent bulk deformations (or backgrounds)
of the membrane theory $\bos{S}_{o}$ preserving the ghost
number symmetry $U$.

{}From now on we assume that $\g$ is a solution of \nba.
We denote by $Q_{\g}$ the degree $U=1$
odd nilpotent Hamiltonian  vector of $h+\g$;
\eqn\qwc{
\{h + \g, \ldots\}_{P} = Q_{\g}.
}
Explicitly 
\eqn\qwd{
\eqalign{
Q_{\g} = 
&\left( \p^{I} + \Fr{\rd \g}{\rd F_{I}}\right)\Fr{\rd}{\rd u^{I}}
+\left(F_{I} + \Fr{\rd \g}{\rd \p^{I}}\right)\Fr{\rd}{\rd \c_{I}}
\cr
& +\Fr{\rd\g}{\rd u^{I}}\Fr{\rd}{\rd F_{I}}
+\Fr{\rd\g}{\rd \c_{I}}\Fr{\rd}{\rd\p^{I}}.
}
}
Then the action functional $\bos{S}_{\g}$ in \moa\  has a
BRST symmetry generated by the odd nilpotent vector $\bos{Q}_{\g}$
with $U=-1$ defined by $\left(\bos{S}_{\g},\ldots\right)_{BV} =
\bos{Q}_{\g}$. Equivalently\foot{We should emphasize
that the BV master equation is equivalent to the
condition that $\bos{Q}_\g^2=0$, which implies that
$Q_\g^2 =0$ and the first equation of \moc.
Thus the moduli space of the theory is defined {\it not} by
the non-linear cohomology of $Q_\g$ but
by the non-linear cohomology of the BV BRST charge $\bos{Q}_\g$.
The difference is precisely encoded in the boundary
degrees due to the relation
$$
\bos{Q}_\g \int_N d^3\th \bos{\phi}^*(\a)
- \int_N d^3\th \bos{\phi}^*(Q_\g\a)
= \int_{\rd N}d^2\th \bos{\phi}^*(\a)|_{\rd N}.
$$
}
\eqn\qwe{
\bos{Q}_{\g} = D + \bos{\phi}^{*}(Q_{\g}).
}
Denoting  $\bos{\g}=\bos{\phi}^{*}(g(u^{I},\c_{I}, F_{I},\p^{I})
=\g(   \bos{u}^{I},\bos{\c}_{I}, \bos{F}_{I},\bos{\p}^{I})$
we have
\eqn\qwee{
\eqalign{
\bos{Q}_{\g} = 
&\left(D\bos{u}^{I} +\bos{ \p}^{I} + \Fr{\rd \bos{\g}}{\rd
\bos{F}_{I}}\right)\Fr{\rd}{\rd \bos{u}^{I}}
+\left(D\bos{\c}_{I} +\bos{F}_{I} + \Fr{\rd \bos{\g}}{\rd
\bos{\p}^{I}}\right)\Fr{\rd}{\rd \bos{\c}_{I}}
\cr
& +\left(D\bos{F}_{I} +\Fr{\rd\g}{\rd \bos{u}^{I}}\right)\Fr{\rd}{\rd
\bos{F}_{I}}
+\left(D\bos{\p}^{I} +\Fr{\rd\bos{\g}}{\rd
\bos{\c}_{I}}\right)\Fr{\rd}{\rd\bos{\p}^{I}}.
}
}

The action functional $\bos{S}_{\g}$ also has a
fermionic symmetry generated
by
$\CK_{\m}=-\Fr{\rd}{\rd \th^{\m}}$, acting on superfields, with $U=-1$.
Together with $\bos{Q}_{\g}$ they satisfy the following anti-commutation
relations,
\eqn\qwf{
\{ \bos{Q}_{\g},\bos{Q}_{\g}\} =0,\qquad \{\bos{Q}_{\g},
\CK_{\m}\} = -\rd_{\m},\qquad \{\CK_{\m},\CK_{\n}\}=0.
}
Given a $\bos{Q}_{\g}$ invariant function $A^{(0)}$  on $\CA$, which is
a $0$-form on $N$, the $n$-form  $A^{(n)}:=
\Fr{1}{n!}\left(\CK_{\m_{1}}\ldots\CK_{\m_{n}}
A^{(0)}\right)dx^{\m_{1}}\wedge \ldots \wedge dx^{\m_{n}}$
for $n=1,2,3$ give the canonical set of solutions of the descent
equations
\eqn\qwg{
\eqalign{
\bos{Q}_{\g} A^{(0)}&=0,\cr
\bos{Q}_{\g} A^{(1)}+ d  A^{(0)} &=0,\cr
\bos{Q}_{\g}  A^{(2)}+ d A^{(1)}&=0,\cr
\bos{Q}_{\g}  A^{(3)}+ d A^{(2)}&=0,\cr
d A^{(3)} &=0.
}
}
The above is a direct consequence of \qwe.

A BV observable in general is a function $\CO$ on $\CA$
satisfying $(-\hbar\bos{\triangle} +\bos{Q}_{\g})\CO=0$.
A zero-dimensional  observable $\CO^{(0)}(x)$ can be
inserted at a point in the interior or at the boundary of $N$.
Since $\CO^{(0)}$ 
is a scalar on $N$ it is a function of the scalar components
of the superfields only. It follows that $\bos{\triangle}\CO^{(0)}=0$
and $\bos{Q}_{\g}\CO^{(0)}=0$. The latter, together
with \qwe, implies that $\CO^{(0)}(x)$, $x\notin \rd\S$,
must derive from a $Q_{\g}$ cohomology class among functions on
$\Pi T\Pi(T^{*}X)$.
Recall that the boundary condition for $\p^{I}(x^{\m})$
and  $F_{I}(x^{\m})$ is that they vanish identically at the boundary.
Thus a zero observable inserted at a boundary point
originates from a function $f(u^{I},\c_{I})$ of the base $\Pi(T^{*}X)$ of
$\Pi T(\Pi(T^{*}X))$.
Note that no such function can be used to deform the
bulk action functional  due to the first condition  in \nba.
We may take for  $N$ a three-dimensional disk with boundary
$\rd N = S^{2}$, having a  number of marked points $x_{i}$ where
zero-dimensional observables are inserted. We can also
consider a small circle $C_{i}\subset \rd\S$ surrounding a marked point
$x_{i}$.
Then $\oint_{C_{i}}d^\th f(\bos{u}^{I},\bos{\c}_{I})$ defines
an observable, whose expectation value does not depend on the
contour. We have other observables $\int_{\rd N}d^{2}\th\;
f(\bos{u}^{I},\bos{\c}_{I})$, which may be viewed as boundary
interactions.

Now our goal is to determine the general $\g$ with $U=3$ satisfying
\nba. 
The explicit form of the above equation is easy to write down,
though complicated,
by the general form of a degree $U=3$ function $\g$ on
on $\Pi T(\Pi T^{*}X)$.

\subsubsection{Turning on a $C$-field}

Now we consider a bulk deformation leading
to the BV  quantized version of the action functional  of the
membrane with a flat $3$-form $C$-field \mbm.
For a $3$-form
$C = \Fr{1}{3!} c_{IJK}(u^L)du^I\wedge du^J\wedge du^K$ on
$X$, we obtain  a degree $U=3$ function $c$ on $\Pi T(\Pi T^{*}(X))$
\eqn\grb{
c = \Fr{1}{3!} c_{IJK}(u^L)\p^I\p^J\p^K,
}
satisfying $c|_{\Pi T^{*}X}=0$.
The condition $Q_{o} c +\Fr{1}{2}\{c,c \}_P=0$
is  equivalent to $d C =0$.
Thus we obtain the following action functional
satisfying the BV master equation
\eqn\gra{
\bos{S}_{c} =
\int_{N}d^{3}\th\left(
D \bos{\c}_{I}\bos{\p}^{I}
+\bos{F}_{I}D\bos{u}^{I}
+\bos{F}_{I}\bos{\p}^{I}\right)
+\Fr{1}{3!}\int_N d^3\th\left(
\bos{c}_{IJK} \bos{\p}^{I}\bos{\p}^{J}\bos{\p}^{K}
\right).
}
This is the BV quantized action functional of \mbm.
The action functional $\bos{S}_{c}$ is the Hamiltonian function,
$(\bos{S}_{c},\ldots)_{BV} =\bos{Q}_{c}$,
of the  odd vector $\bos{Q}$ given by (in terms of BRST
transformation laws)
\eqn\grc{
\eqalign{
\bos{Q}_{c}\bos{u}^{I}
=&D\bos{u}^{I}+ \bos{\p}^{I},
\cr
\bos{Q}_{c}\bos{\c}_{I}
=&D\bos{\c}_{I} + \bos{F}_{I}
+\Fr{1}{2}\bos{c}_{IJK}\bos{\p}^J\bos{\p}^{K},
\cr
\bos{Q}_{c}\bos{\p}^{I}
= &D\bos{\p}^{I} 
\cr
\bos{Q}_{c}\bos{F}_{I}
= &D\bos{F}_{I} 
-\Fr{1}{3!}\rd_{I}\bos{c}_{JKL}\bos{\p}^{J}\bos{\p}^{K}\bos{\p}^{L}.
}
}
On-shell we can eliminate $\bos{F}_{I}$ and $\bos{\p}^{I}$
using the $\bos{F}_{I}$ equation of motion;
\eqn\grd{
\eqalign{
\bos{S}_{c}|_{\hbox{on shell}} =& -\Fr{1}{3!}
\int_{N}d^{3}\th\left(
\bos{c}_{IJK}D\bos{u}^I D\bos{u}^{J}D\bos{u}^{K}
\right)
+\int_{\rd N}d^{2}\th\;\bos{\c}_{I}D\bos{u}^{I}.
}
}

Now consider an arbitrary function $f(u^{I},\c_{I})$ on the base
$\Pi T^{*} X$ of $\Pi T(\Pi T^{*}X)$. The corresponding function
$\bos{f}$ of the superfields has an expansion
\eqn\gre{
\bos{f} = f^{(0)}(x^{\m}) + \ldots.
}
Inserting $f^{(0)}(x_{i})$ at a boundary puncture $x_{i}$ we find
$\bos{Q}_{c} 
f^{(0)}=0$. It also follows that $\oint_{C_{i}}d\th \bos{f}$ and
$\int_{\rd N}d\th \bos{f}$ are BV observables.

Now we discuss more general deformations.
We consider the following non-linear transformations
of the fiber coordinates of $\Pi T(\Pi T^*X)\rightarrow \Pi T^*X$,
\eqn\cqb{
\eqalign{
F_I &\rightarrow F^\pr_I := F_I -\Fr{\rd \pi}{\rd u^I},\cr
\p^I&\rightarrow \p^{\pr I} := \p^I -\Fr{\rd \pi}{\rd \c_I},\cr
}
}
with $u^I$ and $\chi$ unchanged.
Here $\pi$ is an arbitrary degree $U=2$ function on the base space
$\Pi T^{*}X$ of $\Pi T(\Pi T^*X)$, i.e.,
$\pi = \Fr{1}{2}b^{IJ}(u^L)\c_I\c_J$.
Under the above transformation we have
$$
h +c
\rightarrow h+ \g,
$$ where
\eqn\cpd{
\eqalign{
\g  =&
-F_I \Fr{\rd \pi}{\rd\c_I} -\Fr{\rd \pi}{\rd u^I}\p^I
+\Fr{1}{2}[\pi,\pi]_S\cr
&
+\Fr{1}{3!}c_{IJK} \left(\p^I -\Fr{\rd \pi}{\rd\c_I} \right)
\left(\p^J -\Fr{\rd \pi}{\rd\c_J} \right)
\left(\p^K -\Fr{\rd \pi}{\rd\c_K} \right)
}
}
It can be easily shown that
$\{h+\g ,h +\g\}_P =0$ if and only if  $d C=0$.
The above class of solutions of \nba\ is determined
by an element of $H^3(X)$ and
a bi-vector on $X$.
Now the condition $\g|_{\Pi T^{*}X}=0$
implies that
\eqn\yra{
\Fr{1}{2}[\pi,\pi]_{S}
- \Fr{1}{3!}c_{IJK}\Fr{\rd \pi}{\rd\c_{I}}\Fr{\rd
\pi}{\rd\c_{I}}\Fr{\rd \pi}{\rd\c_{I}}=0.
}
More explicitly
\eqn\yrb{
h^{LMN} - c_{IJK}b^{IL}b^{JM}b^{KN} =0,
}
where ${h}^{IJK} = {b}^{LI}\rd_{L}{b}^{JK} +cyclic$.
Such a pair $(c,\pi)$ leads to the
following degree $U=3$ function on $T^{*}\Pi (T^{*}X)$
satisfying \nba,
\eqn\yyy{
\eqalign{
\g = &
-{b}^{IJ}{F}_{I}{\c}_{J}
-\Fr{1}{2}\left(\rd_{I}{b}^{JK}
+{c}_{IMN}{b}^{MJ}{b}^{NK}\right) {\c}_{J}{\c}_{K}{\p}^{I}
\cr
&
-\Fr{1}{2}{c}_{IJK}{b}^{IL}{\c}_{L}{\p}^{J}{\p}^{K}
+\Fr{1}{3!}{c}_{IJK}{\p}^I{\p}^{J}{\p}^{K}.
}
}
It follows that
the action functional
\eqn\mnc{
\eqalign{
\bos{S}_{\g} =
\int_{N}&\!\!d^{3}\th\biggl(
D \bos{\c}_{I}\bos{\p}^{I}
+\bos{F}_{I}D\bos{u}^{I}
+\bos{F}_{I}\bos{\p}^{I}
+\Fr{1}{3!}\bos{c}_{IJK} \bos{\p}^I\bos{\p}^{J}\bos{\p}^{K}
-\bos{b}^{IJ}\bos{F}_{I}\bos{\c}_{J}
\cr
&
-\Fr{1}{2}\bos{c}_{IJK}\bos{b}^{IL}\bos{\c}_{L}\bos{\p}^{J}\bos{\p}^{K}
-\Fr{1}{2}\left(\rd_{I}\bos{b}^{JK}
+\bos{c}_{IMN}\bos{b}^{MJ}\bos{b}^{NK}\right)
\bos{\c}_{J}\bos{\c}_{K}\bos{\p}^{I}
\biggr)
}
}
satisfies the quantum BV master equation
provided that $dC =0$.

The BV BRST transformation laws, with the condition \yrb,
are given by
\eqn\mnca{
\eqalign{
\bos{Q}\bos{u}^{I}
=&D\bos{u}^{I}+ \bos{\p}^{I} -\bos{b}^{IJ}\bos{\c}_{J},
\cr
\bos{Q}\bos{\c}_{I}
=&D\bos{\c}_{I} + \bos{F}_{I}
-\Fr{1}{2}\rd_{I}\bos{b}^{JK}\bos{\c}_{J}\bos\c_{K}
+\Fr{1}{2}\bos{c}_{IJK}\left(\bos{\p}^J -\bos{b}^{JM}\bos{\c}_M\right)
\left(\bos{\p}^K - \bos{b}^{KN}\bos{\c}_N\right),
\cr
\bos{Q}\bos{\p}^{I}
= &D\bos{\p^{I}} 
-\bos{b}^{IJ}\bos{F}_{J}
-\rd_{K}\bos{b}^{IJ}\bos{\p}^{K}\bos{\c}_{J}
-\Fr{1}{2}\bos{h}^{IJK}\bos{\c}_{J}\bos{\c}_{K}
\cr
&
-\Fr{1}{2}  \bos{c}_{LJK} \bos{b}^{LI}
\left(\bos{\p}^J -\bos{ b}^{JM}\bos{\c}_M\right)
\left(\bos{\p}^K - \bos{b}^{KN}\bos{\c}_N\right),
\cr
\bos{Q}\bos{F}_{I}
= &D\bos{F}_{I} 
+\rd_{I}\bos{b}^{JK}\bos{F}_{J}\bos{\c_{K}}
+\Fr{1}{2}\rd_{I}\rd_{K}\bos{b}^{LJ}\bos{\p}^{K}\bos{\c}_{L}\bos{\c}_{J}
+\Fr{1}{3!}\rd_{I}\bos{h}^{JKL}
\bos{\c}_{J}\bos{\c}_{K}\bos{\c}_{L}
\cr
& - \Fr{1}{3!}
\rd_I\bos{c}_{PJK} \left(\bos{\p}^P - \bos{b}^{PL}\bos{\c}_L\right)
\left(\bos{\p}^J -\bos{ b}^{JM}\bos{\c}_M\right)
\left(\bos{\p}^K - \bos{b}^{KN}\bos{\c}_N\right)
\cr
& +\Fr{1}{2}\bos{c}_{PJK} \rd_I \bos{b}^{PL}\bos{\c}_L
\left(\bos{\p}^J -\bos{ b}^{JM}\bos{\c}_M\right)
\left(\bos{\p}^K - \bos{b}^{KN}\bos{\c}_N\right).
}
}
We have $\bos{Q}_{\g}^2=0$, which follows from
$(\bos{S}_{\g},\bos{S}_{\g})_{BV}=0$.

Now we examine the use of
the above deformation.
{}From the $\bos{F}_{I}$ equation of motion
\eqn\mnd{
D\bos{u}^{I} - \bos{b}^{IJ}\bos{\c}_{J} +\bos{\p}^{I}=0
}
we can eliminate $\bos{\p}^{I}$ from the action functional,
leading to the following on-shell action functional
\eqn\mne{
\eqalign{
\bos{S}_{\g}|_{\hbox{on shell}} =& -\Fr{1}{3!}
\int_{N}d^{3}\th\left(
\bos{c}_{IJK}D\bos{u}^I D\bos{u}^{J}D\bos{u}^{K}
\right)
\cr
&
+\int_{\rd N}d^{2}\th\left(\bos{\c}_{I}D\bos{u}^{I}
+\Fr{1}{2}\bos{b}^{IJ}\bos{\c}_{I}\bos{\c}_{J}
\right).
}
}
We obtain a boundary string theory
in an arbitrary bivector background, while a closed $3$-form $C$-field
is coupled to the membrane in the bulk.
Note that the boundary action functional may be viewed
as the closed string version of the
Catteneo-Felder model. Recall that such an action functional
satisfies the quantum master equation (but with a different BV bracket)
if and only if $\pi$ is Poisson.
We may identify the open membrane theory defined by
the action functional $\bos{S}_{\g}$ \mnc\ together with
the condition \yrb\
as the off-shell closed string theory coupled to an arbitrary B-field.
%In the physical
%situation $\bos{b}^{IJ}$ is non-degenerated.
%Performing the Gaussian integral over $\bos{\c}_I$, we
%have
%\eqn\mne{
%\eqalign{
%\bos{S}|^\pr_{\hbox{on shell}} =& -\Fr{1}{3!}
%\int_{N}d^{3}\th\left(
%\bos{c}_{IJK}D\bos{u}^I D\bos{u}^{J}D\bos{u}^{K}
%\right)
%-\Fr{1}{2}\int_{\rd N}d^{2}\th\;\bos{b}_{IJ}D\bos{u}^{I}D\bos{u}^J.
%}
%}

Now we assume that the bi-vector is non-degenerate, and
therefore has an inverse. We then have a corresponding $2$-form
or anti-symmetric
tensor field on $X$,  $B = \Fr{1}{2}b_{IJ} du^I\wedge du^J$.
Then the condition \yrb\ implies that $C = dB$. Hence the $3$-form
$C$ is the field strength of the $B$-field.

\subsec{The first approximation}

In this subsection we discuss the first approximation of
the path integral  for a manifold $X$ with $c_{1}(X)=0$.
Our presentation will be indirect and the
actual path integral calculations
will appear elsewhere \cite{HLP}.

The first order problem
can be viewed as a ``quantization'' of $\Pi T (\Pi T^{*}X)$ viewed as
a classical phase space with respect to the even symplectic
structure $\o$ in \mab. Here we regard $(u^{I},\c_{I})$ as
the canonical coordinates and $(F_{I}, \p^{I})$ as the conjugate
momenta.
Now recall that the bulk term is determined by
the function $h +\g$ on $\Pi T\Pi (T^{*}X)$
\eqn\gva{
\eqalign{
h +\g = &
-{b}^{IJ}{F}_{I}{\c}_{J}
-\Fr{1}{2}\left(\rd_{I}{b}^{JK}
+{c}_{IMN}{b}^{MJ}{b}^{NK}\right) {\c}_{J}{\c}_{K}{\p}^{I}
\cr
&
+{F}_{I}{\p}^{I}
-\Fr{1}{2}{c}_{IJK}{b}^{IL}{\c}_{L}{\p}^{J}{\p}^{K}
+\Fr{1}{3!}{c}_{IJK}{\p}^I{\p}^{J}{\p}^{K}
}
}
satisfying the condition \yra.
By ``quantization'' of   $\Pi T (\Pi T^{*}X)$
we mean the following replacements
\eqn\gvb{
F_{I} \rightarrow   -\hbar \Fr{\rd}{\rd u^{I}},
\qquad
\p^{I} \rightarrow  -\hbar \Fr{\rd}{\rd \c_{I}},
}
where $\hbar$ is regarded as a formal parameter with $U=2$.
{}From $(h+\g)/\hbar$
we obtain the following differential operators acting on functions
on $\Pi T^{*}X$;
\eqn\gvc{
\CD = \CD_{1} +\hbar \CD_{2} +\hbar^{2} \CD_{3},
}
where
\eqn\gvd{
\eqalign{
\CD_{1} &= b^{IJ}\c_{J}\Fr{\rd}{\rd u^{I}} + \Fr{1}{2}\left(\rd_{I}{b}^{JK}
+{c}_{IMN}{b}^{MJ}{b}^{NK}\right) {\c}_{J}{\c}_{K}\Fr{\rd}{\rd\c_{I}},\cr
\CD_{2} &= \Fr{\rd^{2}}{\rd u^{I}\rd\c_{I}} -\Fr{1}{2}c_{IJK}b^{IL}\c_{L}
\Fr{\rd^{2}}{\rd\c_{J}\rd\c_{K}},\cr
\CD_{3}  &= \Fr{1}{3!}c_{IJK}\Fr{\rd^{3}}{\rd\c_{I}\rd \c_{J}\rd \c_{K}}.
}
}
Note that $\CD_{i}$ is  an order $i$ differential operator and has degree
$U=3-2i$. 
Now the conditions to satisfy the BV master equation,
$dC=0$ and \yra, imply that $\CD^{2}=0$;
\eqn\gvda{
\eqalign{
\CD_{1}^{2}=0,\cr
\CD_{1}\CD_{2} +\CD_{2}\CD_{1}=0,\cr
\CD_{1}\CD_{3} + \CD_{3}\CD_{1} + \CD_{2}^{2}=0,\cr
\CD_{2}\CD_{3} +\CD_{3}\CD_{2}=0,\cr
\CD_{3}^{2}=0.
}
}

The differential operators above defines various structures
on the algebra $\CO(\Pi T^{*}X)$ of functions on $\Pi T^{*}X$
(multivectors on $X$).
For $c=\pi =0$, therefore $\g=0$,
we have $\CD_{1}=\CD_{3}=0$, and $\CD_{2}=\triangle$
generates the Schouten-Nijenhuis bracket $[,.,]_{S}$
on functions on $\Pi T^{*}X$.
Together with the ordinary product we get the cohomological
$2$-algebra or GBV-algebra of $X$.
For $c=0$  the condition \yra\ implies that $[\pi,\pi]_{s} =0$.
Now we have $\CD_{1} =Q_{\pi}\equiv [\pi,\ldots]_{S}$,
$\CD_{2}= \triangle$
and $\CD_{3}=0$. Then $\CD_{1}$ induces a differential on
the cohomological $2$-algebra. Forgetting the product we have
a structure of dgLa on $\CO(\Pi T^{*}X)$, or in general a structure
of dGBV algebra. 

Akman \cite{A} (see also \cite{BDA}), motivated by VOSA and
generalizing Koszul, introduced the concept of higher
order differential operators on a general superalgebra.
Using such a differential operator, say $D$, he considered
the following recursive definition of higher
brackets, 
\eqn\gve{
\eqalign{
\Phi_{1}(a)=& D(a),\cr
\Phi_{2}(a,b)=& D(a \cdot b) - (D a)\cdot b - (-1)^{|a||D|}
a\cdot \CD(b),\cr
\Phi_{3}(a,b,c)=& \Phi_{2}(a, b\cdot c) - \Phi_{2}(a, b)\cdot c -
(-1)^{|b|(|a|+|D|}
b \cdot \Phi_{2}(a,c),\cr
&\vdots
\cr
\Phi_{r+1}(a_{1},\ldots, a_{r+1}) =&  \Phi_{r}(a_{1},\ldots,
a_{r}\cdot a_{r+1})
- \Phi_{r}(a_{1}, \ldots,a_{r})\cdot a_{r+1}\cr
&
- (-1)^{|a_{r}|(|a_{1}|+\ldots +|a_{r-1}+|D|} a_{r}\cdot
\Phi_{r}(a_{1},\ldots, a_{r-1}, a_{r+1}),\cr
}
}
such that for $D$ of order $\leq r$  $\Phi_{r+1}$ vanishes
identically. He examined the general properties
of those higher brakets.

Using the above
we define the following $2$- and $3$-brackets associated with $\CD$
among functions on $\Pi T^{*}X$
\eqn\gvf{
\eqalign{
[a,b] &= (-1)^{|a|}\Phi_{2}(a,b),\cr
[a,b,c] & = (-1)^{|b|}\Phi_{3}(a,b,c),\cr
}
}
while the higher brackets all vanish.
Explicitly
\eqn\brdj{
\eqalign{
[a,b] =& \hbar\Fr{\rd a}{\rd \c_{I}}\Fr{\rd b}{\rd u^{I}}
-\Fr{\hbar }{2}c_{IJK}b^{IL}\c_{L}\Fr{\rd a}{\rd\c_{J}}\Fr{\rd b}{\rd\c_{K}}
\cr
&
%+\Fr{\hbar^{2}}{2}c_{IJK} \Fr{\rd^{2} a}{\rd \c_{I}\rd\c_{J}}
%\Fr{\rd b}{\rd\c_{K}} 
- (-1)^{(|a|+1)(|b|+1)}\times ( a\leftrightarrow b),
\cr
[a,b,c] &= \hbar^{2}  c_{IJK}\Fr{\rd a}{\rd \c_{I}}
\Fr{\rd b}{\rd \c_{J}}\Fr{\rd c}{\rd \c_{K}}.
}
}
These brackets have the following properties:

(i) super-commutativity,
\eqn\gvg{
[a,b] = -(-1)^{(|a|+1)(|b|+1)}[b,a]
}

(ii) deformed Leibniz law,
\eqn\gvgg{
[a,bc] -[a,b]c - (-)^{(|a|+1)|b|}b[a,c] = (-1)^{|a|+|b|}[a,b,c].
}

(ii) derivation,
\eqn\gvh{
\CD [a,b] = [\CD a,b] + (-1)^{|a|+1}[a,\CD b].
}

(iii) Jacobi identity up to homotopy,
\eqn\gvi{
\eqalign{
[[a,b],c]  +&(-1)^{(|a|+1)(|b|+1)}[b,[a,c]] - [a,[b,c]]\cr
=& \CD[a,b,c] 
+(-1)^{|b|}[\CD a,b,c]
\cr
&
+(-1)^{|a|+|b|}[a,\CD b,c]
+(-1)^{|a|}[a,b,\CD c].
}
}

Now we return to some special cases. For $c=\pi=0$
the 2-bracket $[.,.]$ becomes the Schouten-Nijenhuis bracket.
Together with the product we have
\eqn\gvj{
\eqalign{
[a,b] - (-1)^{|a|}\triangle (a b) +(-1)^{|a|}(\triangle a) b +
a(\triangle b)=0,\cr
[a,b] +(-1)^{(|a|+1)(|b|+1)}[b,a]=0,\cr
[a,bc] -[a,b]c - (-)^{(|a|+1)|b|}b[a,c] =0,\cr
[[a,b],c]  +(-1)^{(|a|+1)(|b|+1)}[b,[a,c]] -
[a,[b,c]]=0,\cr
\triangle [a,b] - [\triangle a,b]
- (-1)^{|a|+1}[a,\triangle b]=0.
}
}
For $c=0$ with Poisson bi-vector $\pi$ we have
the Schouten-Nijenhuis bracket above and the
differential $Q_{\pi}$ satisfying;
\eqn\gvk{
\eqalign{
Q_{\pi}^{2}&=0,\cr
Q_{\pi} [a,b] - [Q_{\pi} a,b] - (-1)^{|a|+1}[a, Q_{\pi} b]&=0.
}
}
Thus in the $c=0$ case we have the usual structue of a dGVB algebra
for a Poisson manifold $X$ with
$c_{1}(X) =0$ on $\CO(\Pi T^{*}X)$.

Hence the actual deformation comes from non-zero
$c$.  We remark that Kravchenko defined a $BV_{\infty}$-algebra
by a sum of differentials of degree  $3-2i$ and
order $i$, whose square is zero.
Hence the deformed algebra $\CO(\Pi T^{*}X)$
of functions on $\Pi T^{*}X$ induced by the differential
operator $\CD$ in \gvc\ has a structure of $BV_{\infty}$-algebra.
It has a $L_{\infty}$ structure with brackets and a first order
differential.

\subsection{Generalization}

We can relax the requirement that the deformation
terms should have ghost number $U=0$, and
allow them to have any even ghost number.
Now we let $C_{k}=\Fr{1}{(2k+1)!}c_{I_{1}\ldots {I_{2k+1}}}
du^{I_{1}}\wedge\ldots\wedge du^{I_{2k+1}}$ 
be a closed $(2k+1)$-form, $k=0,1,\ldots$ on $X$;
\eqn\gdaa{
d C_k =0.
}
Then we have the corresponding degree $U=2k+1$ function $c_k$ on
$\Pi (T\Pi T^{*}X)$;
\eqn\gdab{
c_{k}=\Fr{1}{(2k+1)!}c_{I_{1}\ldots {I_{2k+1}}}\p^{I_{1}}\ldots
\p^{I_{2k+1}},
}
satisfying 
\eqn\gda{
\eqalign{
\sum_{k}c_{k}|_{\Pi T^{*}X}=0,\cr
Q_{o}\sum_{k}c_{k} 
+\Fr{1}{2}\left\{\sum_{k}c_{k},\sum_{k}c_{k}\right\}_{P}=0,\cr
}
}
Now we let $\pi_{\ell}$ be an arbitrary degree $U=2\ell$ function
on the base
of
$\Pi T(\Pi T^*X)\rightarrow \Pi T^*X$, i.e.,
$\pi_{\ell} = \Fr{1}{(2\ell )!}b^{I_{1}\ldots
I_{2\ell}}(u^L)\c_{I_{1}}\ldots\c_{I_{2\ell}}$.
Then we consider the following non-linear transformations
of the fiber coordinates of $\Pi T(\Pi T^*X)\rightarrow \Pi T^*X$,
\eqn\gdb{
\eqalign{
F_I &\rightarrow F^\pr_I
:= F_I -\sum_{\ell}\Fr{\rd \pi_{\ell}}{\rd u^I},\cr
\p^I&\rightarrow \p^{\pr I} := \p^I -\sum_{\ell}
\Fr{\rd \pi_{\ell}}{\rd
\c_I},\cr
}
}
with $u^I$ and $\chi$ unchanged.
Under the above transformation we have
\eqn\gdc{
h + \sum c_{k }
\rightarrow h+ \G,
}
where
\eqn\gdd{
\eqalign{
\G  =&
-\sum_{k}\left(  F_I \Fr{\rd \pi_{k}}{\rd\c_I} + \Fr{\rd \pi_{k}}{\rd
u^I}\p^I\right)
+\Fr{1}{2}\left[\sum_{k}\pi_{k},\sum_{k}\pi_{k}\right]_S\cr
&
+\sum_{k}\Fr{1}{(2k+1)!} c_{ I_{1}\ldots I_{2k+1}}
\left(\p^{I_{1}} -\sum_{\ell}\Fr{\rd \pi_{\ell} }{ \rd\c_{I_{1}}} \right)
\cdots
\left(\p^{I_{2k+1}} -\sum_{\ell}\Fr{\rd \pi_{\ell}}{\rd\c_{I_{2k+1}}}
\right)
}
}
It can be easily shown that
$Q_{o}\G +\Fr{1}{2}\left\{\G ,\G\right\}_P =0$
if and only if  $d C_{k}=0$ for all $k$.
Now the condition $\G|_{\Pi T^{*}X}=0$
implies that
\eqn\gde{
\Fr{1}{2}\left[\sum_{\ell}\pi_{\ell},\sum_{\ell}\pi_{\ell}\right]_{S}
-\sum_{k}  \Fr{1}{(2k+1)!}c_{I_{1}\ldots I_{2k+1}}
\left(\Fr{\sum_{\ell}\rd \pi_{\ell}}{\rd\c_{I_{1}}}\right)
\cdots\left( \Fr{\sum_{\ell }\rd \pi_{\ell}}{\rd\c_{I_{2k+1}}}\right)=0.
}

Assuming the above conditions
we get the following action functional
\eqn\gdf{
\bos{S}_{\G} =\bos{S}_{o} + \int_{N}d^{3}\th\;\bos{\phi}^{*}(\G),
}
satisfying the BV master equation.
{}Using the $\bos{F}_{I}$ equation of motion
\eqn\gdg{
D\bos{u}^{I} - \sum_{\ell}\Fr{\rd \pi_{\ell}}{\rd\c_{I}}
+\bos{\p}^{I}=0,
}
we have
the following on-shell action functional
\eqn\gdh{
\eqalign{
\bos{S}_{\G}|_{\hbox{on shell}} =& -\sum_{k}\Fr{1}{(2k+1)!}
\int_{N}d^{3}\th\left(
\bos{c}_{I_{1}\ldots I_{2k+1}}D\bos{u}^{I_{1}}\ldots
D\bos{u}^{I_{2k+1}}
\right)
\cr
&
+\int_{\rd N}d^{2}\th\left(\bos{\c}_{I}D\bos{u}^{I}
+\sum_{\ell}\Fr{1}{(2\ell)!}\bos{b}^{I_{1}\ldots I_{2\ell}}\bos{\c}_{I_{1}}
\ldots \bos{\c}_{I_{2\ell}}
\right).
}
}

We may also allow for arbitrary ghost numbers
and the most general form for the solutions.
Such a case may be used to determine explicit quasi-isomorphism
of the $3$-algebra.

\subsection{Other boundary conditions}

In this subsection we consider some variants of the
topological open membrane by changing the boundary
conditions. There can be more general boundary conditions
than mentioned here.  We pick two of them, which
are relevant to mirror symmetry.

\subsubsection{A boundary conditions}

Now we consider the following general action functional
\eqn\gga{
\bos{S}_\G = \int_{N}\!\!d^{3}\th\biggl(
\bos{\p}^{I}D \bos{\c}_{I}
+\bos{F}_{I}D\bos{u}^{I}
+\bos{F}_{I}\bos{\p}^{I}
+\bos{\phi}^*(\G)\biggr),
}
where $\G$ is defined by \gdd\ with the condition \gdaa.
Then the above action functional satisfies the BV master equation
if and only if
\eqn\ggb{
\int_{N}\!\!d^{3}\th\; D\biggl(
\bos{\p}^{I}D \bos{\c}_{I}
+\bos{F}_{I}D\bos{u}^{I}
+\bos{F}_{I}\bos{\p}^{I}
+\bos{\phi}^*(\G)\biggr)=0.
}
Thus we may exchange the boundary conditions of $\bos{\c}_I$ and
$\bos{\p}^I$, such that $\c_I(x)=0$, and $H_I(x)=0$
along the direction tangent to $\rd\S$,
while $A^I(x)=0$ along the direction normal to $\rd\S$
for $x \in \rd\S$.
Then we must set $c_k=0$ for all $k$ instead of \gde\
to satisfy
the master equation.
Hence the following action functional satisfies the BV master
equation
\eqn\ggc{
\eqalign{
\bos{S} =& \int_{N}\!\!d^{3}\th\biggl(
\bos{\p}^I D\bos{\c}_{I}
+\bos{F}_{I}D\bos{u}^{I}
+\bos{F}_{I}\bos{\p}^{I}\biggr)\cr
&-\sum_{k} \int_{N}\!\!d^{3}\th\biggl(
\bos{F}_I \Fr{\rd \bos{\pi}_{k}}{\rd\bos{\c}_I}
+ \Fr{\rd \bos{\pi}_{k}}{\rd \bos{u}^I}\bos{\p}^I\biggr)
\cr
&
+\Fr{1}{2} \int_{N}\!\!d^{3}\th\biggl(\bos{\phi}^*\left(
\left[\sum_{k}\pi_{k},\sum_{k}\pi_{k}\right]_S\right).\cr
}
}
%In the above we performed a integration by part to make $\bos{\p}^I$
%looks like ``canonical coordinates''.
{}From the point of view of the target superspace, we are replacing $\Pi
T(\Pi T^*X)$
with $\Pi T^*(\Pi T X)$. In general $T^*X \rightarrow TX$ is an isomorphism
only when $X$ is a symplectic manifold. In our case
$\Pi T(\Pi T^*X) \rightarrow \Pi T^*(\Pi TX)$ is an isomorphism
since our even symplectic structure $\o$ does not see the
difference. Now the boundary observables of the theory
are derived from functions on $\Pi TX$, which are differential forms on $X$.
We denote by $\CO(\Pi TX)$ the algebra of functions on  $\Pi TX$,
which has the ordinary product (the wedge product).

Now we consider the role of the bulk deformations
to first approximation.
For simplicity we only consider the deformations preserving the
ghost number symmetry;
\eqn\ggea{
\eqalign{
\bos{S} =& \int_{N}\!\!d^{3}\th\biggl(
\bos{\p}^{I}D\bos{\c}_{I}
+\bos{F}_{I}D\bos{u}^{I}
\biggr)\cr
&+ \int_{N}\!\!d^{3}\th\biggl(
\bos{F}_{I}\bos{\p}^{I}
-\bos{b}^{IJ}\bos{F}_I\bos{\c}_J
+\Fr{1}{2}\rd_I \bos{b}^{JK}\bos{\p}^I\bos{\c}_J\bos{\c}_K
+\Fr{1}{3!}\bos{h}^{IJK}\bos{\c}_I\bos{\c}_J\bos{\c}_K
\biggr).
}
}
Then we obtain the following differential operators
\eqn\ggd{
\tilde\CD = \tilde \CD_1 + \hbar \tilde\CD_2 +\hbar^2 \tilde \CD_3
}
where
\eqn\gge{
\eqalign{
\tilde \CD_1 &= \p^I\Fr{\rd}{\rd u^I},\cr
\tilde \CD_2 &= b^{IJ}\Fr{\rd^2}{\rd u^I\rd\p^J}
+\Fr{1}{2}\rd_I b^{JK}\p^I\Fr{\rd^2}{\rd\p^J\rd\p^K},\cr
\tilde \CD_3 &= \Fr{1}{3!}h^{IJK}\Fr{\rd^3}{\rd\p^I\rd\p^J\rd\p^K}.\cr
}
}
It is not difficult to show that $\tilde\CD^2=0$ for any bi-vector;
\eqn\ggf{
\eqalign{
\tilde\CD_{1}^{2}=0,\cr
\tilde\CD_{1}\tilde\CD_{2} +\tilde\CD_{2}\tilde\CD_{1}=0,\cr
\tilde\CD_{1}\tilde\CD_{3} + \tilde\CD_{3}\tilde\CD_{1}
+\tilde \CD_{2}^{2}=0,\cr
\tilde\CD_{2}\tilde\CD_{3} +\tilde\CD_{3}\tilde\CD_{2}=0,\cr
\tilde\CD_{3}^{2}=0.
}
}
Thus we obtain another $BV_\infty$ structure.

We encountered the operator  $\tilde \CD_1$
and $\tilde \CD_2$ in Sect.~$2.4$
where we used the notations $Q$ and $\triangle_\pi$.
The latter generated, for $[\pi,\pi]_S=0$,
the covariant Schouten-Nijenhuis bracket on
the differential forms.  Those operators induced
a dGBV structure on the algebra $\CO(\Pi TX)$
of functions on $\Pi TX$.
%Assuming the the bivector has inverse we can eliminate
%$\bos{F}_{I}$ and $\bos{\c}_{I}$ to obtain the following on-shell
%action functional
%\eqn\ggg{
%\bos{S}|_{on shell} =
%}
We remark that Manin associated a Frobenius structure
to a special kind of dGBV algebra \cite{Ma}.
It is shown that the above dGBV algebra defines a Frobenius
structure when $X$ is symplectic manifold satisfying
the strong Lefschetz condition \cite{Me}.
Any K\"ahler manifold is such a manifold, and
the above construction reproduces the so called AKS theory
\cite{BS}. AKS theory is the A model version of Kodaria-Spencer gravity
\cite{BCOV}.

In general $[\pi,\pi]_S\neq 0$ and
we find a homotopy version of a dGBV algebra.
It will be interesting to examine the corresponding
deformation of the Frobenius structure.
We note that the relation between this subsection
and Sect.~$3.3$
can be seen as the relation of cochains versus chains.
We remark that there is also a non-commutative version of the
above differential operators at least for $h^{IJK}=0$
\cite{Ts,TT}. In such a case the cohomology of $\tilde\CD$
is closely related to the periodic cyclic homology
of the Hochschild chain complex of the deformed algebra
of functions on $X$ with the star product.
It will be interesting to see if $\tilde\CD$ in general
has a non-commutative version, which is not necessarily
associative. We also note that the periodic cyclic homology
is the non-commutative version of the de Rham cohomology.
Can there be an A model on non-commutative space?
If so the extended moduli space of the theory may be
identified with the periodic cyclic homology.

\subsubsection{B boundary conditions}

Now we assume that the target space $X$ is a complex
Calabi-Yau space. We introduce a complex structure
and consider the open membrane theory without background,
\eqn\ggeaz{
\eqalign{
\bos{S}_{o} =& \int_{N}\!\!d^{3}\th\biggl(
\bos{\p}^{i}D\bos{\c}_{i}
+\bos{\p}^{\bar i}D\bos{\c}_{\bar i}
+\bos{F}_{i}D\bos{u}^{i}
+\bos{F}_{\bar i}D\bos{u}^{\bar i}
+\bos{F}_{i}\bos{\p}^{i}
+\bos{F}_{\bar i}\bos{\p}^{\bar i}
\biggr).
}
}
Now consider the original boundary conditions.
We can exchange the boundary conditions of
$\bos{\c}_{\bar i}$ and $\bos{\p}^{\bar i}$,
while maintaining the original boundary conditions
for $\bos{\c}_{i}$ and $\bos{\p}^{i}$.
Then $\c_{\bar i}(x)=\p^{i}(x) =0$ and the $1$-forms
$H_{\bar i}(x)$ and $A^{i}(x)$ vanish along the direction
tangent to $\rd\S$ for $x\in \rd\S$.
Then the action functional \ggeaz\ satisfies the master
equation. 
Now the boundary observables are derived from functions
on $\Pi\CT^{*}(\Pi \bar\CT X)$, namely the elements of
$\oplus\O^{0,\bullet}(X, \wedge^{\bullet}\CT X)$.
The general form of the ghost number
$U=2$ function 
on $\Pi\CT^{*}(\Pi \bar\CT X)$
is
\eqn\rob{
\b := a +\k + b,
\qquad
\left\{
\eqalign{
a &:=a_{\bar i}{}^{j}\c_{j}\p^{\bar i},\cr
\k&:=\Fr{1}{2}\k_{\bar i\bar j}\p^{\bar i}\p^{\bar j},\cr
b&:= \Fr{1}{2}b^{ij}\c_{i}\c_{j}.
}\right.
}
We note that the BRST transformation laws
restricted to boundary are
\eqn\roba{
\eqalign{
\bos{Q}_{o}u^{i} &= 0,\cr
\bos{Q}_{o}u^{\bar i} & =\p^{\bar i},\cr
}\qquad
\eqalign{
\bos{Q}_{o}\c_{i} &=0,\cr
\bos{Q}_{o}\p^{\bar i}&=0.
}
}
Hence the BRST charge acts like the $\bar \rd$-operators.

The undeformed theory induces the following
differential operators acting on functions
on $\Pi\CT^{*}(\Pi \bar\CT X)$ -- which can be identified with the elements
of
$\oplus\O^{0,\bullet}(X, \wedge^{\bullet}\CT X)$ --
\eqn\roc{
\CD_{o} = \bar\rd + \hbar \triangle_{T},
\qquad
\left\{
\eqalign{
\bar\rd &= \p^{\bar i}\Fr{\rd}{\rd u^{\bar i}},\cr
\triangle_{T} & = \Fr{\rd^{2}}{\rd u^{i}\rd\c_{i}}.
}\right.
}
We have $\CD_{o}^{2}=0$ and
\eqn\roe{
\bar\rd^{2} 
= \bar\rd\triangle_{T}+\triangle_{T}\bar\rd = \triangle_{T}^{2}=0.
}
The operators together with the wedge product endow
$\oplus\O^{0,p}(X, \wedge^{q}\CT X)$ with
the structure of dGBV algebra
\cite{BK,Ma,B1}. The operator $\triangle_{T}$ generates
a bracket (Tian-Todorov lemma) $[.,.]_{T}$, whose
holomorphic Schouten-Nijenhuis bracket together with the
wedge product on forms define a dGBV algebra.
This forms the classical algebra
of observables.  We note that the solution space of
the MC equation of this classical algebra modulo equivalences
defines the so-called extended moduli space of complex structures
on $X$ \cite{BK,B1}. This also induces a Frobenius structure
\cite{BK,Ma}, generalizing \cite{BCOV}, which is relevant to mirror
symmetry.

Now we consider a degree $U=3$ function $\g$ on $\Pi T (\Pi T^{*}X)$
satisfying
\eqn\nbba{
\eqalign{
\g|_{\p^{i}=\c_{\bar i}=F_{I}=0}& =0,\cr
Q_{o}\g +\Fr{1}{2}\{\g,\g\}_{P} &=0,
}
}
compare with \nba.
Then the action functional
\eqn\rof{
\bos{S}_{\g} = \bos{S}_{o} + \int_{N}d^{3}\th\; \bos{\phi}^{*}(\g)
}
satisfies the BV master equation.
The deformation term will deform the dGBV algebra (the
classical algebra of observables) as a $2$-algebra.
The deformation is controlled by the MC equation \nbba\
of the $3$-algebra.
It will be interesting to examine the corresponding
deformation of the Frobenius structure.

Here we determine a class of solutions of \nbba.
The desired $\g$ is given by
\eqn\eua{
\eqalign{
\g = &F_{i}\Fr{\rd\b}{\rd\c_{i}}
+ \p^{i}\Fr{\rd\b}{\rd u^{i}}
+\p^{\bar i}\Fr{\rd\b}{\rd u^{\bar i}}
+ \Fr{1}{2}[\b,\b]_{T}
\cr
&
+\Fr{1}{3!}c_{ijk}
\left(\p^{i} +\Fr{\rd\b}{\rd\c_{i}}\right)
\left(\p^{j} +\Fr{\rd\b}{\rd\c_{j}}\right)
\left(\p^{k} +\Fr{\rd\b}{\rd\c_{k}}\right),
}
}
satisfying
\eqn\eub{
\eqalign{
\bar\rd\b
+ \Fr{1}{2}[\b,\b]_{T}
+ \Fr{1}{3!}[\b,\b,\b]_{T}=0,\cr
\bar\rd c + [\b, c]_{T} =0.
}
}
Here we used the following definitions
\eqn\eud{
\eqalign{
&c := \Fr{1}{3!}c_{ijk}\p^{i}\p^{j}\p^{k},\cr
&[\b,\b,\b]_{T}:= c_{ijk}
\Fr{\rd\b}{\rd\c_{i}}\Fr{\rd\b}{\rd\c_{j}}\Fr{\rd\b}{\rd\c_{k}}.
}
}
The first condition in \eub\ comes from the first condition in \nbba,
while the second condition in \eub\ comes from the second condition
in \nbba. 

We note that the first equation in \eub\ is the flatness
condition $\bar\rd^{2}_{\b}=0$
of the ``covariant'' derivative $\bar\rd_{\b}$;
\eqn\eue{
\bar\rd_{\b} := \bar \rd + [\b,.]_{T} + \Fr{1}{2}[\b,\b,.]_{T}.
}
Thus the equations \eub\ become
\eqn\euf{
\bar\rd_{\b}^{2}=0,\qquad
\bar\rd_{\b} c=0,
}
where we used $[\b,\b,c]_{T}=0$.
The equations in \eub\ can be written as follows;
\eqn\euc{
\eqalign{
{\bar \rd} a + \Fr{1}{2}[a,a]_{T}=0,\cr
[b,\k]_{T}+\Fr{1}{2}[b,a,a]_{T}=0,\cr
\Fr{1}{2}[b,b]_{T} +\Fr{1}{3!}[b,b,b]_{T} =0,\cr
}\qquad
\eqalign{
{\bar \rd} \k + [a,\k]_{T} +\Fr{1}{3!}[a,a,a]_{T}
=0,\cr
{\bar \rd} b + [a,b]_{T} +\Fr{1}{2}[a,b,b]_{T}=0,\cr
{\bar \rd} c + [a,c]_{T}=0.\cr
}
}

Now the deformed theory induces the following
differential operators acting on functions on
$\Pi\CT^{*}(\Pi \bar\CT X)$, identified with the elements of
$\oplus\O^{0,\bullet}(X, \wedge^{\bullet}\CT X)$;
\eqn\roc{
\CD_{\g} = \CD_{1} + \hbar \CD_{2}+\hbar \CD_{3},
}
where
\eqn\rocc{
\eqalign{
\CD_{1} &= \p^{\bar i}\Fr{\rd}{\rd u^{\bar i}}
+\Fr{\rd\b}{\rd \c_{i}}\Fr{\rd}{\rd u^{i}}
+\Fr{\rd\b}{\rd u^{i}}\Fr{\rd}{\rd \c_{i}}
+\Fr{1}{2}c_{ijk}\Fr{\rd\b}{\rd\c_{i}}
\Fr{\rd\b}{\rd\c_{j}}\Fr{\rd}{\rd\c_{k}}
,\cr
\CD_{2} & = \Fr{\rd^{2}}{\rd u^{i}\rd\c_{i}}
+ \Fr{1}{2}c_{ijk}\Fr{\rd\b}{\rd\c_{i}}\Fr{\rd^{2}}{\rd\c_{j}\rd\c_{k}}
,\cr
\CD_{3} & = c_{ijk}\Fr{\rd^{3}}{\rd\c_{i}\rd\c_{j}\rd\c_{k}}.
}
}
Once again we have $\CD_{\g}^{2}=0$ from \eub;
\eqn\gvdab{
\eqalign{
\CD_{1}^{2}=0,\cr
\CD_{1}\CD_{2} +\CD_{2}\CD_{1}=0,\cr
\CD_{1}\CD_{3} + \CD_{3}\CD_{1} + \CD_{2}^{2}=0,\cr
\CD_{2}\CD_{3} +\CD_{3}\CD_{2}=0,\cr
\CD_{3}^{2}=0.
}
}
The operators together with the wedge product endow
$\oplus\O^{0,p}(X, \wedge^{q}\CT X)$
with a structure of $BV_\infty$ algebra.

We may consider more general boundary observables of the theory
than $\b$ in \rob.
For this we consider an arbitrary function $\a(u^{i}, u^{\bar
i}, \c_{i}, \p^{\bar i})$ of $(u^{I},\c_{i},\p^{\bar i})$ on
$\Pi\CT^{*}\Pi (\bar\CT X)$, which is an element of
$\oplus \O^{0,\bullet}(X, \wedge^{\bullet}\CT X)$.
Now we can replace $\b$ with $\a$ in the equations
\eua\ and \eub\ to obtain a more general solution
of the BV master equation \nbba. We can also replace $\b$ with
$\a$ in \rocc.
Then the BV master equation is given by
\eqn\eubrs{
\eqalign{
\bar\rd\a
+ \Fr{1}{2}[\a,\a]_{T}
+ \Fr{1}{3!}[\a,\a,\a]_{T}=0,\cr
\bar\rd c + [\a, c]_{T} =0,
}
}
while the differentials acting on functions on
$\Pi\CT^{*}(\Pi \bar\CT X)$ are
\eqn\roccrs{
\eqalign{
\CD_{1} &= \p^{\bar i}\Fr{\rd}{\rd u^{\bar i}}
+\Fr{\rd\a}{\rd \c_{i}}\Fr{\rd}{\rd u^{i}}
+\Fr{\rd\a}{\rd u^{i}}\Fr{\rd}{\rd \c_{i}}
+\Fr{1}{2}c_{ijk}\Fr{\rd\a}{\rd\c_{i}}
\Fr{\rd\a}{\rd\c_{j}}\Fr{\rd}{\rd\c_{k}}
,\cr
\CD_{2} & = \Fr{\rd^{2}}{\rd u^{i}\rd\c_{i}}
+ \Fr{1}{2}c_{ijk}\Fr{\rd\a}{\rd\c_{i}}\Fr{\rd^{2}}{\rd\c_{j}\rd\c_{k}}
,\cr
\CD_{3} & = c_{ijk}\Fr{\rd^{3}}{\rd\c_{i}\rd\c_{j}\rd\c_{k}}.
}
}

Now we consider the special case that $c=0$.
Then we are left with only the boundary degrees of freedom
and the master equation \nbba\ or equivalently \eubrs\ reduces to
\eqn\eufds{
\bar \rd\a + \Fr{1}{2}[\a,\a]_T =0.
}
The differential operators $\CD$ acting on functions on
$\Pi\CT^{*}(\Pi \bar\CT X)$ become
\eqn\rocxc{
\eqalign{
\CD_{1} &= \p^{\bar i}\Fr{\rd}{\rd u^{\bar i}}
+\Fr{\rd\a}{\rd \c_{i}}\Fr{\rd}{\rd u^{i}}
+\Fr{\rd\a}{\rd u^{i}}\Fr{\rd}{\rd \c_{i}}
,\cr
\CD_{2} & =\triangle_T= \Fr{\rd^{2}}{\rd u^{i}\rd\c_{i}},\cr
\CD_{3} & = 0.
}
}
We see that the boundary deformations correspond
to deformations of the first order differential operator.
We shall see that those correspond to deformations of
the BRST operator of the B model of the topological closed
string theory. We shall also see that the
solution space of \eufds\ modulo equivalences
defines the extended moduli space of the B model of
the topological string theory or the extended moduli
space of complex structures. Now turning on the
$c$-field gives rise to bulk deformations beyond
the extended moduli space of the boundary theory.

\newsec{Back to the strings}

In this section we discuss some applications to
homological mirror symmetry. In the previous section
we already discussed about related issues, though
on a level one 
step up (the open membrane).

\subsec{Relation with $N_{ws}=(2,2)$ supersymmetric sigma models}

We consider the following two functions carrying $U=3$
\eqn\xpxc{
\eqalign{
h &= F_{I}\p^{I},\cr
\tilde h &= b^{IJ}F_{I}\c_{J} + \Fr{1}{2}\rd_{K}b^{IJ}\p^{k}\c_{I}\c_{J},
}
}
where $\pi =\Fr{1}{2}b^{IJ}\c_{I}\c_{J}$ is a Poisson bivector,
i.e. $[\pi,\pi]_{S}=0$.
Then
it is easy to check that
\eqn\pxd{
\{h, h\}_{P}=0,\qquad
\{h,\tilde h\}_{P}=0,\qquad
\{\tilde h,\tilde h\}_{P}=0.
}
Thus the Hamiltonian vectors $\{h.\ldots\}_{P} = Q_{o}$ and
$\{\tilde h,\ldots\}_{P} = \tilde Q_{o}$
satisfy the following relations
\eqn\pxd{
\{Q, Q\}=0,\qquad 
\{Q,\tilde Q\}=0,\qquad
\{\tilde Q,\tilde Q\}=0.
}
Explicitly
\eqn\xpxb{
\eqalign{
Q_{o} = & \p^{I}\Fr{\rd}{\rd u^{I}} + F_{I}\Fr{\rd}{\rd \c_{I}},\cr
\tilde Q_{o} = &b^{IJ}\c_{J}\Fr{\rd}{\rd u^{I}} +
\Fr{1}{2}\rd_{I}b^{JK}\c_{J}\c_{K}\Fr{\rd}{\rd \c_{I}}
- \left(b^{IJ}F_{J} +\rd_{K}b^{IJ}\p^{k}\c_{J}\right)\Fr{\rd}{\rd \p^{I}}
\cr
&
-\left(\rd_{I} b^{JK}F_{J}\c_{K} +
\Fr{1}{2}\rd_{I}\rd_{J}b^{LM}\p^{J}\c_{L}\c_{M}\right)\Fr{\rd}{\rd
F^{I}}.
}
}

We note that the two odd vectors $Q_{o}$ and $\tilde Q_{o}$
are the supercharges of a two dimensional $N_{ws}=(1,1)$
supersymmetric non-linear
sigma model after dimensional reductions to zero dimension.
Then the commuting coordinate field $F_{I}$ correspond to the
auxiliary fields. Now we assume that $X$ is a K\"{a}hler manifold
and the bivector is the inverse of the K\"{a}hler form. We introduce
a complex structure and decompose the coordinates according to
$u^{I}=u^{i} + u^{\bar i}$. We decompose the tangent vectors,
$\p^{I} =\p^{i} + \p^{\bar i}$, and cotangent vectors,
$\c_{I} = \c_{\bar i}+\c_{i}$ and
$F_{I}=F_{\bar i}+F_{i}$, accordingly.
Then the odd vectors $Q_{o}$ and $\tilde Q_{o}$ are decomposed
into holomorphic and anti-holomorphic vectors;
$Q_{o} = Q^{\pr}_{o} + Q^{\ppr}_{o}$,
$\tilde Q_{o} = \tilde Q^{\pr}_{o} + \tilde Q^{\ppr}_{o}$
and all four odd vectors are mutually nilpotent.
Again the four odd vectors correspond to
the supercharges of a two dimensional $N_{ws}=(2,2)$
supersymmetric non-linear
sigma model after dimensional reduction to zero dimensions.
Now we imagine undoing the dimensional reduction. Then our four
odd vectors transform as the left or right moving spinors under the
two-dimensional rotation group of a Riemann surface $\S$.
They also carry so called $U(1)_{\CR}$
charges.

As was originally proposed \cite{D, LVW}, the mirror symmetry
is very natural from a physical viewpoint. On the other hand
the symmetry becomes quickly mystifying once we translate it
into mathematical language in terms of Gromov-Witten invariants
and the variation of Hodge structures.
An application of mirror
symmetry in adopting the latter viewpoint
marked the first grand  success of the
mirror conjecture in algebra-arithmetic geometry \cite{COGP},
after
the first construction of a mirror pair \cite{GP}.
Later Witten proposed
a unified and effective viewpoint based on twisted versions (A and B
models) of $\CN=(2,2)$ supersymmetric sigma models \cite{W3}.
Additionally, Witten also argued that both the original A and B models
should be extended and  the mirror symmetry would be
more obvious in the extended models. In this language the mirror
symmetry is the physical equivalence between the extended A model
$A^{e}(X)$ on $X$  and the extended B model $B^{e}(Y)$ on the mirror
$Y$.

The twisting procedure is
explained in detail in \cite{W3, BCOV}. There are two different
twistings, leading to the A and the B model, respectively.
There is also a half-twisted
model which we will not consider here.
Note that the B model makes sense
iff $X$ is a Calabi-Yau space with Ricci-flat metric.
After twisting two supercharges
transform as scalars on $\S$. The other two supercharges
transform as vectors. The auxiliary fields $F_{i}$ and
$F_{\bar i}$ transform as a vector for the A model,
while decomposing into a scalar and
a two-form in the B model.
For the A model, the two anti-commuting scalar supercharges (BRST charges)
are given by
\eqn\mia{
\eqalign{
\CQ^{\pr}_{o} &= \p^{i}\Fr{\rd}{\rd u^{i}} +\ldots,\cr
\CQ^{\ppr}_{o}&= \p^{\bar i}\Fr{\rd}{\rd u^{\bar i}} +\ldots,
}
}
where the omitted the part involving non-scalar fields.
For the B model the two BRST charges are given by
\eqn\mib{
\eqalign{
\CQ^{\ppr}_{o} &= \p^{\bar i}\Fr{\rd}{\rd u^{\bar i}}
+F_{i}\Fr{\rd}{\rd \c_{i}}+\ldots,\cr
\tilde\CQ^{\ppr}_{o}&=
b^{\bar i j}\c_{j}\Fr{\rd}{\rd u^{\bar i}}
- \left(b^{\bar i j}F_{j} +\rd_{\bar k}b^{\bar i
j}\p^{k}\c_{j}\right)\Fr{\rd}{\rd \p^{\bar i}}
+\ldots,
}
}
where we again omitted the part involving non-scalar fields.

At this level we already see the sharp distinction between the A
and the B model.  As a general principle observables of the
theory are constructed by BRST cohomology classes.
The zero-dimensional observables, transforming as scalars,
are constructed from scalar fields. For the A model the answer
is obvious.  The two BRST charges originated
from  the $\rd$ and $\bar \rd$ differentials of the target space $X$.
Thus the zero-dimensional observables are the pull-backs
of de Rham cohomology classes of $X$ after a parity change.

On the other hand the $\CQ^{\ppr}_{o}$ charge can better
be interpreted as $\bar\rd$-operator on the total superspace
$\Pi T^{*}X$. Furthermore, the $\rd$ part of BRST charge
does not exist. The other BRST charge $\tilde\CQ^{\ppr}_{o}$
can be interpreted as the holomorphic part of the Poisson differential
on the superspace $\Pi T^{*}X$. Consequently the zero-dimensional
observables of the B model should come from the Dolbeault cohomology
of the superspace $\Pi T^{*}X$. However it is not clear
how it can be defined. The usual approach is based on an
on-shell formalism\foot{On-shell the term
$(b^{\bar i j}F_{j} +\rd_{\bar k}b^{\bar i j}\p^{k}\c_{j})$ vanishes
due to an equation of motion.} and taking the diagonal.
The zero-dimensional observables are elements of the
sheaf cohomology $H^{0,\bullet}(X, \wedge^{\bullet}T^{1,0}X)$,
modulo equations of motion. Then one solves the descent equations
to construct two dimensional observables.
However some problems appear in constructing higher
dimensional observables by solving the descent equations.
Two dimensional observables are very important since
one can add them by integrating over $\S$ to the action
functional to define a family of the theory.
Witten gave a recipe for the B model with some first order
analysis and shows that the classical moduli space $\CM_{cl}$
of complex structures should be extended \cite{W3}. He argued that
the tangent space of the extended moduli space $\CM$
at a classical point is
$\oplus H^{0,\bullet}(X, \wedge^{\bullet}T^{1,0}X)$.
A noble point of the above analysis is that the BRST transformation
laws should be changed recursively. This is a perfect lead
to the method of BV quantization.

The question on the extended moduli space $\CM$
was the starting point of the homological mirror
conjecture of Kontsevich \cite{K1}. Roughly there are
three questions. (i) define (or find equations for)
the moduli space $\CM$. (ii) define generalized
periods. (iii)  what is the meaning of $\CM$ or which
kind of object does it parameterize?. The first and second
questions are answered by Barannikov and Kontsevich
\cite{BK, B1} using purely mathematical techniques
based on modern deformation theory. They also
constructed explicit formulas for the tree level potential, which
is the generalization of the Kodaira-Spensor
theory of Bershadsky et. al \cite{BCOV} (see also \cite{Ma}).
However the
corresponding generalization for string loops is not yet
achieved. The third question remains still mysterious.
Kontsevich conjectured that $\CM$ parameterizes the $A_{\infty}$
deformation of $D^{b}(Coh(X))$. This conjecture was answered
affirmatively in \cite{B1},
based on the formality theorem of Kontsevich.

The homological mirror conjecture seems to be closely
related to the SYZ conjecture \cite{SYZ,Va,KS2}. In many
respects the SYZ construction makes mirror symmetry
obvious as a physical equivalence \cite{SYZ}.
Recently Hori et. al. \cite{HV,HQV} developed
another physically natural approach which has a
highly constructive
power.

\subsubsection{Cohomological ABC model}

It is possible to describe the A, B and Catteneo-Felder (C in short) models
in a unified fashion.
For this we consider a theory based on the following map
\eqn\mic{
\bos{\w}: \Pi T\S \rightarrow \Pi T^{*}(\Pi (TX)).
}
We let $\CA_{\S}$ denotes the space of all maps.
We denote a system of local coordinates on $\Pi TX$
by $(\{u\}^{I},\{\p^{I}\})$, while by $(\{\c_{I}\},\{F_{I}\})$
for the fiber of   $\Pi T^{*} (TX)$.
We assign the ghost numbers of the coordinates by
$U(u^{I},\p_{I}, \c_{I}, F_{I})= (0,1,1,0)$.
We may use the same component notations
as for the membrane case \pya\
by simply setting the $3$-form parts to zero.
We describe a map \mic\ by the superfields
\eqn\mid{
(\bos{u}^{I},\bos{\p}^{I},\bos{\c}_{I},\bos{F}_{I})
:= (u^{I},\c_{I},\p^{I}, F_{I})(x^{\m},\th^{\m}),
}
where $(\{x^{\m}\},\{\th^{\m}\})$, $\m=1,2$
denote a set of local coordinates on $\Pi T\S$.
On 
$\Pi T^{*}\Pi (T X)$ we have the following
odd symplectic form with $U=1$,
\eqn\mie{
\o_{\S} = du^{I}d\c_{I} + d\p^{I}d F_{I}.
}
The corresponding odd Poisson bracket is an extension
of the Schouten-Nijenhuis bracket.
On $\CA_{\S}$ we have an induced odd symplectic form
$\bos{\o}_{\S}$ with $U=-1$
\eqn\mif{
\bos{\o}_{\S} = \int_{\S}d^{2}\th\left(\d\bos{u}^{I}\d\bos{\c}_{I}
+ \d\bos{\p}^{I}\d \bos{F}_{I}\right).
}
Thus the BV bracket of the theory originates form the extended
Schouten-Nijenhuis bracket.
Now we may follow the usual steps, as in the previous sections.

We start from the following action functional,
\eqn\mif{
\bos{\CS}_{o}  = \int_{\S}d^{2}\th
\biggl( \bos{\c}_{I}D\bos{u}^{I}
+\bos{F}_{I}D\bos{\p}^{I}
+ \bos{\c}_{I}\bos{\p}^{I}\biggr).
}
The above action functional satisfies the BV master equation.
The BRST transformation laws can be obtained by
the odd Hamiltonian vector $\bos{\CQ}_{o}$ of $\bos{\CS}_{o}$;
\eqn\mig{
\bos{\CQ}_{o} = \left( D \bos{u}^{I} +
\bos{\p}^{I}\right)\Fr{\rd}{\rd \bos{u}^{I}}
+D \bos{\p}^{I} \Fr{\rd}{\rd \bos{u}^{I} }
+\left(D \bos{F}_{I} +\bos{\c}_{I}\right) \Fr{\rd}{\rd \bos{F}_{I}}
+D \bos{\c}_{I} \Fr{\rd}{\rd \bos{\c}_{I}}.
}

Now we consider the following general deformation preserving
the ghost number
\eqn\mih{
\eqalign{
\bos{\CS}_{\G} &=
 \bos{\CS}_o  +\int_{\S}d^{2}\th \bos{\w}^*(\G)\cr
&= \bos{\CS}_o  +\int_{\S}d^{2}\th
\biggl( 
{a}_{I}{}^{J}(\bos{u}^{L})\bos{\c}_{J}\bos{\p}^{I}
+ \Fr{1}{2}{\k}_{IJ}(\bos{u}^{L})\bos{\p}^{I}\bos{\p}^{J}
+\Fr{1}{2}{b}^{IJ}(\bos{u}^{L})\bos{\c}_{I}\bos{\c}_{J}
\biggr),
}
}
where $\G = A +K +\pi \in \O^{2}(X) \oplus \O^{1}(X, TX) \oplus
\O^{0}(X, \wedge^{2}TX)$ with suitable parity changes;
\eqn\mihx{
\eqalign{
A &=a_{I}{}^{J}(u^L)\c_I\p^J,\cr
K &=\Fr{1}{2} \k_{IJ}(u^L)\p^I\p^J,\cr
\pi&=\Fr{1}{2}b^{IJ}(u^L)\c_I\c_J.
}
}
The above deformation terms do not have any dependence on $\bos{F}_I$,
which simplifies our analysis a lots.\foot{Hence we are using $\bos{F}_I$
as auxiliary devices.}
The above action functional satisfies the BV master equation
if and only if
\eqn\mig{
\eqalign{
d K  +[A,K]_{S}&=0,\cr
d A +\Fr{1}{2}[A,A]_{S}+[\pi,K]_{S}&=0,\cr
d \pi +[A,\pi]_{S}&=0,\cr
[\pi,\pi]_{S}&=0,
}
}
where $d:= \p^I\Fr{\rd}{\rd u^I}$ is the (parity changed)
exterior derivative on $X$
and the bracket above denotes the Schouten-Nijenhuis
bracket together with the wedge product on forms.
The above equation can be summarized into a single
equation,
\eqn\mih{
d \G + \Fr{1}{2}[\G,\G]_{S} =0.
}
It seems reasonable to relate the moduli space defined by the above
equations to the moduli space of $N =(2,2)$ superconformal
field theory in two-dimensions.

Now we consider the special cases of $\bos{\CS}_{\g}$.
The C model\foot{We note that the relation between the
C model and the ABC model is analogous to the relation between
cohomological field theory and balanced cohomological
field theory \cite{DM}.}
can be obtained by setting $\bos{\p}^{I}=\bos{F}_{I}=0$
for all $I$
\eqn\mii{
\bos{S}^{C} = \int_{\S}d^{2}\th\;
\left(\bos{\c}_{I}D\bos{u}^{I} +
\Fr{1}{2}\bos{b}^{IJ}\bos{\c}_{I}\bos{\c}_{J}\right).
}
The A model in Sect.~$2.4.$ can be obtained by setting 
$\bos{\c}_{I}=\bos{F}_{I}=0$
for all $I$;
\eqn\mij{
\bos{S}^{A} = \Fr{1}{2}\int_{\S}d^{2}\th\;
\left(\bos{\k}_{IJ}\bos{\p}^{I}\bos{\p}^{J}\right).
}
We may also consider the open string version 
of the A model \cite{W4,FO,F1,FO3}.
It is worth mentioning that the BV master equation
does not tell anything on the possible boundary
conditions, since the master action functional does
not have a kinetic term before gauge fixing.

Next we turn to the B model.

\subsection{Extended B model}

The B model is more complicated. First of all $X$ must be
a Calabi-Yau manifold. We pick a complex structure on
$X$ and $\bos{\p}^{i}= \bos{F}_{\bar i}=0$ for all $i$.
Then $\bos{\CS}_{\g}$ in \mih\ becomes
\eqn\mik{
\eqalign{
\bos{S}^{B}_{\b} &= \bos{S}^{B}_{o} + \int_{\S}\bos{\w}^{*}(\b),\cr
}
}
where
\eqn\mil{
\eqalign{
\bos{S}^{B}_{o}&=\int_{\S}d^{2}\th
\biggl( 
\bos{\c}_{i}D\bos{u}^{i}
+\bos{\c}_{\bar i}D\bos{u}^{\bar i}
+ \bos{\c}_{\bar i}\bos{\p}^{\bar i}\biggr),\cr
}
}
and
\eqn\mim{
\eqalign{
\int_{\S}\bos{\w}^{*}(\b)&=\int_{\S}d^{2}\th
\biggl( \bos{a}_{\bar i}{}^{j}\bos{\c}_{j}\bos{\p}^{\bar i}
+ \Fr{1}{2}\bos{\k}_{\bar i\bar j}\bos{\p}^{\bar i}\bos{\p}^{\bar i}
+\Fr{1}{2}\bos{b}^{ij}\bos{\c}_{i}\bos{\c}_{j}
\biggr).
}
}
The authors of \cite{AKSZ}
showed that the standard B model action functional
can be obtained by a suitable gauge fixing of the
master action functional $\bos{S}^{B}_{o}$ given by \mil.\foot{
They actually start from the action functional $\bos{\CS}_{o}$ in
\mif. They regard the conditions $\bos{\p}^{i}= \bos{F}_{\bar i}=0$
as part of the gauge fixing.}
For us it suffices to mention that $H_{i}$, which
is the $1$-form component of $\bos{\c}_{i}$, is
replaced by
\eqn\hisr{
H_{i} = g_{i\bar i} * du^{\bar i} + \ldots.
}
We will use this later. We also take the
only non-vanishing components
of the superfields $\bos{\c}_{\bar i}$ to be the two-form parts.
Then the kinetic term 
$\int_\S d^2\th\; \bos{\c}_{\bar i} D\bos{u}^{\bar i}$
in \mil\ vanishes.
Thus the undeformed master action functional for the B model
is actually given by
\eqn\mill{
\eqalign{
\bos{S}^{B}_{o}&=\int_{\S}d^{2}\th
\biggl( 
\bos{\c}_{i}D\bos{u}^{i}
+ \bos{\c}_{\bar i}\bos{\p}^{\bar i}\biggr).\cr
}
}
We note that the conditions
$\bos{\p}^{i}= \bos{F}_{\bar i}=\c_{\bar i}= H_{\bar i}=0$
are equivalent to the B boundary conditions for the open membrane.

Now we regard the terms in \mim\ as deformations of the theory
preserving the ghost number symmetry.
The BV master equation reduces
to 
\eqn\minn{
\bar \rd \b + \Fr{1}{2}[\b,\b]_{T} =0,
}
where
$\b = \k +a +b  \in \O^{0,2}(X)\oplus \O^{0,1}(X, \CT X)\oplus
\O^{0,0}(X, \wedge^{2}\CT X)$. Note that $\b$ is defined
as \rob.
The bracket denotes a holomorphic version of the Schouten-Nijenhuis
bracket and we have the wedge product on forms. By the Tian-Todorov lemma
the bracket can be obtained from the holomorphic top-form on $X$,
etc. \cite{BCOV,BK}.

We may consider a more general deformation of the theory
\mil. A zero-dimensional observable is a function
of the scalar components of the superfields \mid.
For this we consider an arbitrary function $\a(u^{i}, u^{\bar
i}, \c_{i}, \p^{\bar i})$ of $(u^{I},\c_{i},\p^{\bar i})$ on
$\Pi\CT^{*}\Pi (\bar\CT X)$, which is an element of
$\oplus \O^{0,\bullet}(X, \wedge^{\bullet}\CT X)$.
Then $\a(x^{\m})$
is a BV BRST observable if and only if
$\bos{Q}_{o} \a = \p^{\bar i}\rd_{\bar i} \a=0$.
Thus $\bos{Q}_{o}$ acts on $\a$ as the $\bar\rd$ operators on $X$;
$\a$ is an element of $\oplus H^{0,\bullet}(X,
\wedge^{\bullet}\CT X)$.
We consider a basis $\{\a_{a}\}$ of
$\oplus H^{0,\bullet}(X, \wedge^{\bullet}\CT X)$.
Then we have the following family of B models
\eqn\mio{
\bos{S}^{B}(\{t^{a}\}) = \bos{S}^B_{o} + \sum t^{a}\int_{\S}d^{2}\th\;
\bos{\w}^{*}(\a_{a}).
}
The above action functional satisfies the BV master
equation if and only if
\eqn\mio{
\bar \rd(t^{a}\a_{a}) + \Fr{1}{2}\left[\sum t^{a}\a_{a},
\sum t^{b}\a_{b}\right]_{T} =0.
}
It follows that the BV master equation holds up to first order in
$t^{a}$. 
To go beyond the first order we consider a certain function $\a(t)$,
which is an element of $\oplus \O^{0,\bullet}(X, \wedge^{\bullet}\CT
X)$, having the
the following formal expansion:
\eqn\mip{
\a(t) = t^{a}\a_{a} + \sum_{n > 1}t^{a_{1}}\ldots
t^{a_{n}}\a_{a_{1}\ldots a_{n}}.
}
Then we consider the family of action functionals
$\bos{S}(t)$ defined by
\eqn\miq{
\bos{S}^{B}(t) = \bos{S}^{B}_{o} + \int_{\S}d^{2}\th\;\bos{\w}^{*}(\a(t)).
}
The above deformation is said to be well-defined if $\bos{S}(t)$
satisfies the BV master equation
\eqn\mir{
\left(\bos{S}^{B}(t),\bos{S}^{B}(t)\right)_{BV} =0.
}
Equivalently,
\eqn\mis{
\bos{Q}_{o}
\int_{\S}d^{2}\th\;\bos{\w}^{*}(\a(t)) +\Fr{1}{2}\left(
\int_{\S}d^{2}\th\;\bos{\w}^{*}(\a(t)),\int_{\S}d^{2}\th\;\bos{\w}^{*}(\a(t)
)\right)_{BV}=0.
}

The above condition is equivalent to the following MC equation,
\eqn\mis{
\bar \rd \a(t) + \Fr{1}{2}\left[\a(t),\a(t)\right]_{T} =0.
}
Thus the success of BV quantization is equivalent to
having solutions to the above MC equation in the form \mip.
We note that the BV BRST transformation laws for the deformed
action functional \miq\ are
\eqn\misr{
\eqalign{
\bos{Q}(t) \bos{u}^i &= D \bos{u}^i + \Fr{\rd \bos{\a}(t)}{\rd
\bos{\c}_i},\cr
\bos{Q}(t) \bos{u}^{\bar i}&= \bos{\p}^{\bar i},\cr
\bos{Q}(t) \bos{\p}^{\bar i}&=0,\cr
\bos{Q}(t) \bos{\c}_i&= D\bos{\c}_i + \Fr{\rd \bos{\a}(t)}{\rd
\bos{u}^i},\cr
\bos{Q}(t) \bos{\c}_{\bar i} & = \Fr{\rd \bos{\a}(t)}{\rd \bos{u}^{\bar
i}}.\cr
}
}
The condition that $\bos{Q}(t)^2 =0$ is equivalent to the master
equation \mis. 
The MC equation or master equation \mis\ is precisely
the equation for the extended moduli space of complex structures on $X$,
defined by \cite{BK}.
Here we derived the result as the consistency condition
for quantization, completing the original analysis of Witten in \cite{W3}.

We also obtained a family of B models, which is actually parametrized by
the extended moduli space. We call the resulting theory the
extended B model.  We note that the deformations of the B model
leading to the extended B model are deformations of the differential
or the BRST charge of the theory.
On the other hand the topological open membrane theory
with B boundary conditions in Sect.~$3.5.2$ deforms
the bracket as well by the bulk deformations.
We also showed that  the topological open membrane theory
with B boundary conditions contains all the information
of the extended B model of topological closed string theory.

Now we turn to the open string version
of the extended B model.

\subsubsection{Towards physical descriptions of
$A_{\infty}$-deformations\\ of $D^{b}Coh(X)$}

According to a conjecture of Kontsevich
the extended moduli space $\CM$ of complex
structures parameterizes the $A_{\infty}$
deformations of $D^{b}(Coh(X))$ \cite{K1, B1}.
In this section we sketch a program constructing
$D^{b}(Coh(X))$ and its $A_{\infty}$-deformations  by
path integral methods. For the A model side of
the story we refer to \cite{FO3} for a state of art
construction.
For this we consider the theory on a disc.

We begin by comparing the B model with the C model in Sect.~$2$.
The undeformed C model is given by
\eqn\mkjb{
\bos{S}^{C}_{o} = \int_{\S}d^{2}\th\;
\biggl( 
\bos{\c}_{I}D\bos{u}^{I}
\biggr)
}
while the undeformed B model is given by
\eqn\mkjc{
\eqalign{
\bos{S}^{B}_{o}&=\int_{\S}d^{2}\th
\biggl( 
\bos{\c}_{i}D\bos{u}^{i}
+ \bos{\c}_{\bar i}\bos{\p}^{\bar i}\biggr).\cr
}
}
In both cases the boundary condition, in order to satisfy
the BV master equation, should be
that $\bos{\c}_{I}(x)$ vanish along the direction tangent
to $\rd\S$ for $x\in \rd\S$. We note that
the B model is just the holomorphic version of the
C-model but with
one additional term compared to
the B model.  Recall that the boundary observables
of C model are given by arbitrary functions on $X$.
The algebra $\CO(X)$ of functions (cohomological
$1$-algebra of $X$) on $X$ is determined only
by the commutative and associative product.
As we explained earlier, the undeformed action
functional has only the kinetic term, accordingly.

For the B model above a boundary observable
is given by a function on $\Pi \bar\CT X$, i.e., $\oplus\O^{0,\bullet}(X)$.
The algebra $\CO(\Pi\bar \CT X)$ of functions on $\Pi \bar\CT X$
has a supercommutative, an associative product,
and a differential.
The undeformed B model action functional tells us
that the differential is the $\bar \rd$ operator.
This can be seen by the BRST transformation laws
restricted to the boundary;
\eqn\mkjk{
\eqalign{
\bos{Q}_{o} \bos{u}^{i} &= D \bos{u}^{i},\cr
\bos{Q}_{o} \bos{u}^{\bar i} &= \bos{\p}^{\bar i},\cr
\bos{Q}_{o}\bos{\p}^{\bar i} &= 0.
}
}
Then $\bos{Q}_{o}u^{i}=0$, $\bos{Q}_{o}u^{\bar i}=\p^{\bar i}$ and
$\bos{Q}_{o}\p^{\bar i}=0$, which defines the $\bar\rd$ operator.
Thus the algebra of classical observables is $\CO(\Pi \bar\CT X)$,
with a supercommutative and associative product (wedge product)
and a differential $\bar \rd$, which equip $\CO(\Pi\bar \CT X)$ with
a structure of dgLa without bracket.

Now we turn on a deformation preserving the ghost number symmetry.
For the C model we have
\eqn\mkjl{
\d\bos{S}^{C} =\int_{\S}d^{2}\th\; \bos{\w}^{*}(\pi),
}
where $\pi = \Fr{1}{2}b^{IJ}\c_{I}\c_{J}$ denotes a bi-vector on $X$.
The bivector is an element of the $2$nd Hochschild cohomology of $\CO(X)$,
i.e. $\pi \in H^{2}({\tt Hoch}(\CO(X)))$. Now the bulk deformation
above deforms the product in a non-commutative direction.
The master equation
\eqn\mkjm{
[\pi,\pi]_{S} =0,
}
implies that the deformed product is associative.

For the B model we have
\eqn\mkjn{
\d\bos{S}^{B} =\int_{\S}d^{2}\th\;\bos{\w}^{*} (a +b),
}
where 
\eqn\mkjo{
\eqalign{
a &= a_{\bar i}{}^{j}\c_{j}\p^{\bar i},\cr
b &=\Fr{1}{2}b^{ij}\c_{i}\c_{j}.\cr
}
}
Note that the deformation by $\Fr{1}{2}\k_{\bar i\bar j}\p^{\bar
i}\p^{\bar j}$ in \mim\ is not allowed due to the master equation.
As before it is not a bulk term. The two objects above correspond
to elements of the $2$nd  Hochschild cohomology of $\CO(\Pi \bar\CT X)$,
i.e. $a,b \in H^{2}({\tt Hoch}(\CO(\bar\CT X))$.
According
to the mathematical definition  $\Fr{1}{2}\k_{\bar i\bar j}\p^{\bar
i}\p^{\bar j}\in \O^{0,2}(X)$ also belongs to
$H^{2}({\tt Hoch}(\CO(\Pi\CT X)))$ \cite{K1}.
However it is physically unnatural.  In general
we identify $H^{*}({\tt Hoch}(\CO(\Pi\bar\CT X)))$ with
$\oplus_{p\geq 1, q} \O^{0,q}(X, \wedge^{p}\bar\CT X)$.
The meaning of a deformation by $a$ is obvious; it is a deformation
of the complex structure. Comparing with the C model we see that
the $b$ term in the bulk generates a non-commutative deformation
of the product. The master equation is
\eqn\mkjp{
\eqalign{
\bar \rd a + \Fr{1}{2}[a,a]_{T} =0,\cr
\bar\rd b + [a, b]_{T}=0,\cr
[b,b]_{T} =0.
}
}
The first equation implies that the deformation of the complex structure
is integrable. The last equation implies that the non-commutative
deformation is associative. The middle equation may be
viewed as a compatibility of the two deformations.
The mirror picture  for Abelian varieties
is discussed in \cite{F2} (see also \cite{B1} for a non-commutative
version of the variation of Hodge structures more for closed string
case).

Note that the deformation by $a$ changes the BRST transformation
laws restricted to the boundary as follows
\eqn\mkjq{
\eqalign{
\bos{Q}\bos{u}^{i} &= D \bos{u}^{i} + \bos{a}_{i}{}^{\bar
j}\bos{\p}_{\bar j},\cr
\bos{Q} \bos{u}^{\bar i} &= \bos{\p}^{\bar i},\cr
\bos{Q}\bos{\p}^{\bar i} &= 0.
}
}
Thus the boundary observables should be changed accordingly.
In the above case it is just the change of complex structure.
We may consider the general deformation
\eqn\mkir{
\d \bos{S}^{B} = \int_{\S}d^{2}\th \bos{\w}^{*}(\g(u^{i},u^{\bar i},
\c_{i},\p^{\bar i})
}
where $\g \in \oplus_{p\geq 1, q} \O^{0,q}(X, \wedge^{p}\CT X)
\equiv H^{*}({\tt Hoch}(\CO(\Pi\bar\CT X))$.
Then the boundary BRST transformation
laws are changed as follows
\eqn\mkjs{
\eqalign{
\bos{Q} \bos{u}^{i} &= D \bos{u}^{i} +\Fr{\rd \bos{\g}}{\rd
\bos{\c}_{i}}|_{\c_{i}=0},\cr
\bos{Q} \bos{u}^{\bar i} &= \bos{\p}^{\bar i},\cr
\bos{Q}\bos{\p}^{\bar i} &= 0.
}
}
Thus only the element in $\oplus_{p}\O^{0,p}(X, \CT X)$
affects the boundary BRST transformation laws.
Such elements correspond to the classical deformation
of complex structures. Thus, without loss of generality,
we say that the boundary observables are determined
by the $\bar\rd$ cohomology among functions on $\Pi \bar\CT X$.

Kontsevich's formality theorem implies that the dgLa on
$\oplus_{p, q} \O^{0,q}(X, \wedge^{p}\CT X)$ or, equivalently,
on $H^{*}({\tt Hoch}(\CO(\Pi\bar\CT X)))$ is quasi-isomorphic to
the Lie algebra of local Hochschild cochains
${\tt Hoch}(\CO(\Pi\bar\CT X))$.
According to \cite{B1}
this also implies that the extended
moduli space parameterizes $A_{\infty}$-deformations of
$D^{b}(Coh(X))$ \cite{B1}. The solutions of the MC equation
\eqn\mkjt{
\bar\rd \g + \Fr{1}{2}[\g,\g]_{T} =0,
}
also parameterizes the  consistent deformations of the B model,
through the solutions of BV master equation. The path integral
of the theory can be viewed as a morphism from
$H^{*}({\tt Hoch}(\CO(\Pi\bar\CT X)))$ to
${\tt Hoch}(\CO(\Pi\bar\CT X))$.
A path integral proof that it is a quasi-isomorphism
can be done as for the C model.
An explicit computation of the path integral will be useful
even for very simple manifolds. We expect that both the product
and differential of $\CO(\Pi\bar\CT X)$ deform.

Now we make some remarks on adding gauge fields.
We regard the target space $X$ as the D-brane world-volume,
which has gauge fields on it.
The background gauge field can be coupled to the theory
by adding the following term to the action
\eqn\mkju{
\oint_{\rd\S} d u^{I} A_{I}(u^{L}).
}
For the C model an arbitrary gauge field $A$ can be added without
destroying the master equation since $\bos{Q} \bos{u}^{I} =D \bos{u}^{I}$
on the boundary. Seiberg-Witten showed that the effective theory is governed
by a non-commutative gauge theory, at least for constant B field
\cite{SW}.
For the B model the situation is different,
since $\bos{Q}\bos{u}^{\bar i} = D\bos{u}^{\bar i} + \bos{\p}^{\bar i}$ on
the boundary.
Following Witten we consider
\eqn\mkjv{
\oint_{\rd\S}\left( d u^{I} A_{I}(u^{L}) - \p^{\bar i}F_{\bar i
j}(A)\r^{i}\right)
}
where $\r^{i}$ is the $1$-form component of $\bos{u}^{i}$.
Using $\bos{Q} \p^{\bar i} =0$ and $\bos{Q}\r^{i} = - d u^{i}$
we see that the above is $\bos{Q}$-invariant if the $(0,2)$ part
of the curvature of the connection $1$-form $A$ vanish.
In general we can couple connection of $U(N)$ (Hermitian)
holomorphic bundles $E$ on $X$.
Similar to the C model case
we expect that the D-brane world-volume effective theory
is governed by a non-commutative gauge theory due to
the bulk term 
\eqn\mkjw{
\Fr{1}{2}\int_{\S}d^{2}\th \bos{b}^{ij}\bos{\c}_{i}\bos{\c}_{j}
= \Fr{1}{2}\int_{\S}\left( b^{ij} H_{i}\wedge H_{j} + \ldots\right).
}
The above bulk term, after the gauge fixing \hisr, contains
\eqn\mkjx{
\int_{\S} \Fr{1}{2}b_{\bar i\bar j}du^{\bar i}\wedge
du^{\bar j} := \int_{\S} u^{*}(B^{0,2})
}
where $b_{\bar i\bar j}:= b^{ij}g_{i\bar i}g_{j\bar j}$ and $g_{i\bar j}$
is a Ricci-flat K\"{a}hler metric.
The presence of the above term implies, due to the gauge invariance,
that the $(0,2)$ part of the curvature $2$-form $F^{0,2}$ should be
replaced by $F^{0,2}+ B^{0,2}$ everywhere  in the effective
theory.

For $X$ a Calabi-Yau $3$-fold, without the bulk deformation
by $b$, Witten determined the effective theory to be the holomorphic
Chern-Simons theory,
\eqn\mkjy{
I_{HCS} = \int_{X}\o^{3,0}\wedge\tr \left(A\wedge\bar\rd A +
\Fr{2}{3}A\wedge A\wedge A\right).
}
Now by turning on the bulk deformation by $b$ we know that
the wedge product among elements of $\O^{0,\bullet}(X)$ 
must be deformed to a suitable
non-supercommutative product, say $\star$. We also expect
that the differential $\bar \rd$ will be deformed in general,
say to $\bar Q$.
Thus the effective theory would look like
\eqn\mkjz{
I = \int_{X}\o^{3,0}\wedge\tr\left(A\star\left(\bar Q A + B^{0,2}
\mathbb{I}\right)
+\Fr{2}{3}A\star A\star A\right),
}
which is more similar to the open string field theory \cite{W}.

We may consider arbitrary deformations and more general
boundary interactions by including first descendents
($1$-dimensional
observables) of observables.  Being inserted in cyclic order
in the boundary with punctures, the general correlation functions
are no-longer expected to show strict associativity, but will do so
only up to homotopy (See a general discussion in \cite{HM}).
Recall that all the boundary observables originate
from $\oplus H^{0,q}(X)$.
We also let those $1$-dimensional observables
take values in $(E_{i}^{\bullet})^{*}\otimes E_{i+1}^{\bullet}$,
i.e. $Ext(E_{i}^{\bullet}, E_{i+1}^{\bullet})$.
Those may be viewed as the $A_{\infty}$-category structure
on $D^{b}(Coh(X)$, whose compositions are given by
correlation functions.\foot{
It is rather correlation maps than correlation
functions, though we will stick to the original terminology}.
Being an $A_{\infty}$-category
(as the collection of all $A_{\infty}$ modules over the
$A_{\infty}$-algebra $(\CO(\Pi \bar\CT X),\bar \rd)$)
it satisfies a certain MC equation.
Now turning on all bulk deformations leads to certain deformations
of $D^{b}(Coh(X))$.  The BV master equation, via the
general structure of the BV Ward identity, may imply that
the deformed category is again an $A_{\infty}$ category.
Since the extended moduli space $\CM$ parameterizes all
consistent bulk deformations of the extended B-model,
we see that $\CM$ indeed parameterizes the $A_{\infty}$
deformations
of $D^{b}(Coh(X))$.

We expect that the above discussions can be made more precise
and explicit by carefully studying the extended B-model.
We also expect to have explicit results on correlation functions
at least for simple cases. The final effective theory will be
much more complicated than \mkjz, resembling the
$A_{\infty}$-version
of open string field theory \cite{GZ}.
In principle the path integral of the extended B-model can lead
to explicit answers for all higher compositions (string products).
Then it will be the first explicit construction of open string field
theory introduced in \cite{GZ}.

\subsubsection{Topological open $p$-branes and
generalized homological mirror conjectures}

We showed that there is a unified description of A, B and $C$ models
based on essentially the same geometrical structure of
the topological open membrane. Those models are
special limits (or projections or gauge fixing) of the same
(ABC) model.  More fundamentally we showed that topological open membrane
theory can describes and control A, B and C models of topological
closed string as different boundary theories.
The relation between A and B models
are much like different twistings of the same superconformal
field theory. Furthermore the topological open membrane
theory leads to further (bulk) deformations of the
mathematical or physical structure of the boundary
string theory.
Then it is natural to conjecture that
the homological mirror symmetry
can be generalized to the category of homotopy $2$-algebra.

We  may further generalize the topological membrane
theory the way we generalized topological strings.
The natural candidate is to use essentially the
same geometrical structure as the topological open $3$-brane.
The resulting theory will have a quite large internal symmetry group,
and many topological membrane theories in different limits.
Then we may use open membrane versions of the
resulting theories to state a
homotopy $2$-algebraic mirror conjecture.
Note that the mirror symmetry between the A and B models is a
remnant of the duality between type IIA and IIB strings.
It would be interesting to examine if the internal symmetry
of the generalized topological membrane theory is related
to duality in M theory.
In general we may consider topological open $p$-branes on $X$,
based on the $p$-th iterated superbundle
$$
M_p \rightarrow (M_{p-1}\rightarrow (M_{p-1}\rightarrow
\ldots\rightarrow (M_1\rightarrow X)\ldots).
$$ 
We recall that the total space $M_p$ has a degree $U=p$
symplectic structure $\o_p$ and the base space $M_{p-1}$
is a Lagrangian submanifold with respect to $\o_p$.
The space $M_p$ has various discrete symmetries preserving
$\o_p$. We may take different Lagrangian subspaces
and corresponding boundary conditions.
Then we may obtain different theories, which can be
physically equivalent in a suitable sense. We conjecture
that the homological mirror symmetry can be generalized
accordingly, at least for Calabi-Yau spaces $X$.
More specifically, we conjecture that for a given
homological mirror symmetry
we have a corresponding mirror symmetry in the
category of $p$-algebras. Then the mirror symmetry
is generalized to physical equivalences of all topological
open p-brane theories.

\section*{Acknowledgement}

I am grateful to Robbert Dijkgraaf,
Kenji Fukaya, Brian Greene, Christiaan Hofman,
Seungjoon Hyun, Daniel Kabat,
Calin Lazaroiu, Sangmin Lee,
Yong-Geun Oh, Jongwon Park, Richard Thomas, David Tong
and  Herman Verlinde for discussions or communications
in the various stage of this paper.
I am also grateful to Christiaan Hofman,
Calin Lazaroiu, Sangmin Lee
and Jongwon Park for collaborations in related subjects.
I am grateful to Christiaan Hofman for a proof reading
of this manuscript as well as for useful  suggestions.
Some results in this paper have been announced in the conference
``Symplectic geometry and mirror symmetry, 2000''  held in KIAS.
I am grateful for the organizers and KIAS for financial support and
hospitality.
This research is supported by DOE Grant \# DE-FG02-92ER40699.

\end{document}